\documentclass[11pt,a4paper]{article}
\usepackage{jheppub}
\usepackage[T1]{fontenc}
\usepackage{lmodern,microtype,booktabs,mathtools,amssymb,bm}
\usepackage{graphicx,xcolor,array,placeins,etoolbox,tikz}
\usetikzlibrary{arrows.meta,calc}
\makeatletter
\patchcmd{\maketitle}{\PackageWarningNoLine{\jname}{E-mails are missing.\MessageBreak Plese use \protect\emailAdd\space macro to provide e-mails.}}{}{}{}
\makeatother
\hypersetup{pdftitle={Exceptional Quivers: Massive phi21 Deformations of E Series Minimal Models},pdfauthor={},colorlinks=true,linkcolor=blue!45!black,citecolor=blue!45!black,urlcolor=blue!45!black}
\newcommand{\ket}[1]{\left\lvert #1\right\rangle}
\newcommand{\bra}[1]{\left\langle #1\right\rvert}
\newcommand{\Ec}{E_{\rm c}}
\newcommand{\Eop}{E_{\rm op}}
\newcommand{\tr}{\operatorname{tr}}
\newcommand{\ii}{\mathrm i}
\newcommand{\e}{\mathrm e}
\newcommand{\sinc}{\operatorname{sinc}}
\providecommand{\quivercircleloop}[5][particle loop]{%
  \begingroup
  \pgfmathsetmacro{\quiverLoopOffset}{#3+#5}%
  \pgfmathsetmacro{\quiverLoopStart}{#4+150}%
  \draw[#1]
    ($(#2)+(#4:\quiverLoopOffset)+(\quiverLoopStart:#3)$)
    arc[start angle=\quiverLoopStart,delta angle=-300,radius=#3];%
  \endgroup
}
\definecolor{exceptionalBlue}{HTML}{226390}
\definecolor{exceptionalInk}{HTML}{172E42}
\definecolor{exceptionalNode}{HTML}{F5F8FA}
\providecommand{\exceptionalreferencecircleloop}[5][particle loop]{%
  \begingroup
  \pgfmathsetmacro{\referenceLoopOffset}{#3+#5}%
  \pgfmathsetmacro{\referenceLoopStart}{#4+180-.34*180/pi}%
  \pgfmathsetmacro{\referenceLoopSweep}{-360+.68*180/pi}%
  \draw[#1] ($(#2)+(#4:\referenceLoopOffset)+(\referenceLoopStart:#3)$)
    arc[start angle=\referenceLoopStart,delta angle=\referenceLoopSweep,radius=#3];%
  \endgroup
}
\tikzset{
  exceptional reference body/.style={x=1cm,y=1cm,
    particle line/.style={draw=exceptionalBlue,line width=1.05pt},
    kink arrow/.style={particle line,-{Stealth[length=2.6mm,width=2.0mm]}},
    particle loop/.style={kink arrow},
    vacuum/.style={circle,draw=exceptionalInk,fill=exceptionalNode,
      text=exceptionalInk,line width=.85pt,minimum size=.44cm,inner sep=0pt},
    every node/.style={font=\small}},
  exceptional reference overview/.style={x=1cm,y=1cm,
    particle line/.style={draw=exceptionalBlue,line width=.85pt},
    kink arrow/.style={particle line,-{Stealth[length=2.1mm,width=1.55mm]}},
    particle loop/.style={kink arrow},
    vacuum/.style={circle,draw=exceptionalInk,fill=exceptionalNode,
      text=exceptionalInk,line width=.7pt,minimum size=.36cm,inner sep=0pt},
    every node/.style={font=\small}},
  exceptional body/.style={x=1cm,y=1cm,
    particle line/.style={draw=exceptionalBlue,line width=1.05pt},
    kink arrow/.style={particle line,-{Stealth[length=2.6mm,width=2.0mm]}},
    particle loop/.style={kink arrow},
    vacuum/.style={circle,fill=black,inner sep=1.8pt},
    every node/.style={font=\small}},
  exceptional overview/.style={x=1cm,y=1cm,
    particle line/.style={draw=exceptionalBlue,line width=.85pt},
    kink arrow/.style={particle line,-{Stealth[length=2.1mm,width=1.55mm]}},
    particle loop/.style={kink arrow},
    vacuum/.style={circle,fill=black,inner sep=1.65pt},
    every node/.style={font=\small}}
}
\preprint{USTC-ICTS/PCFT-26-95}

\title{Exceptional Massive Quivers
}

\author{Ting-Kai Chen$^{a}$}
\author{and Jin Chen$^{a,\,b,\,\dagger}$}

\affiliation{$^{a}$Department of Physics, Xiamen University, Xiamen, 361005, China}
\affiliation{$^{b}$Peng Huanwu Center for Fundamental Theory, Hefei, Anhui 230026, China}

\emailAdd{elysianchan13@gmail.com}
\emailAdd{zenofox@gmail.com}

\abstract{We study the ordered and disordered massive phases of the
$\phi_{2,1}$-deformations of the $E_6$, $E_7$ and $E_8$ Virasoro minimal
models using the truncated conformal space approach on a defect circle. 
We first explain how to obtain the quiver representations of
their kink and particle multiplets from the corresponding $A$-type deformed theories via generalized orbifoldings. We then reconstruct these quivers from
finite volume spectra and local operator form factors. In particular, we find invertible intertwiners between the deformed Hamiltonians of the ordered and disordered phases in $E_6$ and $E_8$, respectively. The two phases therefore have identical energy and momentum spectra on both the periodic and defect circles. Our form factor method nevertheless distinguishes the two phases by reconstructing their distinct massive quivers. This work extends the defect circle method introduced in our previous paper \href{https://arxiv.org/abs/2611.34129}{2611.34129} to the exceptional series. }

\begin{document}
\begin{NoHyper}
\note[$\dagger$]{Corresponding author.}
\end{NoHyper}
\maketitle
\pagenumbering{roman}
\clearpage
\pagenumbering{arabic}

\section{Introduction and summary}
\label{sec:introduction}

Topological defects provide a general framework for describing symmetries
and their constraints on quantum field theory~\cite{Gaiotto:2014kfa,McGreevy:2022oyu,Luo:2023ive,Shao:2023gho}.
In two dimensions, topological defect lines (TDLs) extend the notion of group symmetries
to fusion categories, in which objects are not necessarily invertible and their fusion can produce a direct sum of simple TDLs~\cite{Frohlich:2004ef,Bhardwaj:2017xup,Chang:2018iay}. When such a categorical symmetry is
spontaneously broken, its action connects particles above individual
vacua to kinks interpolating between different vacua~\cite{Chang:2018iay,Cordova:2024vsq,Cordova:2024iti}.
The resulting multiplet of particle and kink degenerate states is conveniently collected in a quiver: Its vertices
are the vacua, a loop is a particle, and an oriented edge is a kink.
An irreducible multiplet fixes a common mass for its kink and particle components, while
the dynamics determines which multiplets occur in the spectrum.
Related constraints on particle and soliton spectra in fermionic theories
are described by superstrip algebras~\cite{Chen:2025qub,Chen:2026wyn}.

The massive deformations of Virasoro minimal models provide a natural battlefield
in which both the kinematic constraints and dynamics can be studied. The massive theories deformed from diagonal $A$-type minimal model have long been known for their
integrable scattering matrix in terms of a restricted solid on solid (RSOS) description of the vacua and the kinks interpolating between them~\cite{Lassig:1990xy,Colomo:1991gw,Mussardo:2009ja}. However,
the vacuum structure and particle and kink spectra of massive deformations
of the exceptional $E$-type models remain much less understood. Their
noninvertible symmetries constrain the possible infrared phases, as
illustrated by the analysis of the $E_6$ model in~\cite{Chang:2018iay}, while the
multiplets realized by the IR dynamics still need to be determined.
Recent developments in generalized orbifolding provide a systematic
framework for relating the diagonal and exceptional theories by gauging
noninvertible symmetries~\cite{Diatlyk:2023fwf,Chen:2024ulc,Tanaka:2025qou}. This gives a route to the
exceptional massive quivers, starting from the quivers of the diagonal
models and transporting them through the generalized orbifold. On the
other hand, these predictions can be tested by a direct numerical study
of the deformed Hamiltonians. Recent
results for the bulk operator product coefficients of the exceptional
models make such Hamiltonian calculations possible~\cite{Nivesvivat:2025odb,Lhoste:2026qce}.

In this paper, we study the $\phi_{2,1}$-deformations of the
$(A_{10},E_6)$, $(A_{16},E_7)$ and $(A_{28},E_8)$ minimal models to fill this gap.
We first construct the kink and particle multiplets of these exceptional
models by applying generalized orbifolding to those of the corresponding
$A$-type theories. We then employ the truncated conformal space approach (TCSA)~\cite{Yurov:1989yu,James:2017cpc,Horvath:2022zwx} on a circle
connected by a preserved non-invertible defect along the time direction~\cite{Bajnok:2013waa,Chen:2026ngp}.
Such a defect can close the two different vacua of a
single kink, making it accessible as an individual finite volume state. We establish the relation between the eigenstates obtained from the deformed Hamiltonian, termed the \emph{momentum basis}, and the physical kink or particle states in a quiver multiplet, termed the \emph{circle basis}~\cite{Chen:2026ngp}. We further compute the form factors of local operators to measure, in each momentum state, the total probability of circle basis states with a given pair of vacua. Summing these probabilities over the momentum states of the multiplet gives the number of defect circle states associated with each pair of vacua.

For the exceptional models, a given defect circle state can have more than one
close junction if the supported vacua can be connected by the defect with multiplicities greater than one. We determine the junction multiplicities by computing a connection matrix defined by
$C_{ab}=\bra{\Omega_a}y\ket{\Omega_b}$, where $y$ is the chosen defect
and $\ket{\Omega_a}$ are the reference vacua identified on the periodic
circle. Dividing the
probability sum by the corresponding nonzero multiplicity
will give the correct numbers of kinks and particles of the quiver. This step extends the reconstruction method
of~\cite{Chen:2026ngp} to vacuum modules with nontrivial junction
multiplicities. For $E_6$ and $E_8$, we also find unitary intertwiners between the
deformed Hamiltonians at opposite deformation couplings $\lambda$, both on the periodic
circle and on the defect circle.
They give a one-to-one correspondence between states with the same
energy and momentum, so the two phases are \emph{isospectral}, but have different massive quivers from each other. We summarize the reconstructed
quivers of all six phases in Table~\ref{tab:exceptional-summary}.

\begin{table}[!t]
\centering
\begingroup
\small
\setlength{\tabcolsep}{5pt}
\renewcommand{\arraystretch}{1.12}
\tikzset{
  spectrum vertex/.style={circle,fill=black,inner sep=1.35pt},
  spectrum arrow/.style={draw=exceptionalBlue,
    -{Stealth[length=1.45mm,width=1.05mm]},line width=.60pt}
}
\begin{tabular*}{\textwidth}{@{\extracolsep{\fill}}ccccc@{}}
\toprule
Model & Phase & Vacua & Quiver &
\shortstack{Kinks, particles\\$(n_K,n_P)$} \\
\midrule
$E_6$ & \shortstack{Ordered\\$\lambda<0$} & 3 &
\begin{tikzpicture}[x=.65cm,y=.65cm,
  baseline={([yshift=-.5ex]current bounding box.center)}]
\path[use as bounding box] (-3.13,-0.85000) rectangle (3.13,0.85000);
\draw[spectrum arrow] (-1.86000,0.04500) .. controls (-1.53360,0.23500) and (-0.59640,0.23500) .. (-0.27000,0.04500);
\draw[spectrum arrow] (-0.27000,-0.04500) .. controls (-0.59640,-0.23500) and (-1.53360,-0.23500) .. (-1.86000,-0.04500);
\draw[spectrum arrow] (0.27000,0.04500) .. controls (0.59640,0.23500) and (1.53360,0.23500) .. (1.86000,0.04500);
\draw[spectrum arrow] (1.86000,-0.04500) .. controls (1.53360,-0.23500) and (0.59640,-0.23500) .. (0.27000,-0.04500);
\node[spectrum vertex] (v1) at (-2.13000,0.00000) {};
\node[spectrum vertex] (v3) at (0.00000,0.00000) {};
\quivercircleloop[spectrum arrow]{v3}{.23}{90}{.27}
\quivercircleloop[spectrum arrow]{v3}{.23}{-90}{.27}
\node[spectrum vertex] (v5) at (2.13000,0.00000) {};
\end{tikzpicture} & $(4,2)$ \\
\addlinespace[2pt]
 & \shortstack{Disordered\\$\lambda>0$} & 3 &
\begin{tikzpicture}[x=.65cm,y=.65cm,
  baseline={([yshift=-.5ex]current bounding box.center)}]
\path[use as bounding box] (-3.13,-1.16232) rectangle (3.13,1.27232);
\draw[spectrum arrow] (-0.79500,0.96732) .. controls (-0.46860,1.15732) and (0.46860,1.15732) .. (0.79500,0.96732);
\draw[spectrum arrow] (0.79500,0.87732) .. controls (0.46860,0.68732) and (-0.46860,0.68732) .. (-0.79500,0.87732);
\draw[spectrum arrow] (-0.89103,0.71099) .. controls (-0.56328,0.52332) and (-0.09468,-0.28832) .. (-0.09603,-0.66599);
\draw[spectrum arrow] (-0.17397,-0.71099) .. controls (-0.50172,-0.52332) and (-0.97032,0.28832) .. (-0.96897,0.66599);
\draw[spectrum arrow] (0.96897,0.66599) .. controls (0.97032,0.28832) and (0.50172,-0.52332) .. (0.17397,-0.71099);
\draw[spectrum arrow] (0.09603,-0.66599) .. controls (0.09468,-0.28832) and (0.56328,0.52332) .. (0.89103,0.71099);
\node[spectrum vertex] (v2) at (-1.06500,0.92232) {};
\quivercircleloop[spectrum arrow]{v2}{.23}{180}{.27}
\node[spectrum vertex] (v4) at (1.06500,0.92232) {};
\quivercircleloop[spectrum arrow]{v4}{.23}{0}{.27}
\node[spectrum vertex] (v6) at (0.00000,-0.92232) {};
\end{tikzpicture} & $(6,2)$ \\
\midrule
$E_7$ & \shortstack{Ordered\\$\lambda<0$} & 4 &
\begin{tikzpicture}[x=.65cm,y=.65cm,
  baseline={([yshift=-.5ex]current bounding box.center)}]
\path[use as bounding box] (-3.13,-1.23593) rectangle (3.13,1.34593);
\draw[spectrum arrow] (1.33367,0.82196) .. controls (1.17950,0.51355) and (0.42050,0.07534) .. (0.07633,0.09603);
\draw[spectrum arrow] (0.03133,0.17397) .. controls (0.18550,0.48238) and (0.94450,0.92059) .. (1.28867,0.89990);
\draw[spectrum arrow] (1.59000,0.72593) .. controls (1.78000,0.43821) and (1.78000,-0.43821) .. (1.59000,-0.72593);
\draw[spectrum arrow] (1.50000,-0.72593) .. controls (1.31000,-0.43821) and (1.31000,0.43821) .. (1.50000,0.72593);
\draw[spectrum arrow] (-0.45000,-0.04500) .. controls (-0.75120,-0.23500) and (-1.64880,-0.23500) .. (-1.95000,-0.04500);
\draw[spectrum arrow] (-1.95000,0.04500) .. controls (-1.64880,0.23500) and (-0.75120,0.23500) .. (-0.45000,0.04500);
\draw[spectrum arrow] (0.07633,-0.09603) .. controls (0.42050,-0.07534) and (1.17950,-0.51355) .. (1.33367,-0.82196);
\draw[spectrum arrow] (1.28867,-0.89990) .. controls (0.94450,-0.92059) and (0.18550,-0.48238) .. (0.03133,-0.17397);
\node[spectrum vertex] (v2) at (1.54500,0.99593) {};
\quivercircleloop[spectrum arrow]{v2}{.23}{0}{.27}
\node[spectrum vertex] (v4) at (-0.18000,0.00000) {};
\quivercircleloop[spectrum arrow]{v4}{.23}{94}{.27}
\node[spectrum vertex] (v6) at (-2.22000,0.00000) {};
\node[spectrum vertex] (v7) at (1.54500,-0.99593) {};
\end{tikzpicture} & $(8,2)$ \\
\addlinespace[2pt]
 & \shortstack{Disordered\\$\lambda>0$} & 3 &
\begin{tikzpicture}[x=.65cm,y=.65cm,
  baseline={([yshift=-.5ex]current bounding box.center)}]
\path[use as bounding box] (-3.13,-0.85000) rectangle (3.13,0.85000);
\draw[spectrum arrow] (-1.86000,0.04500) .. controls (-1.53360,0.23500) and (-0.59640,0.23500) .. (-0.27000,0.04500);
\draw[spectrum arrow] (-0.27000,-0.04500) .. controls (-0.59640,-0.23500) and (-1.53360,-0.23500) .. (-1.86000,-0.04500);
\draw[spectrum arrow] (0.27000,0.04500) .. controls (0.59640,0.23500) and (1.53360,0.23500) .. (1.86000,0.04500);
\draw[spectrum arrow] (1.86000,-0.04500) .. controls (1.53360,-0.23500) and (0.59640,-0.23500) .. (0.27000,-0.04500);
\node[spectrum vertex] (v1) at (-2.13000,0.00000) {};
\node[spectrum vertex] (v3) at (0.00000,0.00000) {};
\quivercircleloop[spectrum arrow]{v3}{.23}{90}{.27}
\quivercircleloop[spectrum arrow]{v3}{.23}{-90}{.27}
\node[spectrum vertex] (v5) at (2.13000,0.00000) {};
\quivercircleloop[spectrum arrow]{v5}{.23}{0}{.27}
\end{tikzpicture} & $(4,3)$ \\
\midrule
$E_8$ & \shortstack{Ordered\\$\lambda<0$} & 4 &
\begin{tikzpicture}[x=.65cm,y=.65cm,
  baseline={([yshift=-.5ex]current bounding box.center)}]
\path[use as bounding box] (-3.13,-0.85000) rectangle (3.13,0.85000);
\draw[spectrum arrow] (-2.43000,0.04500) .. controls (-2.19600,0.23500) and (-1.40400,0.23500) .. (-1.17000,0.04500);
\draw[spectrum arrow] (-1.17000,-0.04500) .. controls (-1.40400,-0.23500) and (-2.19600,-0.23500) .. (-2.43000,-0.04500);
\draw[spectrum arrow] (-0.63000,0.04500) .. controls (-0.39600,0.23500) and (0.39600,0.23500) .. (0.63000,0.04500);
\draw[spectrum arrow] (0.63000,-0.04500) .. controls (0.39600,-0.23500) and (-0.39600,-0.23500) .. (-0.63000,-0.04500);
\draw[spectrum arrow] (1.17000,0.04500) .. controls (1.40400,0.23500) and (2.19600,0.23500) .. (2.43000,0.04500);
\draw[spectrum arrow] (2.43000,-0.04500) .. controls (2.19600,-0.23500) and (1.40400,-0.23500) .. (1.17000,-0.04500);
\node[spectrum vertex] (v1) at (-2.70000,0.00000) {};
\node[spectrum vertex] (v3) at (-0.90000,0.00000) {};
\quivercircleloop[spectrum arrow]{v3}{.23}{90}{.27}
\quivercircleloop[spectrum arrow]{v3}{.23}{-90}{.27}
\node[spectrum vertex] (v5) at (0.90000,0.00000) {};
\quivercircleloop[spectrum arrow]{v5}{.23}{85}{.27}
\node[spectrum vertex] (v7) at (2.70000,0.00000) {};
\end{tikzpicture} & $(6,3)$ \\
\addlinespace[2pt]
 & \shortstack{Disordered\\$\lambda>0$} & 4 &
\begin{tikzpicture}[x=.65cm,y=.65cm,
  baseline={([yshift=-.5ex]current bounding box.center)}]
\path[use as bounding box] (-3.13,-1.23593) rectangle (3.13,1.34593);
\draw[spectrum arrow] (1.33367,0.82196) .. controls (1.17950,0.51355) and (0.42050,0.07534) .. (0.07633,0.09603);
\draw[spectrum arrow] (0.03133,0.17397) .. controls (0.18550,0.48238) and (0.94450,0.92059) .. (1.28867,0.89990);
\draw[spectrum arrow] (1.59000,0.72593) .. controls (1.78000,0.43821) and (1.78000,-0.43821) .. (1.59000,-0.72593);
\draw[spectrum arrow] (1.50000,-0.72593) .. controls (1.31000,-0.43821) and (1.31000,0.43821) .. (1.50000,0.72593);
\draw[spectrum arrow] (-0.45000,-0.04500) .. controls (-0.75120,-0.23500) and (-1.64880,-0.23500) .. (-1.95000,-0.04500);
\draw[spectrum arrow] (-1.95000,0.04500) .. controls (-1.64880,0.23500) and (-0.75120,0.23500) .. (-0.45000,0.04500);
\draw[spectrum arrow] (0.07633,-0.09603) .. controls (0.42050,-0.07534) and (1.17950,-0.51355) .. (1.33367,-0.82196);
\draw[spectrum arrow] (1.28867,-0.89990) .. controls (0.94450,-0.92059) and (0.18550,-0.48238) .. (0.03133,-0.17397);
\node[spectrum vertex] (v2) at (1.54500,0.99593) {};
\quivercircleloop[spectrum arrow]{v2}{.23}{0}{.27}
\node[spectrum vertex] (v4) at (-0.18000,0.00000) {};
\quivercircleloop[spectrum arrow]{v4}{.23}{94}{.27}
\node[spectrum vertex] (v6) at (-2.22000,0.00000) {};
\quivercircleloop[spectrum arrow]{v6}{.23}{180}{.27}
\node[spectrum vertex] (v8) at (1.54500,-0.99593) {};
\end{tikzpicture} & $(8,3)$ \\
\bottomrule
\end{tabular*}
\endgroup
\caption{Reconstructed quivers of the light multiplets in the
$\phi_{2,1}$-deformations of the exceptional minimal models.
Vertices denote vacua, oriented edges between distinct vertices
denote kinks and antikinks, and loops denote particles.
The sign of coupling constant $\lambda$ determines the different IR phases.}
\label{tab:exceptional-summary}
\end{table}

The paper is organized as follows: In Section~\ref{sec:orb} we establish the exceptional quivers from their corresponding massive $A$-type cousins, explain the defect junction multiplicities and construct the isospectral
intertwiner (Section~\ref{iso}). Section~\ref{sec:measurement} describes the TCSA
Hamiltonians, the local form factor measurements, as well as the connection matrices for junction multiplicities. In
Section~\ref{sec:results} we apply this procedure to both phases of
$E_6$, $E_7$ and $E_8$, displaying their spectra, and reconstructed quivers.
\section{Massive exceptional theories and their light quivers}
\label{sec:orb}

\subsection{From the diagonal theories to the exceptional quivers}

\begin{figure}[tbp]
\centering
\definecolor{ExceptionalRGWall}{RGB}{181,43,114}
\begin{tikzpicture}[
  >=Stealth, line width=0.8pt,
  theory/.style={inner sep=5pt,align=center},
  topological/.style={->,draw=green!45!black},
  rg/.style={->,draw=ExceptionalRGWall,line width=1pt},
  flow label/.style={font=\small,fill=white,inner sep=2pt,align=center}
]
  \node[theory] (Auv) at (-3.7,0) {$A$-type CFT};
  \node[theory] (Euv) at (3.7,0) {$E$-type CFT};
  \node[theory] (Air) at (-3.7,-3.7)
    {Massive $A$ theory\\(RSOS description)};
  \node[theory] (Eir) at (3.7,-3.7)
    {Massive $E$ theory\\(generalized orbifold)};

  \draw[topological] (-2.25,0.17) --
    node[above=4pt,font=\small] {Topological interface} (2.25,0.17);
  \draw[topological] (2.25,-0.17) -- (-2.25,-0.17);
  \draw[topological] (-1.65,-3.53) --
    node[above=4pt,font=\small] {Topological interface} (1.65,-3.53);
  \draw[topological] (1.65,-3.87) -- (-1.65,-3.87);

  \draw[rg] (Auv.south) -- (Air.north);
  \draw[rg] (Euv.south) -- (Eir.north);
  \node[flow label,anchor=east] at (-3.95,-1.85) {RG flow\\$\lambda\,\phi_{2,1}$};
  \node[flow label,anchor=west] at (3.95,-1.85) {RG flow\\$\lambda\,\phi_{2,1}$};
\end{tikzpicture}
\caption{Generalized orbifolding relates the $A$-type and $E$-type
models before and after the massive deformation. Each pair of green
arrows denotes the two orientations of a topological interface.
The vertical arrows denote the two RG flows. Perturbing fields and
couplings are matched across the interface.}
\label{fig:exceptional-orbifold-rg}
\end{figure}
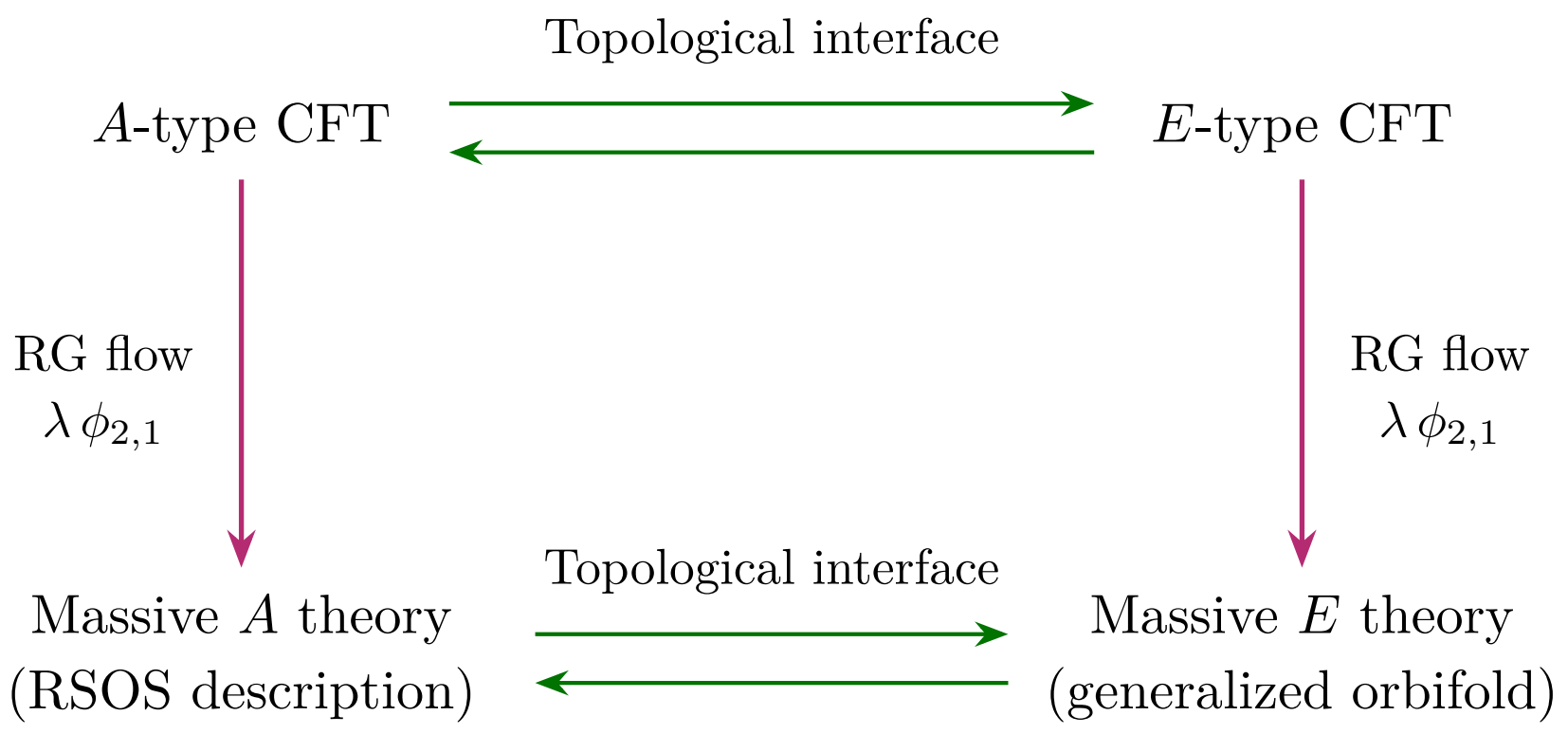

To identify the massive vacua and excitations of the three $\phi_{2,1}$-deformed unitary exceptional minimal models, we
start with the diagonal $A$-type theories of the same central charges and their $\phi_{2,1}$-deformations. Since the gauging algebra, also known as Frobenius algebras, from the $A$-type to $E$-type CFTs are preserved under the $\phi_{2,1}$-deformation, we have the commutative diagram in Figure~\ref{fig:exceptional-orbifold-rg}, in which the topological interface is defined with respect to the Frobenius algebra that we will explain below.

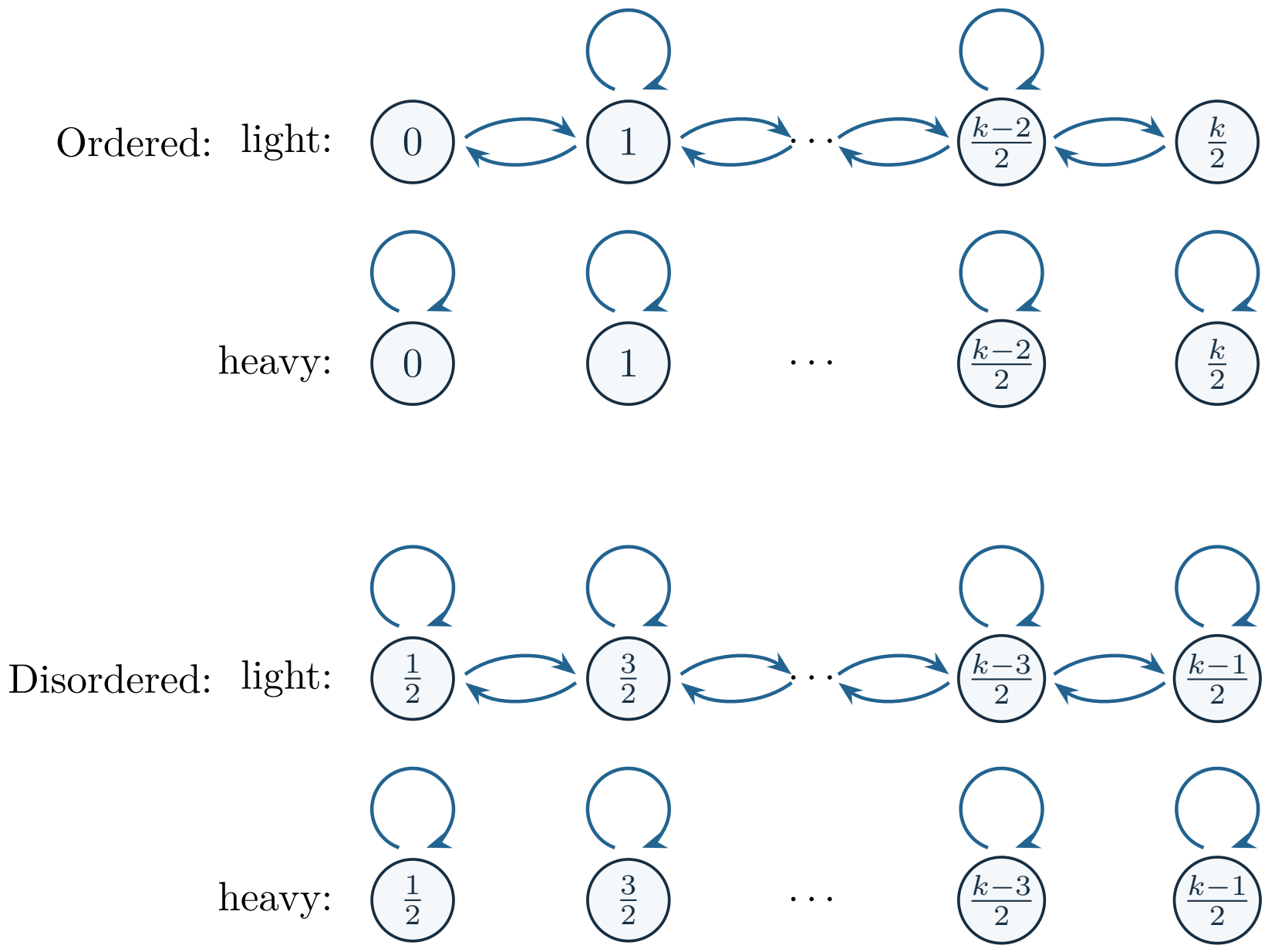
\begin{figure}[!htb]
\centering
\begin{tikzpicture}[exceptional reference overview,
  particle line/.style={draw=exceptionalBlue,line width=.85pt},
  vertex/.style={vacuum,minimum size=.70cm,inner sep=0pt}]
\node[anchor=east] at (-1.58,0) {Ordered:};
\node[anchor=east] at (-.55,0) {light:};
\node[anchor=east] at (-.55,-1.90) {heavy:};
\node[anchor=east] at (-1.58,-4.60) {Disordered:};
\node[anchor=east] at (-.55,-4.60) {light:};
\node[anchor=east] at (-.55,-6.50) {heavy:};
\foreach \prefix/\row in {a/0,c/-1.90}{
 \node[vertex] (\prefix0) at (0,\row) {$0$};
 \node[vertex] (\prefix1) at (1.85,\row) {$1$};
 \node (\prefix d) at (3.45,\row) {$\cdots$};
 \node[vertex] (\prefix2) at (5.05,\row) {$\frac{k-2}{2}$};
 \node[vertex] (\prefix3) at (6.90,\row) {$\frac{k}{2}$};
}
\foreach \prefix/\row in {b/-4.60,d/-6.50}{
 \node[vertex] (\prefix0) at (0,\row) {$\frac{1}{2}$};
 \node[vertex] (\prefix1) at (1.85,\row) {$\frac{3}{2}$};
 \node (\prefix d) at (3.45,\row) {$\cdots$};
 \node[vertex] (\prefix2) at (5.05,\row) {$\frac{k-3}{2}$};
 \node[vertex] (\prefix3) at (6.90,\row) {$\frac{k-1}{2}$};
}
\foreach \leftnode/\rightnode/\leftgap/\rightgap in {
  a0/a1/.45/.45,a1/ad/.45/.20,ad/a2/.20/.45,a2/a3/.45/.45,
  b0/b1/.45/.45,b1/bd/.45/.20,bd/b2/.20/.45,b2/b3/.45/.45}{
 \coordinate (pairleft) at ($(\leftnode)+(\leftgap,0)$);
 \coordinate (pairright) at ($(\rightnode)+(-\rightgap,0)$);
 \draw[kink arrow] ($(pairleft)+(0,.035)$)
 .. controls ($(pairleft)+(.26,.235)$) and ($(pairright)+(-.26,.235)$)
 .. ($(pairright)+(0,.035)$);
 \draw[kink arrow] ($(pairright)+(0,-.035)$)
 .. controls ($(pairright)+(-.26,-.235)$) and ($(pairleft)+(.26,-.235)$)
 .. ($(pairleft)+(0,-.035)$);
}
\foreach \loopnode in {a1,a2,b0,b1,b2,b3,c0,c1,c2,c3,d0,d1,d2,d3}{
 \exceptionalreferencecircleloop{\loopnode}{.35}{90}{.43}
}
\end{tikzpicture}
\caption{The light and heavy multiplets in the ordered and disordered
phases of the diagonal $A$-type theories with even $k=10,16,28$. The light quiver $Q_1^A$ contains kinks in both orientations between adjacent vacua, with one particle at each interior vacuum in the ordered phase and at every vacuum in the disordered phase. The heavy quiver $Q_0^A$ contains one particle at each vacuum.}
\label{fig:parent-quivers}
\end{figure}

First, for the $A$-type minimal model $\mathcal M_{m,m+1}$, we collect its central charges and primaries in Kac labels~\cite{Cappelli:1987xt},
\begin{align}
 c&=1-\frac{6}{m(m+1)}\,, &
 h_{r,s}&=\frac{\bigl((m+1)r-ms\bigr)^2-1}{4m(m+1)}\,,
 \label{eq:theory-kac}
\end{align}
where $1\leq r<m$, $1\leq s\leq m$ and
$(r,s)\sim(m-r,m+1-s)$. The $A$-type minimal models are perturbed by the diagonal operator $\phi_{2,1}$ with the action,
\begin{align}
 S_\lambda&=S_{\rm CFT}+\lambda\int\mathrm d^2x\,\phi_{2,1}(x)\,,
 &h_{2,1}=\bar h_{2,1}&=\frac{m+3}{4m}\,,
 \label{eq:theory-action}
\end{align}
where $\lambda$ is the coupling constant, whose sign leads the theories flowing to massive, but different IR phases. For the exceptional massive theories, we call the phase \emph{ordered} for $\lambda<0$, and \emph{disordered} for $\lambda>0$ as is defined in their $A$-type counterparts, where the ordered phase is corresponding to the RSOS models with integer heights, and disordered with half-integer heights, respectively~\cite{Mussardo:2009ja}. Since the exceptional minimal models correspond to $m=11,17,29$, we only consider $k=m-1$ even. In the $A$-type theory, we denote the line with
Kac label $(1,2j+1)$ by $X_j$, with
$j=0,\frac{1}{2},\ldots,\frac{k}{2}$. These labels obey the fusion rules of
$\mathrm{SU}(2)_k$. It is straightforward to show that the lines with integer $j$, collected as a fusion category $\mathcal C_k={\rm SO}(3)_k$, pass through the perturbing operator $\phi_{2,1}$ and remain topological and thus preserved along the RG flow~\cite{Chang:2018iay, Chen:2026ngp}. In the deep massive IR, the fusion category $\mathcal C_k$ is spontaneously broken and leads to multiple degenerate vacua. The vacua are labeled by the integer spin and half-integer ${\rm SU}(2)$ isospin for ordered and disordered phases, respectively,
\begin{align}
    \text{Ordered:}\  \mathcal M^A_{\rm ord.}= \left\{v_j\bigg|j=0,1,\dots\frac k2\right\}\,,\quad
    \text{Disordered:}\  \mathcal M^A_{\rm dis.}= \left\{v_j\bigg|j=\frac12,\frac32,\dots\frac {k-1}2\right\}\,.   
\end{align}
Above the vacua, each of the phases contains a light multiplet consisting of kinks and particles that can be organized in a connected quiver encoded in a quiver matrix $Q^A_1$,
\begin{align}
    \left(Q^A_1\right)_{j\,j'}=N_{1j}^{j'}\,,\quad
    \text{for $N_{1j}^{j'}$ satisfies}\quad
    X_j\otimes X_1=\sum_{j'}N_{1j}^{j'}\,X_{j'}=\bigoplus_{j'=|j-1|}^{\min{1+j,\,k-1-j}}X_{j'}\,,
    \label{eq:light_m}
\end{align}
where $N_{1j}^{j'}$ is simply the ${\rm SU}(2)$ fusion coefficients. Besides, the two phases also contain a heavy multiplet consisting of only particles as the bound states of the light multiplet in Eq.~\eqref{eq:light_m}. These particle excitations are described by the identity quiver
matrix $Q_0^A$, with one particle at each vacuum.
\begin{align}
    Q^A_0=\mathbf 1\,.
\end{align}
The two types of multiplets, denoted by $R_1$ and $R_0$, are also known as the quiver representations of the strip algebra $\mathbf{Str}(\mathcal C_k)$ explained in details in~\cite{Cordova:2024iti, Chen:2026ngp}. We summarize the quiver diagrams for the $\phi_{2,1}$-deformed $A$-type theory in both ordered and disordered phases in Figure~\ref{fig:parent-quivers}, where the vertices
are the vacua, a loop is a particle, and an oriented edge is a kink.

\enlargethispage{\baselineskip}

We now aim to transport these vacuum modules and their particle multiplets to
the exceptional theories. Let's first consider the generalized orbifolding from the diagonal minimal models to the exceptional ones. The topological interface and the Frobenius
algebra provide two equivalent descriptions of this generalized
orbifold construction~\cite{Fuchs:2002cm,Frohlich:2009gb,Diatlyk:2023fwf,Chen:2026byx}. Let $\mathcal I$ be the
gauging interface connected between the $A$-type and $E$-type CFTs,
and $\mathcal I^\vee$ denote its orientation reversal. Fusing them
back on the diagonal side define a Frobenius algebra $\mathcal A$ as
\begin{align}
 \mathcal A\simeq\mathcal I^\vee\circ\mathcal I\in\mathcal C_k\,,
 \label{eq:orbifold-interface-algebra}
\end{align}
Conversely, gauging this algebra on half of spacetime produces
$\mathcal I$ at its boundary~\cite{Frohlich:2009gb,Diatlyk:2023fwf}. For the three exceptional theories, the underlying objects of $\mathcal A$ are
listed in Table~\ref{tab:theory-models}, with the multiplication
junctions supplied by their exceptional algebra structures~\cite{Fuchs:2002cm,Coquereaux:2001di,Tanaka:2025qou}. Since every summand has integer spin and passes through $\phi_{2,1}$, the
same algebra can be gauged along the massive flow. We fix the gauge of the perturbing operator $\phi_{2,1}$ to make
its transmission phase through $\mathcal I$ trivial, so that the coupling
and the ordered or disordered branch are inherited from the $A$-type
theory.\footnote{Bulk fields admit independent sign choices, accompanied
by corresponding sign changes of their OPE coefficients~\cite{Lhoste:2026qce}.
Relative to the OPE convention of~\cite{Lhoste:2026qce}, our transmitted
$\phi_{2,1}$ has the opposite sign for $E_6$ and $E_8$ and the same sign
for $E_7$. The coupling is transformed accordingly so that the
perturbation $\lambda\,\phi_{2,1}$ remains unchanged.}

\begin{table}[tbp]
\centering\small
\begin{tabular}{@{}lcccl@{}}
\toprule
Invariant & $m$ & $k$ & $h_{2,1}$ & Frobenius algebra\\
\midrule
$(A_{10},E_6)$ & $11$ & $10$ & $\frac{7}{22}$ &
$X_0\oplus X_3$\\[3pt]
$(A_{16},E_7)$ & $17$ & $16$ & $\frac{5}{17}$ &
$X_0\oplus X_4\oplus X_8$\\[3pt]
$(A_{28},E_8)$ & $29$ & $28$ & $\frac{8}{29}$ &
$X_0\oplus X_5\oplus X_9\oplus X_{14}$\\
\bottomrule
\end{tabular}
\caption{The exceptional theories and the underlying Frobenius algebra objects
from the diagonal minimal models. All algebra
summands preserve the $\phi_{2,1}$ perturbation.}
\label{tab:theory-models}
\end{table}

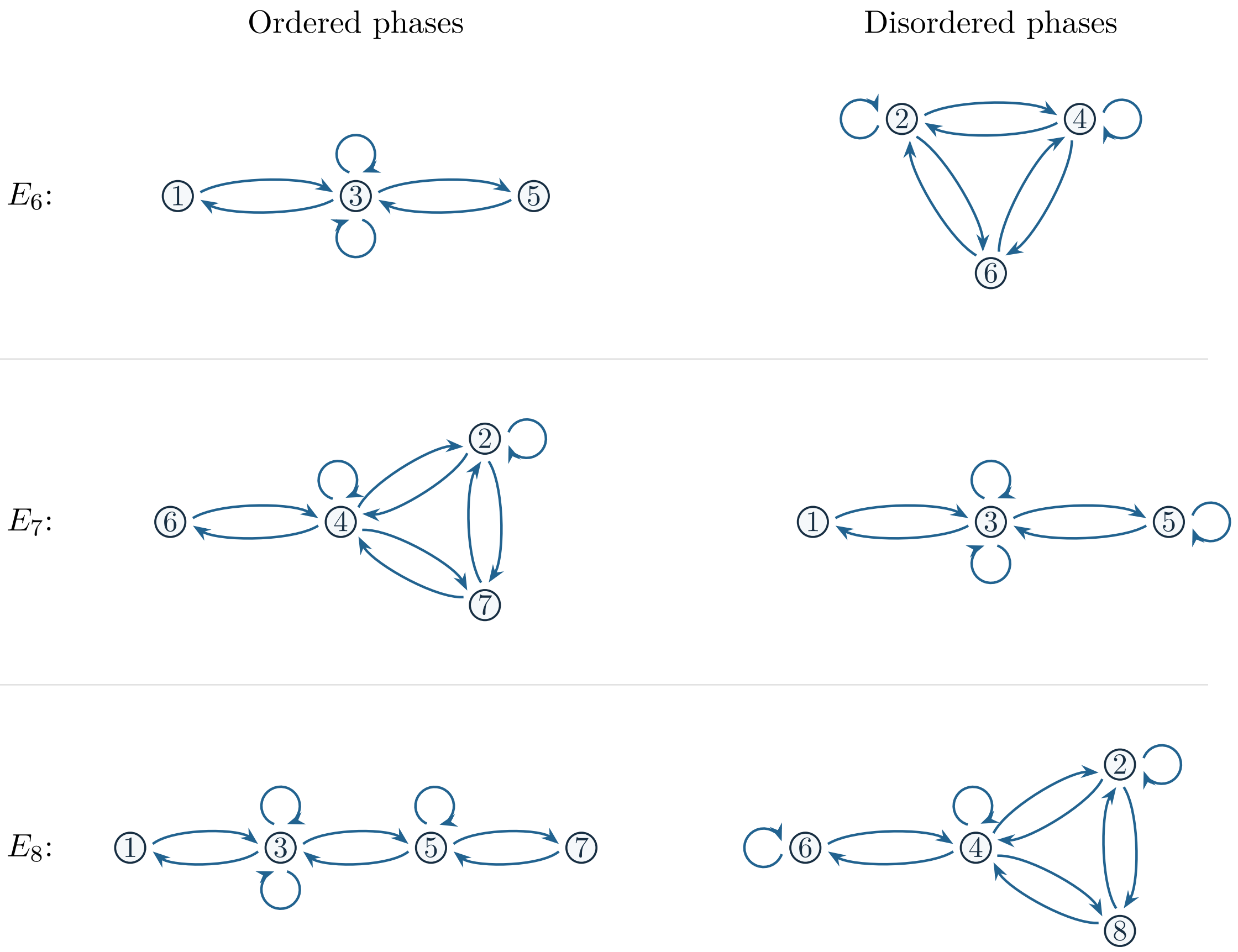
\begin{figure}[!htb]\centering
\begin{tikzpicture}[x=1cm,y=1cm]
\tikzset{exceptional reference overview/.append style={
  particle line/.style={draw=exceptionalBlue,line width=.85pt}}}
\node at (3.9,2.05) {Ordered phases};
\node at (11.5,2.05) {Disordered phases};
\node[anchor=center] at (0,-0.0) {$E_6$:};
\begin{scope}[shift={(3.9,-0.0)},exceptional reference overview]
\draw[kink arrow] (-1.86000,0.04500) .. controls (-1.53360,0.23500) and (-0.59640,0.23500) .. (-0.27000,0.04500);
\draw[kink arrow] (-0.27000,-0.04500) .. controls (-0.59640,-0.23500) and (-1.53360,-0.23500) .. (-1.86000,-0.04500);
\draw[kink arrow] (0.27000,0.04500) .. controls (0.59640,0.23500) and (1.53360,0.23500) .. (1.86000,0.04500);
\draw[kink arrow] (1.86000,-0.04500) .. controls (1.53360,-0.23500) and (0.59640,-0.23500) .. (0.27000,-0.04500);
\coordinate (v0) at (-2.13000,0.00000);
\coordinate (v1) at (0.00000,0.00000);
\exceptionalreferencecircleloop[particle loop]{v1}{0.23}{90}{0.27}
\exceptionalreferencecircleloop[particle loop]{v1}{0.23}{-90}{0.27}
\coordinate (v2) at (2.13000,0.00000);
\node[vacuum] at (-2.13000,0.00000) {$1$};
\node[vacuum] at (0.00000,0.00000) {$3$};
\node[vacuum] at (2.13000,0.00000) {$5$};
\end{scope}
\begin{scope}[shift={(11.5,-0.0)},exceptional reference overview]
\draw[kink arrow] (-0.79500,0.96732) .. controls (-0.46860,1.15732) and (0.46860,1.15732) .. (0.79500,0.96732);
\draw[kink arrow] (0.79500,0.87732) .. controls (0.46860,0.68732) and (-0.46860,0.68732) .. (-0.79500,0.87732);
\draw[kink arrow] (-0.89103,0.71099) .. controls (-0.56328,0.52332) and (-0.09468,-0.28832) .. (-0.09603,-0.66599);
\draw[kink arrow] (-0.17397,-0.71099) .. controls (-0.50172,-0.52332) and (-0.97032,0.28832) .. (-0.96897,0.66599);
\draw[kink arrow] (0.96897,0.66599) .. controls (0.97032,0.28832) and (0.50172,-0.52332) .. (0.17397,-0.71099);
\draw[kink arrow] (0.09603,-0.66599) .. controls (0.09468,-0.28832) and (0.56328,0.52332) .. (0.89103,0.71099);
\coordinate (v0) at (-1.06500,0.92232);
\exceptionalreferencecircleloop[particle loop]{v0}{0.23}{180}{0.27}
\coordinate (v1) at (1.06500,0.92232);
\exceptionalreferencecircleloop[particle loop]{v1}{0.23}{0}{0.27}
\coordinate (v2) at (0.00000,-0.92232);
\node[vacuum] at (-1.06500,0.92232) {$2$};
\node[vacuum] at (1.06500,0.92232) {$4$};
\node[vacuum] at (0.00000,-0.92232) {$6$};
\end{scope}
\draw[draw=black!15,line width=.4pt] (-.45,-1.95) -- (14.1,-1.95);
\node[anchor=center] at (0,-3.9) {$E_7$:};
\begin{scope}[shift={(3.9,-3.9)},exceptional reference overview]
\draw[kink arrow] (1.33367,0.82196) .. controls (1.17950,0.51355) and (0.42050,0.07534) .. (0.07633,0.09603);
\draw[kink arrow] (0.03133,0.17397) .. controls (0.18550,0.48238) and (0.94450,0.92059) .. (1.28867,0.89990);
\draw[kink arrow] (1.59000,0.72593) .. controls (1.78000,0.43821) and (1.78000,-0.43821) .. (1.59000,-0.72593);
\draw[kink arrow] (1.50000,-0.72593) .. controls (1.31000,-0.43821) and (1.31000,0.43821) .. (1.50000,0.72593);
\draw[kink arrow] (-0.45000,-0.04500) .. controls (-0.75120,-0.23500) and (-1.64880,-0.23500) .. (-1.95000,-0.04500);
\draw[kink arrow] (-1.95000,0.04500) .. controls (-1.64880,0.23500) and (-0.75120,0.23500) .. (-0.45000,0.04500);
\draw[kink arrow] (0.07633,-0.09603) .. controls (0.42050,-0.07534) and (1.17950,-0.51355) .. (1.33367,-0.82196);
\draw[kink arrow] (1.28867,-0.89990) .. controls (0.94450,-0.92059) and (0.18550,-0.48238) .. (0.03133,-0.17397);
\coordinate (v0) at (1.54500,0.99593);
\exceptionalreferencecircleloop[particle loop]{v0}{0.23}{0}{0.27}
\coordinate (v1) at (-0.18000,0.00000);
\exceptionalreferencecircleloop[particle loop]{v1}{0.23}{94}{0.27}
\coordinate (v2) at (-2.22000,0.00000);
\coordinate (v3) at (1.54500,-0.99593);
\node[vacuum] at (1.54500,0.99593) {$2$};
\node[vacuum] at (-0.18000,0.00000) {$4$};
\node[vacuum] at (-2.22000,0.00000) {$6$};
\node[vacuum] at (1.54500,-0.99593) {$7$};
\end{scope}
\begin{scope}[shift={(11.5,-3.9)},exceptional reference overview]
\draw[kink arrow] (-1.86000,0.04500) .. controls (-1.53360,0.23500) and (-0.59640,0.23500) .. (-0.27000,0.04500);
\draw[kink arrow] (-0.27000,-0.04500) .. controls (-0.59640,-0.23500) and (-1.53360,-0.23500) .. (-1.86000,-0.04500);
\draw[kink arrow] (0.27000,0.04500) .. controls (0.59640,0.23500) and (1.53360,0.23500) .. (1.86000,0.04500);
\draw[kink arrow] (1.86000,-0.04500) .. controls (1.53360,-0.23500) and (0.59640,-0.23500) .. (0.27000,-0.04500);
\coordinate (v0) at (-2.13000,0.00000);
\coordinate (v1) at (0.00000,0.00000);
\exceptionalreferencecircleloop[particle loop]{v1}{0.23}{90}{0.27}
\exceptionalreferencecircleloop[particle loop]{v1}{0.23}{-90}{0.27}
\coordinate (v2) at (2.13000,0.00000);
\exceptionalreferencecircleloop[particle loop]{v2}{0.23}{0}{0.27}
\node[vacuum] at (-2.13000,0.00000) {$1$};
\node[vacuum] at (0.00000,0.00000) {$3$};
\node[vacuum] at (2.13000,0.00000) {$5$};
\end{scope}
\draw[draw=black!15,line width=.4pt] (-.45,-5.85) -- (14.1,-5.85);
\node[anchor=center] at (0,-7.8) {$E_8$:};
\begin{scope}[shift={(3.9,-7.8)},exceptional reference overview]
\draw[kink arrow] (-2.43000,0.04500) .. controls (-2.19600,0.23500) and (-1.40400,0.23500) .. (-1.17000,0.04500);
\draw[kink arrow] (-1.17000,-0.04500) .. controls (-1.40400,-0.23500) and (-2.19600,-0.23500) .. (-2.43000,-0.04500);
\draw[kink arrow] (-0.63000,0.04500) .. controls (-0.39600,0.23500) and (0.39600,0.23500) .. (0.63000,0.04500);
\draw[kink arrow] (0.63000,-0.04500) .. controls (0.39600,-0.23500) and (-0.39600,-0.23500) .. (-0.63000,-0.04500);
\draw[kink arrow] (1.17000,0.04500) .. controls (1.40400,0.23500) and (2.19600,0.23500) .. (2.43000,0.04500);
\draw[kink arrow] (2.43000,-0.04500) .. controls (2.19600,-0.23500) and (1.40400,-0.23500) .. (1.17000,-0.04500);
\coordinate (v0) at (-2.70000,0.00000);
\coordinate (v1) at (-0.90000,0.00000);
\exceptionalreferencecircleloop[particle loop]{v1}{0.23}{90}{0.27}
\exceptionalreferencecircleloop[particle loop]{v1}{0.23}{-90}{0.27}
\coordinate (v2) at (0.90000,0.00000);
\exceptionalreferencecircleloop[particle loop]{v2}{0.23}{85}{0.27}
\coordinate (v3) at (2.70000,0.00000);
\node[vacuum] at (-2.70000,0.00000) {$1$};
\node[vacuum] at (-0.90000,0.00000) {$3$};
\node[vacuum] at (0.90000,0.00000) {$5$};
\node[vacuum] at (2.70000,0.00000) {$7$};
\end{scope}
\begin{scope}[shift={(11.5,-7.8)},exceptional reference overview]
\draw[kink arrow] (1.33367,0.82196) .. controls (1.17950,0.51355) and (0.42050,0.07534) .. (0.07633,0.09603);
\draw[kink arrow] (0.03133,0.17397) .. controls (0.18550,0.48238) and (0.94450,0.92059) .. (1.28867,0.89990);
\draw[kink arrow] (1.59000,0.72593) .. controls (1.78000,0.43821) and (1.78000,-0.43821) .. (1.59000,-0.72593);
\draw[kink arrow] (1.50000,-0.72593) .. controls (1.31000,-0.43821) and (1.31000,0.43821) .. (1.50000,0.72593);
\draw[kink arrow] (-0.45000,-0.04500) .. controls (-0.75120,-0.23500) and (-1.64880,-0.23500) .. (-1.95000,-0.04500);
\draw[kink arrow] (-1.95000,0.04500) .. controls (-1.64880,0.23500) and (-0.75120,0.23500) .. (-0.45000,0.04500);
\draw[kink arrow] (0.07633,-0.09603) .. controls (0.42050,-0.07534) and (1.17950,-0.51355) .. (1.33367,-0.82196);
\draw[kink arrow] (1.28867,-0.89990) .. controls (0.94450,-0.92059) and (0.18550,-0.48238) .. (0.03133,-0.17397);
\coordinate (v0) at (1.54500,0.99593);
\exceptionalreferencecircleloop[particle loop]{v0}{0.23}{0}{0.27}
\coordinate (v1) at (-0.18000,0.00000);
\exceptionalreferencecircleloop[particle loop]{v1}{0.23}{94}{0.27}
\coordinate (v2) at (-2.22000,0.00000);
\exceptionalreferencecircleloop[particle loop]{v2}{0.23}{180}{0.27}
\coordinate (v3) at (1.54500,-0.99593);
\node[vacuum] at (1.54500,0.99593) {$2$};
\node[vacuum] at (-0.18000,0.00000) {$4$};
\node[vacuum] at (-2.22000,0.00000) {$6$};
\node[vacuum] at (1.54500,-0.99593) {$8$};
\end{scope}
\end{tikzpicture}
\caption{The light multiplet quivers of the $\phi_{2,1}$-deformed massive exceptional theories in ordered and disordered phases.}
\label{fig:theory-quivers}\end{figure}

The vacua in the exceptional theories are then the simple left $\mathcal A$ modules in the parent
vacuum module category. Such a module consists of an object $w$ and an action
$\rho:\mathcal A\otimes w\to w$ compatible with the multiplication of
$\mathcal A$. Thus the two exceptional vacuum categories are
\begin{align}
 \mathcal M^E_{\rm ord.}
 &=\mathcal A\text{-}\operatorname{Mod}(\mathcal M^A_{\rm ord.})\,,&
 \mathcal M^E_{\rm dis.}
 &=\mathcal A\text{-}\operatorname{Mod}(\mathcal M^A_{\rm dis.})\,.
 \label{eq:orbifold-vacuum-modules}
\end{align}
To find their simple objects, we apply the interface to a parent vacuum
$v_j$. Its image is the induced left $\mathcal A$-module, defined by
\begin{align}
 F(v_j)\equiv\mathcal A\otimes v_j\,,
 \label{eq:orbifold-induction}
\end{align}
with $\mathcal A$ acting by multiplication on the first factor. An induced
module can split into several simple modules, each representing an
exceptional vacuum. All simple modules occur in this way. Their
decomposition is determined using Frobenius reciprocity~\cite{Fuchs:2002cm},
\begin{align}
 \operatorname{Hom}_{\mathcal A}\bigl(F(v_i),F(v_j)\bigr)
 &\simeq\operatorname{Hom}_{\mathcal M^A_\eta}
       \bigl(v_i,\mathcal A\otimes v_j\bigr)\,,
 \qquad\eta\in\{\mathrm{ord.},\mathrm{dis.}\}\,,
 \label{eq:orbifold-reciprocity}
\end{align}
where $\operatorname{Hom}_{\mathcal A}$ denotes maps that respect the $\mathcal A$ action.
A one dimensional endomorphism space identifies a simple module,
while the other Hom spaces determine its multiplicities in the
remaining induced modules. For example, consider $E_6$, with $\mathcal A=X_0\oplus X_3$. The underlying
parent object of $F(v_0)$ is $v_0\oplus v_3$, while
Eq.~\eqref{eq:orbifold-reciprocity} gives
$\dim\operatorname{End}_{\mathcal A}(F(v_0))=1$. It therefore represents a single
exceptional vacuum. In the ordered phase, $F(v_0)$, $F(v_1)$ and $F(v_5)$
are three distinct simple
modules, which we denote by $w_1,w_3,w_5$. The full decomposition is
\begin{align}
 \begin{array}{c|@{\quad}c@{\quad}c@{\quad}c@{\quad}c@{\quad}c@{\quad}c}
 j&0&1&2&3&4&5\\\hline
 F(v_j)&w_1&w_3&w_3\oplus w_5&w_1\oplus w_3&w_3&w_5
 \end{array}\,.
 \label{eq:orbifold-e6-ordered-induction}
\end{align}
The subscripts agree with the vacuum labels in Figure~\ref{fig:theory-quivers}. In
the disordered phase, $F(v_{\frac12})$ and $F(v_{\frac92})$ are simple
and give $w_2,w_4$. The module $F(v_{\frac32})$ has a two dimensional
endomorphism space and contains $w_4$ once. Its remaining simple
summand is $w_6$, giving
\begin{align}
 \begin{array}{c|@{\quad}c@{\quad}c@{\quad}c@{\quad}c@{\quad}c}
 j&\frac12&\frac32&\frac52&\frac72&\frac92\\[3pt]\hline
 F(v_j)&w_2&w_4\oplus w_6&w_2\oplus w_4&w_2\oplus w_6&w_4
 \end{array}\,.
 \label{eq:orbifold-e6-disordered-induction}
\end{align}
The two exceptional phases therefore each have three vacua, labelled
by $(w_1,w_3,w_5)$ and $(w_2,w_4,w_6)$, respectively.

Having obtained the nodes, we determine the arrows by transporting
the parent action of $X_1$. Recall that the Frobenius algebra $\mathcal A$ acts on a module from the
left, while the particle label $X_i$ acts from the right~\cite{Cordova:2024iti}. Associativity
therefore gives
\begin{align}
 F(v_j\otimes X_i)\simeq F(v_j)\otimes X_i\,,\qquad i=0,1\,.
 \label{eq:orbifold-particle-transport}
\end{align}
For $i=1$, this transported action $X_i$ is realized by a
symmetry defect $y$, in the exceptional theory, obtained by $\alpha$ induction of
$X_1$~\cite{Frohlich:2009gb,Chui:2002bp,Pearce:2024udz}. It is an $\mathcal A$ bimodule
with underlying object $\mathcal A\otimes X_1$. Its left $\mathcal A$
action is given by multiplication, while its right action uses the
chosen parent braiding to pass $X_1$ through $\mathcal A$. On a vacuum of the exceptional theory, it acts as
\begin{align}
 y\otimes_{\mathcal A}w_a\simeq w_a\otimes X_1\,,
 \label{eq:def_y}
\end{align}
where $\otimes_{\mathcal A}$ denotes fusion over $\mathcal A$.
We suppress this subscript below. Therefore the induced $y$-defect implements
the same action that determines the light quiver, in the exceptional theory, in terms of the simple modules $w_a$. Equivalently, in either phase, we have
\begin{align}
 w_a\otimes X_i
 &=\bigoplus_b (Q_i^E)_{ab}\,w_b\,,&
 (Q_i^E)_{ab}
 &=\dim\operatorname{Hom}_{\mathcal A}(w_a\otimes X_i,w_b)\,,
 \label{eq:orbifold-quiver-action}
\end{align}
where the row $a$ is the initial vacuum and the column $b$ is the final
vacuum. Again, for $E_6$, $y$-defect is the non-invertible generator of a half $E_6$
fusion category~\cite{Pearce:2024udz,Tanaka:2025qou}. In the ordered phase,
the middle vacuum provides a direct example of its action:
\begin{align}
 y\otimes w_3
 \simeq w_3\otimes X_1= F(v_1\otimes X_1)
 = F(v_0\oplus v_1\oplus v_2)
  = w_1\oplus2w_3\oplus w_5\,.
 \label{eq:orbifold-e6-middle-action}
\end{align}
It has one kink to each neighbouring vacuum and two particles.
One loop comes from the parent particle at $v_1$. The other comes
from the parent kink to $v_2$, since its induced endpoint contains
$w_3$. Together with $w_1\otimes X_1\simeq w_3$ and
$w_5\otimes X_1\simeq w_3$, this gives the ordered quiver.
For the disordered phase, the same calculation gives
$w_2\otimes X_1\simeq w_4\otimes X_1
\simeq w_2\oplus w_4\oplus w_6$ and
$w_6\otimes X_1\simeq w_2\oplus w_4$.
In the vacuum orders stated above, the two matrices are
\begin{align}
 \left.Q_1^{E_6}\right|_{\rm ord.}
 &=\begin{pmatrix}0&1&0\\1&2&1\\0&1&0\end{pmatrix}\,,&
 \left.Q_1^{E_6}\right|_{\rm dis.}
 &=\begin{pmatrix}1&1&1\\1&1&1\\1&1&0\end{pmatrix}\,.
 \label{eq:orbifold-e6-quivers}
\end{align}
The heavy family follows by setting $i=0$ in
Eq.~\eqref{eq:orbifold-quiver-action}. Since $X_0$ is the tensor unit,
$Q_0^E=\mathbf1$ in each phase, with one particle at every exceptional
vacuum.

Similarly, we obtain the exceptional vacua and quiver matrices for
$E_7$ and $E_8$. For $E_7$, the ordered and disordered vacua are
$(w_2,w_4,w_6,w_7)$ and $(w_1,w_3,w_5)$, respectively. In these
vacuum orders, the light quiver matrices are
\begin{align}
 \left.Q_1^{E_7}\right|_{\rm ord.}
 &=\begin{pmatrix}1&1&0&1\\1&1&1&1\\0&1&0&0\\1&1&0&0\end{pmatrix}\,,&
 \left.Q_1^{E_7}\right|_{\rm dis.}
 &=\begin{pmatrix}0&1&0\\1&2&1\\0&1&1\end{pmatrix}\,.
 \label{eq:theory-e7-quivers}
\end{align}
For $E_8$, the ordered and disordered vacua are $(w_1,w_3,w_5,w_7)$
and $(w_2,w_4,w_6,w_8)$. In these orders, we find
\begin{align}
 \left.Q_1^{E_8}\right|_{\rm ord.}
 &=\begin{pmatrix}0&1&0&0\\1&2&1&0\\0&1&1&1\\0&0&1&0\end{pmatrix}\,,&
 \left.Q_1^{E_8}\right|_{\rm dis.}
 &=\begin{pmatrix}1&1&0&1\\1&1&1&1\\0&1&1&0\\1&1&0&0\end{pmatrix}\,.
 \label{eq:theory-e8-quivers}
\end{align}
Figure~\ref{fig:theory-quivers} summarizes the light quivers in both
phases of all three exceptional models.
The vacuum labels follow the $E_n$ Dynkin diagram, numbered by a chain
$1,\ldots,n-1$ with node $n$ attached to node $3$.
The integer and half integer parent vacua map to its two alternating
sets of nodes, with adjacent nodes belonging to different sets.

So far, we only discuss the preserved $y$-line and its associated kink and particle multiplet in terms of quiver representations. The full preserved TDLs in the deformed $E$-type models can be found
directly in each exceptional CFT by requiring that they pass through
the perturbing field~\cite{Chang:2018iay,Coquereaux:2001di,Tanaka:2025qou}. Equivalently, one can also gauge
$\mathcal A$ inside the preserved parent category $\mathcal C_k$.
The resulting symmetry category is
\begin{align}
 \mathcal C^E
 \simeq\operatorname{Bimod}_{\mathcal C_k}(\mathcal A)\,,
 \label{eq:orbifold-preserved-category}
\end{align}
where an $\mathcal A$ bimodule carries compatible left and right
$\mathcal A$ actions, describing the attachment of the gauging network
on both sides of a defect~\cite{Fuchs:2002cm,Frohlich:2009gb,Diatlyk:2023fwf}. The two constructions
give the same categories, with $6$, $9$ and $16$ simple TDLs for
$E_6$, $E_7$ and $E_8$, respectively. Their spontaneous symmetry breaking patterns are specified by the
action on the massive vacua~\cite{Thorngren:2019iar,Huang:2021zvu,Bhardwaj:2023idu,Cordova:2024iti}. For example, the six
preserved $E_6$ lines act on three vacua in either phase. Restricting
this action to half $E_6$ gives its regular module in the ordered phase
and a different module with three vacua in the disordered phase. Appendix~\ref{app:preserved-symmetry} collects the preserved categories,
fusion rules for the subcategories generated by the chosen defect $y$,
and their realization in the two massive phases.

\subsection{Junction multiplicities}
\label{sec:closures}

In later defect TCSA setup, We will place the induced defect $y$ to close a circle, with its line
extending in time direction. A kink $k_{ab}$ from vacuum $w_a$ to $w_b$
can then return through the defect from $w_b$ to $w_a$. The number of defect junctions that close $k_{ab}$ is encoded in the \emph{junction multiplicity} $j_{ab}$,
read directly from the fusion
\begin{align}
 y\otimes w_b=\bigoplus_a j_{ab}\,w_a\,,\qquad j_a\equiv j_{aa}\,.
 \label{eq:theory-junctions}
\end{align}
Thus $j_a$ counts the junctions for a particle at vacuum $w_a$, while
$j_{ab}$ with $a\ne b$ counts those for a kink from $w_a$ to $w_b$.
Comparing with Eq.~\eqref{eq:orbifold-quiver-action}, we have
\begin{align}
 j_{ab}=(Q_1^E)_{ba}\,,\qquad j_a=(Q_1^E)_{aa}\,.
 \label{eq:theory-junction-degree}
\end{align}
For a given multiplet with quiver matrix $Q$, each of its $Q_{ab}$
excitations from $w_a$ to $w_b$ can use any of these $j_{ab}$
junctions. The total number of corresponding states on the defect circle is therefore
\begin{align}
 D_{ab}=Q_{ab}\times j_{ab}\,.
 \label{eq:closed-factorization}
\end{align}
In particular, $Q_{aa}$ number of particles at $w_a$ measured on the defect circle will give $Q_{aa}j_a$ states. In contrast, on the periodic circle, the total number of particle states is simply $\sum_a Q_{aa}$. In addition, without an excitation, the vacuum is the same on both sides of the defect, and $w_a$ gives $j_a$ defect vacuum states. So the total number of defect vacua is $\sum_a j_a$. Overall, for the Hilbert space on a defect circle, we have
\begin{align}
    &\#_{\rm vac.}=\sum_a j_a={\rm Tr}\,Q\,,
    &\#_{\text{1-pt states}}=\sum_{a,b}Q_{ab}\,j_{ab}={\rm Tr}\,Q^2
\end{align}
For example, in the ordered phase of $E_6$,
\begin{align}
 y\otimes w_3=w_1\oplus2w_3\oplus w_5\,.
 \label{eq:e6-ordered-junction-example}
\end{align}
The two particles at $w_3$ each have two junctions, giving four
particle states on the defect circle. Each of the four oriented
kinks has one junction, so the light multiplet gives eight states.
In disordered $E_6$, the junction multiplicities at $(w_2,w_4,w_6)$
are $(1,1,0)$. Its two particles and six kinks again give eight
defect states. Both phases have two defect vacua. Similarly, for $E_8$, the ordered and disordered junction multiplicities are
$(0,2,1,0)$ and $(1,1,1,0)$ in the vacuum orders given in Eq.~\eqref{eq:theory-e8-quivers}.
The ordered phase has five particle states and six kink states on
the defect circle, while the disordered phase has three and eight.
Both phases therefore have three defect vacua and eleven light states.
Table~\ref{tab:theory-counts} summarizes these counts, together with
the $E_7$ results.

\begin{table}[tbp]
\centering\small
\setlength{\tabcolsep}{5pt}
\begin{tabular}{@{}llcccccc@{}}
\toprule
&&\multicolumn{2}{c}{Periodic circle}
 &\multicolumn{4}{c}{$y$ defect circle}\\
\cmidrule(lr){3-4}\cmidrule(l){5-8}
Model & Phase & Vacua & Particles & Vacua & Particles & Kinks & Total\\
\midrule
$E_6$ & Ordered    & 3 & 2 & 2 & 4 & 4 & 8\\
      & Disordered & 3 & 2 & 2 & 2 & 6 & 8\\
$E_7$ & Ordered    & 4 & 2 & 2 & 2 & 8 & 10\\
      & Disordered & 3 & 3 & 3 & 5 & 4 & 9\\
$E_8$ & Ordered    & 4 & 3 & 3 & 5 & 6 & 11\\
      & Disordered & 4 & 3 & 3 & 3 & 8 & 11\\
\bottomrule
\end{tabular}
\caption{Vacuum and light excitation counts on the periodic and
$y$ defect circles. Particle and kink counts on the defect circle
include junction multiplicities, and kinks are counted with
orientation. The last column counts all light states, excluding vacua.}
\label{tab:theory-counts}
\end{table}

\FloatBarrier

\subsection{Isospectrality in \texorpdfstring{$E_6$ and $E_8$}{E6 and E8}}
\label{iso}

It is interesting to notice that the ordered and disordered phases of $E_6$ and $E_8$ have exactly the same numbers of vacuum and light multiplet states in both periodic
and defect spectra, despite their different kink and particle quivers.
The heavy family $Q_0^E=\mathbf1$ also has matching multiplicities:
one particle per periodic vacuum and $j_a$ states per vacuum $w_a$
on the defect circle. These matching multiplicities actually can be extended to an exact isomorphism between the full ordered and disordered Hamiltonian spectra.

We now aim to show this \emph{isospectrality} by constructing
a grading of the conformal Hilbert space. First, write
$\mathcal V_{r,p}$ for a Virasoro module, and let $M_{pq}$ denote the
exceptional modular invariant pairing its left label $p$ with the
right label $q$, so that the Hilbert spaces of the $E$-type minimal models spell
\begin{align}
    \mathcal H_{\rm per.}=\bigoplus_{p,q,r}M_{pq}\mathcal V_{r,p}\otimes\overline{\mathcal V}_{r,q}\,.
    \label{eq:ord_hilbert}
\end{align}
Further, since the $y$-defect is induced from $X_1$, the $y$-defect Hilbert
space can be shown as~\cite{Coquereaux:2001di,Chui:2002bp}
\begin{align}
 \mathcal H_y
 &=\bigoplus_{r,p,q,t}M_{pq}N_{3q}^{t}\,
 \mathcal V_{r,p}\otimes\overline{\mathcal V}_{r,t}=\bigoplus_s \mathcal H_y^{(s)}\,,
 \label{eq:theory-hilbert}
\end{align}
where $N_{3q}^{t}\equiv (N_{A_1})^{(t-1)/2}_{1,(q-1)/2}$ is the $\mathrm{SU}(2)_{k}$ fusion coefficient $N_{A_1}$ for isospin $1$, $\frac{q-1}2$ and $\frac{t-1}2$. In the second equality, one can also decompose the full defect Hilbert space in terms of subspace $\mathcal H_y^{(s)}$ of states respect to specific spin $s\pmod 1$. It is useful to count momentum states $\ket{\xi}$ with fractional momentum $P_\xi\propto s\pmod 1$ in defect TCSA measurement in Section~\ref{sec:results}. According to Eq.~\eqref{eq:theory-hilbert}, the states in the defect Hilbert space \eqref{eq:theory-hilbert} can be labeled by $|r,p,q,t;\rho\rangle$ or $|r,p,q,t;\rho\rangle\vert_{t=q}$ for the ordinary Hilbert space \eqref{eq:ord_hilbert}, where $\rho$ is the descendant labels.
We define the dimensionless perturbing operator by averaging the plane field over the unit circle,
\[
 V=\frac{1}{2\pi}\int_0^{2\pi}\mathrm d\theta\,\phi_{2,1}(e^{i\theta},e^{-i\theta})\,.
\]
The nonzero matrix elements of $V$ are constrained by the fusion rule
\begin{align}
    (2,1)\times(r,p,q,t)=(r+1,p,q,t)\oplus(r-1,p,q,t)\,,
    \label{eq:fusion21}
\end{align}
and thus we have
\begin{align}
    \langle r^\prime,p',q',t',\rho^\prime|V|r,p,q,t;\rho\rangle\neq 0
    \qquad\Longrightarrow\qquad
    r'=r\pm1\,,\quad (p',q',t')=(p,q,t)\,.
    \label{eq:selectionrule}
\end{align}
Bearing this in mind, we define a $\mathbb Z_2$-involution operator $\mathcal J$ on the states,
\begin{align}
   \mathcal J|r,p,q,t;\rho\rangle=
   \begin{cases}
       (-1)^r|r,p,q,t;\rho\rangle\,,\quad {\rm for}\ \ p<\frac{m+1}{2}\\[5pt]
       -(-1)^r|r,p,q,t;\rho\rangle\,,\quad {\rm for}\ \ p>\frac{m+1}{2}
   \end{cases}\,.
    \label{eq:involution}
\end{align}
One can easily check that $\mathcal J$ is well-defined respect to the field identification $(r,p,q,t)\sim(m-r,m+1-p,m+1-q,m+1-t)$ for the $E_6$ and $E_8$ case, corresponding to $m=11$ and $m=29$, as the states with labels $(p,q,t)=(\frac{m+1}{2},\frac{m+1}{2},\frac{m+1}{2})$ has been projected out. However $(p,q,t)=(\frac{m+1}{2},\frac{m+1}{2},\frac{m+1}{2})=(9,9,9)$ labels states in $E_7$, so that $\mathcal J$ cannot be defined for the $E_7$ model. Now, applying the $\mathcal J$-involution \eqref{eq:involution} and Eq.\eqref{eq:selectionrule}, it is straightforward to show that
\begin{align}
    &\mathcal J^\dagger H_0\,\mathcal J=H_0\,,\quad
    &\mathcal J^\dagger V\,\mathcal J=-V\,,
\end{align}
for $E_6$ and $E_8$ models, where $H_0$ is the undeformed CFT Hamilton. Therefore, for the deformed Hamiltonian with coupling constant $\lambda$ on a circle of circumference $L$, we have
\begin{align}
    &H(L,\lambda)
 =\frac{2\pi}{L}\left(L_0+\bar L_0-\frac{c}{12}\right)
   +\lambda\, L\left(\frac{2\pi}{L}\right)^{2h_{2,1}} V\,,
 \notag\\[7pt]
 &\mathcal J^\dagger H(L,\lambda)\,\mathcal J=H(L,-\lambda)\,.
 \label{eq:sign-map} 
\end{align}
Moreover, $\mathcal J$ also commutes with momentum operator and with a cutoff on conformal
energy. Eq.~\eqref{eq:sign-map} therefore holds at every given common cutoff and relates levels at the same momentum for $\pm \lambda$ coupling, throughout
the spectrum. The existence of $\mathcal J$ for the deformed $E_6$ and $E_8$ thus explains the isospectrality between the ordered and disordered phases.

However, the different quiver structures between two massive phases can be observed by introducing additional local probes and computing their form factors. Suppose that a pair of states $|i,\mp\rangle$ are the eigenstates of the Hamiltonian in the ordered and disordered phases with a fixed $|\lambda|$, and are related to each other by $\mathcal J$, $\mathcal J|i,-\rangle=|i,+\rangle$. A given local
probe $\mathcal O$ then has matrix elements
\begin{align}
 \bra{i,-}\mathcal O\ket{j,-}
 &=\bra{i,+}\mathcal J^\dagger\mathcal O\,\mathcal J\ket{j,+}\,.
 \label{eq:theory-probe-map}
\end{align}
The $\mathcal J$ transformed operator generically differs from the original one. Measuring these probes in both phases
therefore supplies additional information that is absent from its energy and
momentum spectra, and resolves the different quivers in
Figure~\ref{fig:theory-quivers}.
\section{Reconstructing quivers from local form factors}
\label{sec:measurement}

In our previous work~\cite{Chen:2026ngp}, we reconstructed massive quivers by measuring the local vacua on the two sides of an excitation on a defect circle. Within the light multiplets studied there, each allowed pair of vacua $(a,b)$ specifies a unique circle basis state. In the exceptional theories, however, several particles or kinks can have the same pair of vacua, and each excitation can admit several junctions with the defect. We now extend the method to account for both multiplicities.

\subsection{Vacuum composition on the defect circle}
\label{subsec:measurement-composition}

Consider a massive multiplet with quiver matrix $Q$ on a circle closed by the defect $y$. In Section~\ref{sec:closures}, the number of circle basis states with vacuum pair $(a,b)$ is $D_{ab}=Q_{ab}j_{ab}$, see Eq.~\eqref{eq:closed-factorization}. We choose an orthonormal basis $\ket{ab;\mu}$ for these states, with $\mu=1,\ldots,D_{ab}$. The projector onto this subspace is
\begin{align}
 \Pi_{ab}=\sum_{\mu=1}^{D_{ab}}\ket{ab;\mu}\bra{ab;\mu}\,.
 \label{eq:measurement-pair-projector}
\end{align}
As in~\cite{Chen:2026ngp}, we expand the momentum eigenstates in the circle basis and measure their squared amplitudes,
\begin{align}
 \ket{\xi}=\sum_{a,b}\sum_{\mu=1}^{D_{ab}}U_{ab;\mu,\xi}\ket{ab;\mu}\,,\qquad
 p_{ab,\xi}=\bra{\xi}\Pi_{ab}\ket{\xi}=\sum_{\mu=1}^{D_{ab}}|U_{ab;\mu,\xi}|^2\,.
 \label{eq:measurement-internal}
\end{align}
Here $\xi$ labels the momentum states of the chosen multiplet, where independent states within the same fractional momentum channel have to be counted separately. By computing form factors of local operators, we numerically measure
the probability $p_{ab,\xi}$ of observing the vacuum pair $(a,b)$
in each momentum state $\ket{\xi}$. Summing over the complete momentum basis states and using the unitarity of $U$, we find
\begin{align}
 \sum_\xi p_{ab,\xi}=\sum_{\mu=1}^{D_{ab}}\underbrace{\sum_\xi|U_{ab;\mu,\xi}|^2}_{=1}=\tr\Pi_{ab}=D_{ab}=Q_{ab} \, j_{ab}\,.
 \label{eq:measurement-ideal-trace}
\end{align}
Thus the trace of the vacuum-pair projector counts states on the defect circle, while the physical quiver counts excitations on the infinite line. To obtain the latter, we also need to determine the junction multiplicities. When $D_{ab}=1$, Eq.~\eqref{eq:measurement-internal} reduces to the squared amplitude $|U_{ab,\xi}|^2$ measured in~\cite{Chen:2026ngp}.

\subsection{Local probes and reference vacua}
\label{subsec:calibration}

We compute the finite volume spectra by diagonalizing the Hamiltonian of Eq.~\eqref{eq:sign-map}, with $\Delta_{21}=2h_{2,1}$, on the periodic circle and on the circle with a temporal $y$-defect. The primary matrix elements are obtained from the bulk OPE coefficients of~\cite{Lhoste:2026qce}. In the $y$-defect sector, we use
chiral $F$-moves to express these coefficients in the chosen
defect basis, including the corresponding normalization factors. We retain conformal states with $h+\bar h+n_{\rm L}+n_{\rm R}\leq\Ec$ and diagonalize the Hamiltonian on various fractional momentum, or say spin, sectors.
We denote the conformal energy cutoff of the periodic and defect probe spaces used for readout probabilities by $\Eop$, separately from the Hamiltonian cutoff $\Ec$.
The cutoff dependence of TCSA spectra has also been studied using
renormalization group methods~\cite{Giokas:2011ix}.

\paragraph{Local probe on an interval.} Let $\mathcal O$ be a real scalar field whose expectation values can distinguish the massive vacua. To measure the vacuum in a spatial region, we average the field over a small interval $I$ away from the defect,
\begin{align}
 \mathcal O_I=\frac{1}{|I|}\int_I dx\,\mathcal O_{\rm cyl}(x)\,.
 \label{eq:measurement-interval}
\end{align}
This average preserves a constant vacuum expectation value and suppresses contributions oscillating rapidly across the interval. Write the center and width of $I$ as $c_I L$ and $\ell_I L$, respectively. For a scalar primary of dimension $\Delta_{\mathcal O}$, the map from the plane to the cylinder supplies the factor $(2\pi/L)^{\Delta_{\mathcal O}}$, while translation along the circle supplies a phase determined by the spins $s_\alpha$ and $s_\beta$ of the conformal basis states. The matrix elements of Eq.~\eqref{eq:measurement-interval} are then
\begin{align}
 (\mathcal O_I)_{\alpha\beta}&=\left(\frac{2\pi}{L}\right)^{\Delta_{\mathcal O}}\bra{\alpha}\mathcal O(1,1)\ket{\beta}\frac{1}{\ell_I}\int_{c_I-\ell_I/2}^{c_I+\ell_I/2}du\,\e^{2\pi\ii(s_\alpha-s_\beta)u}\notag\\[5pt]
 &=\left(\frac{2\pi}{L}\right)^{\Delta_{\mathcal O}}\bra{\alpha}\mathcal O(1,1)\ket{\beta}\,\e^{2\pi\ii(s_\alpha-s_\beta)c_I}\sinc\bigl(\pi\ell_I(s_\alpha-s_\beta)\bigr)\,,
 \label{eq:measurement-probe}
\end{align}
where $u=x/L$ and $\sinc z=\tfrac{\sin z}{z}$. If $v_i$ and $v_j$ are the eigenstates of the truncated Hamiltonian, the corresponding finite volume form factor is $v_i^\dagger\mathcal O_I v_j$~\cite{Pozsgay:2007kn}.

\paragraph{Reference vacua and vacuum projectors.} We first identify the $n_{\rm vac}$ nearly degenerate lowest levels of the periodic spectrum, which span the vacuum subspace. We denote their energies by $\varepsilon_i$ and their orthonormal eigenstates by $\ket{e_i}$,
\begin{align}
 H_{\rm per}(L,\lambda)\ket{e_i}=\varepsilon_i\ket{e_i}\,,\qquad
 \langle e_i|e_j\rangle=\delta_{ij}\,,\qquad i,j=1,\ldots,n_{\rm vac}\,.
 \label{eq:measurement-periodic-vacua}
\end{align}
At finite volume these energy eigenstates can be linear combinations of the massive vacua. In the infinite volume limit, a local field has vanishing matrix elements between distinct vacua, and its diagonal entries give the vacuum expectation values. We therefore diagonalize $\mathcal O$ within the vacuum subspace,
\begin{align}
 (\mathcal O_{\rm vac})_{ij}&=\bra{e_i}\mathcal O_{\rm cyl}\ket{e_j}\,,\qquad \mathcal O_{\rm vac}u^{(a)}=\mu_a u^{(a)}\,,\qquad \ket{\Omega_a}=\sum_i u_i^{(a)}\ket{e_i}\,.
 \label{eq:measurement-reference-vacua}
\end{align}
Since the periodic vacuum states have zero momentum, the matrix elements of $\mathcal O$ in this subspace are independent of the position and width of the interval. We label the reference vacua $\ket{\Omega_a}$ in order of increasing $\mu_a$ and match them by overlaps when varying the cutoff. To compare different probes, we keep the reference vacua determined by the original operator and recalibrate the measurement thresholds for each probe. In what follows, we will briefly describe the calibration procedure and refer the reader to~\cite{Chen:2026ngp} for further details.

With the reference vacua fixed, we diagonalize $\mathcal O_I$
on the truncated periodic probe space. Let $\ket{r}$ be its orthonormal eigenstates with eigenvalues $g_r$. A reference vacuum generally produces a distribution of probe
readings. We assign an interval $\mathcal B_a$ of readings to each
vacuum $a$ and classify a reading in this interval as vacuum $a$.
Our aim is to choose these intervals to maximize the average
probability of correctly identifying the reference vacua.
We therefore define the readout distribution $P(g\mid\Omega_a)$,
the probability $A_{aa}$ of correctly identifying vacuum $a$,
and the mean identification accuracy $A$ as
\begin{align}
 P(g\mid\Omega_a)&=\sum_{r:\,g_r=g}|\langle r|\Omega_a\rangle|^2\,,& A_{aa}&=P(g\in\mathcal B_a\mid\Omega_a)\,,& A&=\frac{1}{n_{\rm vac}}\sum_a A_{aa}\,.
 \label{eq:measurement-accuracy}
\end{align}
For a theory with $n_{\rm vac}$ degenerate vacua, we choose $n_{\rm vac}-1$ ordered thresholds that partition the real line into intervals $\mathcal B_a$ and maximize the mean identification accuracy $A$. In a massive vacuum region, the local expectation value of the probe is insensitive to the distant boundary conditions. We therefore use the given intervals $\mathcal B_a$'s determined on the periodic circle to define the vacuum projectors on the defect circle,
\begin{align}
 P_a^I=\mathbf1_{\mathcal B_a}(\mathcal O_I)\equiv \sum_{r:\,g_r\in\mathcal B_a}\ket{r}\bra{r}\,,\qquad \sum_a P_a^I=\mathbf1\,.
 \label{eq:measurement-detectors}
\end{align}

\subsection{Junction multiplicities from the vacuum sector}
\label{subsec:measurement-junctions}

As discussed in Section~\ref{sec:closures}, the junction multiplicity $j_{ab}$ counts the ways in which $y$ closes a kink or particle with vacua $(a,b)$. We determine it by computing the action of the spatial line $y$ on the periodic vacua. First, for the nearly degenerate state $\ket{e_i}$ in Eq.~\eqref{eq:measurement-periodic-vacua} computed in TCSA, one can track back to identify the conformal sector contributing it. Since the perturbation preserves $y$-defect, we choose these states to diagonalize both the Hamiltonian and $y$. Their sector labels then determine the $y$-line eigenvalues,
\begin{align}
 y\ket{e_i}=d_i\ket{e_i}\,,\qquad
 d_i=\frac{\sin\!\left(\frac{3\pi q_i}{m+1}\right)}
 {\sin\!\left(\frac{\pi q_i}{m+1}\right)}\,,
 \label{eq:measurement-line-eigenvalues}
\end{align}
where $q_i$ is a right Kac label of the conformal basis states contributing to $\ket{e_i}$, and $m+1=12,18,30$ for $E_6,E_7,E_8$. Further using the coefficients $u_i^{(a)}$ in Eq.~\eqref{eq:measurement-reference-vacua}, we then can numerically compute this line action in the reference vacuum basis,
\begin{align}
 C_{ab}=\bra{\Omega_a}y\ket{\Omega_b}
 =\sum_i\bigl(u_i^{(a)}\bigr)^*d_i\,u_i^{(b)}
 \simeq j_{ab}\,,
 \label{eq:measurement-closure-matrix}
\end{align}
where we are free to choose appropriate phases of $\ket{\Omega_a}$ so that $C$ approaches the nonnegative integer action in Eq.~\eqref{eq:theory-junctions}~\cite{Huang:2021zvu}. Its entry $C_{ab}$ thus determines the number of junctions through which $y$ returns vacuum $b$ to vacuum $a$.

\subsection{Quiver reconstruction}
\label{subsec:measurement-endpoints}

For measurements on the defect circle, we place the $y$-defect at
$x=0$ and two probes near its two sides. The probes have widths
$\ell_{\rm L}L$ and $\ell_{\rm R}L$ and are centered at
$c_{\rm L}L$ and $c_{\rm R}L$, respectively, with
$0<c_{\rm L}\ll1$ and $0<1-c_{\rm R}\ll1$. When an excitation lies between the two probes and is sufficiently far from both, the probes read out the vacua on its left and right, respectively. In practice, we measure the left vacuum followed by the right, giving the joint probability of a momentum state,
\begin{align}
 p^{\rm L\to R}_{ab,\xi}=\|P_b^{I_{\rm R}}P_a^{I_{\rm L}}\ket{\xi}\|^2=\bra{\xi}P_a^{I_{\rm L}}P_b^{I_{\rm R}}P_a^{I_{\rm L}}\ket{\xi}\,,
 \label{eq:measurement-sequential}
\end{align}
where $P^{I_{\rm R,L}}_a$ are given in Eq.~\eqref{eq:measurement-detectors} with label ``L'' and ``R'' denoting the left and right vacua connected by the defect. Since the projectors need not commute after truncation, following~\cite{Chen:2026ngp}, we average the probabilities from the two orders of measurement,
\begin{align}
 (E_{ab})_{\xi\xi}=\frac12\left(p^{\rm L\to R}_{ab,\xi}+p^{\rm R\to L}_{ab,\xi}\right)=\frac12\bra{\xi}\bigl(P_a^{I_{\rm L}}P_b^{I_{\rm R}}P_a^{I_{\rm L}}+P_b^{I_{\rm R}}P_a^{I_{\rm L}}P_b^{I_{\rm R}}\bigr)\ket{\xi}\,.
 \label{eq:Eab}
\end{align}
$ (E_{ab})_{\xi\xi}$, which estimates $p_{ab,\xi}$ in Eq.~\eqref{eq:measurement-internal}, is the measured probability of reading vacuum $a$ on the left and vacuum $b$ on the right in the momentum state $\ket{\xi}$. Following Eq.~\eqref{eq:measurement-ideal-trace}, we sum over a complete orthonormal set of momentum states of the chosen multiplet to obtain
\begin{align}
 \widehat D_{ab}=\sum_\xi(E_{ab})_{\xi\xi}\,.
 \label{eq:measurement-endpoint}
\end{align}
This sum estimates the number $D_{ab}$ of circle basis states with vacuum pair $(a,b)$, including their junction multiplicities. For a complete set of vacuum assignments, Eq.~\eqref{eq:measurement-detectors} gives
\begin{align}
 \sum_{a,b}\widehat D_{ab}=\dim\mathcal H^{\rm IR}_{y}\big|_{\text{1-pt states}}\,,
 \label{eq:measurement-completeness}
\end{align}
where $\mathcal H^{\rm IR}_{y}\big|_{\text{1-pt states}}$ denotes the IR $y$-defect Hilbert space, restricted onto its one-particle states, along the RG flow from Eq.~\eqref{eq:theory-hilbert}. Overall, Eq.~\eqref{eq:measurement-ideal-trace},~\eqref{eq:measurement-closure-matrix} and \eqref{eq:measurement-endpoint}, we obtain the measured quiver entries
\begin{align}
 \widehat Q_{ab}=\frac{\widehat D_{ab}}{j_{ab}}\,,\quad{\rm for}\quad C_{ab}\gtrsim 1\,.
 \label{eq:measurement-quiver-count}
\end{align}
We have inferred the integer multiplicity $j_{ab}$ from the measured matrix element $C_{ab}$ from Eq.~\eqref{eq:measurement-closure-matrix}. For example, for a given pair of vacua $(a,b)$, $C_{ab}\simeq0$ indicates that the $y$ defect cannot close a kink or particle with these endpoint vacua. In the quiver reconstructed with this defect, we therefore set $\widehat Q_{ab}=0$, regardless of the measured value of $\widehat D_{ab}$. On the other hand, if $\widehat D_{aa}$ is close to four and $C_{aa}$ is close to two, this means that vacuum $a$ has two junctions with $y$-defect and supports two physical particle states.

\section{Reconstructing the exceptional quivers}
\label{sec:results}

In this section we apply the defect TCSA reviewed in Section~\ref{sec:measurement} to reconstruct the massive quivers of the $\phi_{2,1}$-deformed $E_6$, $E_7$, and $E_8$ theories. For each model we proceed in three steps. First, in the periodic spectrum we identify the massive vacua, distinguish them with a local scalar probe $\mathcal O$, and measure the spatial line matrix $C_{ab}\simeq j_{ab}$. Second, on the $y$-defect circle we collect the nearly degenerate momentum states $\ket{\xi}$ at the first excitation level, which encode the information of physical kink and particle states in the light multiplet. With properly chosen probes and their form factors, we measure their vacuum pair probabilities $(E_{ab})_{\xi\xi}$. Third, we sum $(E_{ab})_{\xi\xi}$ over these degenerate states $|\xi\rangle$ to obtain $\widehat D_{ab}$ and divide by $j_{ab}\simeq C_{ab}$ to read off the quiver matrix $\widehat{Q}$. In $E_6$ and $E_8$, Eq.~\eqref{eq:sign-map} in Section~\ref{iso} implies that the different signs of $\lambda$ give the same spectra on both the periodic and defect circles. Indeed, in defect TCSA, we precisely observe this phenomenon. Only by introducing additional local operator form factors can we distinguish the ordered and disordered phases. In contrast, for $E_7$, the two phases already differ in their spectra as we will see later.

Unless stated otherwise, we take $|\lambda|=0.1$ and truncate the Hamiltonian at $\Ec=20$. For a state with momentum $P_\xi=2\pi s_\xi/L$, we compute its rest mass from the gap $\Delta \varepsilon_\xi=\varepsilon_\xi-\varepsilon_0$ as
\begin{equation}
	M_\xi(L)=\sqrt{(\Delta \varepsilon_\xi)^2-\left(\frac{2\pi s_\xi}{L}\right)^2}\,,
	\label{eq:results-mass}
\end{equation}
where $\varepsilon_0$ is the lowest energy in the same geometry at circumference $L$. For all three models, we evaluate the vacuum readout and vacuum pair probabilities at $L=18$, using periodic and defect probe spaces with $\Eop=8$. We project the Hamiltonian states into these spaces and orthonormalize the projected vectors before constructing the reference vacua and evaluating the probe probabilities. The probe intervals have width $0.02L$ and are centered at $0.015L$ and $0.985L$. The reference vacua are labeled $0,1,\ldots$ in order of increasing probe eigenvalue, independently of the Dynkin labels in Figure~\ref{fig:theory-quivers}. The measured vacuum pair probabilities for each channel are listed in Appendix~\ref{app:numerics}.

\subsection{The two phases of \texorpdfstring{$E_6$}{E6}}
\label{subsec:e6-results}

\begin{figure}[!htbp]
	\centering
	\includegraphics[width=0.90\linewidth]{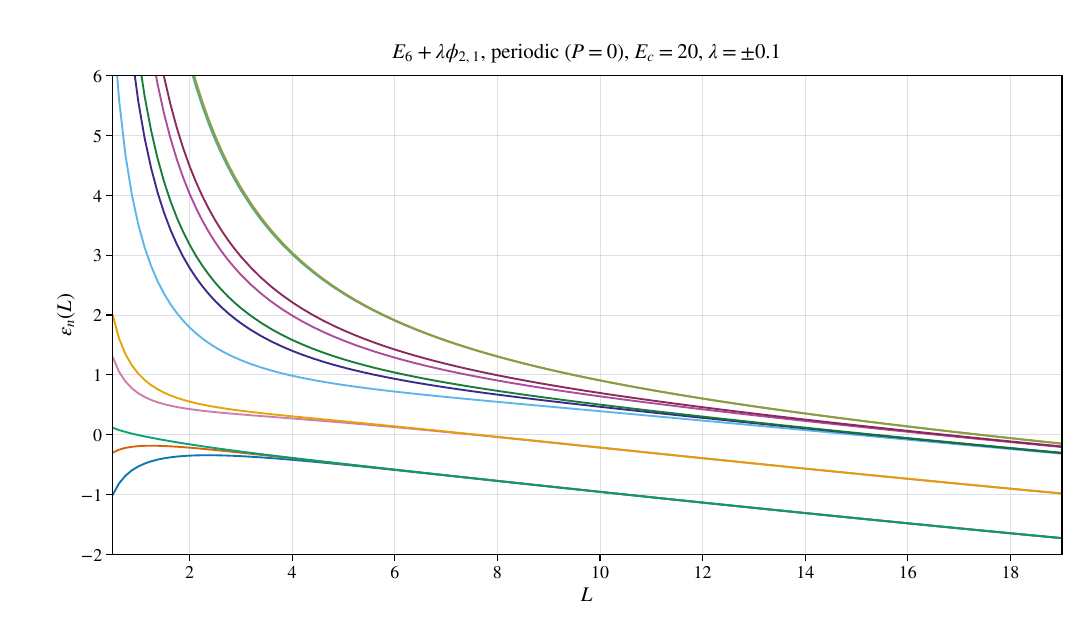}
	\par\medskip
	{\small
		\begin{tabular}{crrrr}\toprule
			$L$ & $\varepsilon_1-\varepsilon_0$ & $\varepsilon_2-\varepsilon_0$ & $\varepsilon_3-\varepsilon_0$ & $\varepsilon_4-\varepsilon_0$ \\ \midrule
			10 & 0.000443 & 0.000484 & 0.739127 & 0.740722 \\
			14 & 0.000517 & 0.000751 & 0.743073 & 0.743622 \\
			18 & 0.000949 & 0.001433 & 0.745662 & 0.746565 \\
			\bottomrule\end{tabular}
	}
	\caption{The zero momentum periodic energies $\varepsilon_n(L)$ of $E_6+\lambda\phi_{2,1}$ at $\Ec=20$ and $\lambda=\pm0.1$. The two signs give the same spectrum. We plot the twelve lowest levels.}
	\label{fig:e6-periodic}
\end{figure}

\paragraph{Periodic spectrum and vacuum labels.}
Consider first $E_6$. Figure~\ref{fig:e6-periodic} shows the periodic spectrum at momentum $P_\xi=0$ of the two phases and tabulates the lowest gaps $\varepsilon_n(L)-\varepsilon_0(L)$. At the two largest circumferences in the table, both $\varepsilon_1-\varepsilon_0$ and $\varepsilon_2-\varepsilon_0$ are less than one percent of $\varepsilon_3-\varepsilon_0$, so we identify the lowest three states as the massive vacua. The gaps $\varepsilon_3-\varepsilon_0$ and $\varepsilon_4-\varepsilon_0$ stay close and indicate two neutral particles at this mass in either phase.

To distinguish the three vacua, we follow the prescription in Section~\ref{subsec:calibration} to diagonalize the interval averaged form factor in Eq.~\eqref{eq:measurement-probe} for operator $\mathcal O=\phi_{4,4}$, with $h_{4,4}=\tfrac{5}{176}$, in the basis of the three lowest periodic states.
At $L=18$, we order these states $\ket{e_1},\ket{e_2},\ket{e_3}$ by their
$y$ eigenvalues $(d_1,d_2,d_3)=(1+\sqrt3,1-\sqrt3,0)$.
Choosing their phases so that the nonzero probe matrix elements are
real and positive, the vacuum matrices in the cylinder normalization of
Eq.~\eqref{eq:measurement-probe} are
\begin{align}
 \mathcal O_{\rm vac}^{\rm ord}&\simeq
 \begin{pmatrix}
 0 & 0 & 0.955154\\
 0 & 0 & 1.846086\\
 0.955154 & 1.846086 & 0
 \end{pmatrix}\,,&
 \mathcal O_{\rm vac}^{\rm dis}&\simeq
 \begin{pmatrix}
 0 & 0 & 0.955154\\
 0 & 0 & 0.493624\\
 0.955154 & 0.493624 & 0
 \end{pmatrix}\,.
 \label{eq:e6-vacuum-probe}
\end{align}
Diagonalizing these matrices gives
\begin{align}
 (\mu_0,\mu_1,\mu_2)_{\rm ord}\simeq (-2.078546,0,2.078546)\,,\notag\\[5pt]
 (\mu_0,\mu_1,\mu_2)_{\rm dis}\simeq (-1.075167,0,1.075167)\,.
 \label{eq:e6-vacuum-readings}
\end{align}
\begin{samepage}
The corresponding normalized eigenvectors, written as columns in the
same energy basis, are
\begin{align}
 \bigl(u^{(0)},u^{(1)},u^{(2)}\bigr)_{\rm ord}
 &\simeq\begin{pmatrix}
 0.324937 & 0.888162 & 0.324937\\
 0.628026 & -0.459530 & 0.628026\\
 -0.707107 & 0 & 0.707107
 \end{pmatrix}\,,\notag\\[6pt]
 \bigl(u^{(0)},u^{(1)},u^{(2)}\bigr)_{\rm dis}
 &\simeq\begin{pmatrix}
 0.628178 & 0.459114 & 0.628178\\
 0.324642 & -0.888378 & 0.324642\\
 -0.707107 & 0 & 0.707107
 \end{pmatrix}\,.
 \label{eq:e6-reference-vectors}
\end{align}
\end{samepage}
Each column defines a reference vacuum
$\ket{\Omega_a}=\sum_{i=1}^{3}u_i^{(a)}\ket{e_i}$, with
$a=0,1,2$ ordered by increasing $\mu_a$.
We now express the spatial line $y$ in this reference vacuum basis.
Its matrix is $y_{\rm vac}=C$, whose entries follow from
Eq.~\eqref{eq:measurement-closure-matrix} as
\begin{align}
 C_{ab}=\bra{\Omega_a}y\ket{\Omega_b}
 =\sum_{i=1}^{3}\bigl(u_i^{(a)}\bigr)^*d_i u_i^{(b)}\,.
 \label{eq:e6-vacuum-line}
\end{align}
Using the eigenvectors before rounding gives
\begin{align}
 C_{\rm ord}\simeq\begin{pmatrix}
 -0.000272 & 0.999728 & -0.000272\\
 0.999728 & 2.000544 & 0.999728\\
 -0.000272 & 0.999728 & -0.000272
 \end{pmatrix}\,,\quad
 C_{\rm dis}\simeq\begin{pmatrix}
 1.000935 & 0.999064 & 1.000935\\
 0.999064 & -0.001869 & 0.999064\\
 1.000935 & 0.999064 & 1.000935
 \end{pmatrix}\,.
 \label{eq:E6_C}
\end{align}
Rounding these entries to the nearest integers gives the junction matrices
\begin{align}
	j_{\rm ord}&=\begin{pmatrix}0&1&0\\1&2&1\\0&1&0\end{pmatrix}\,,&
	j_{\rm dis}&=\begin{pmatrix}1&1&1\\1&0&1\\1&1&1\end{pmatrix}\,.
	\label{eq:e6-closures}
\end{align}
In the ordered phase only vacuum $1$ can close onto itself through $y$, with two independent junctions, while in the disordered phase vacua $0$ and $2$ each admit a single self-junction. Both phases therefore have two defect vacua, but with different local vacuum content.
Using these reference vacua, we calibrate the probe thresholds by maximizing the mean identification accuracy in Eq.~\eqref{eq:measurement-accuracy}. This gives $98.18\%$ in the ordered phase and $98.03\%$ in the disordered phase.

\begin{figure}[!htbp]
	\centering
	\includegraphics[width=0.98\linewidth]{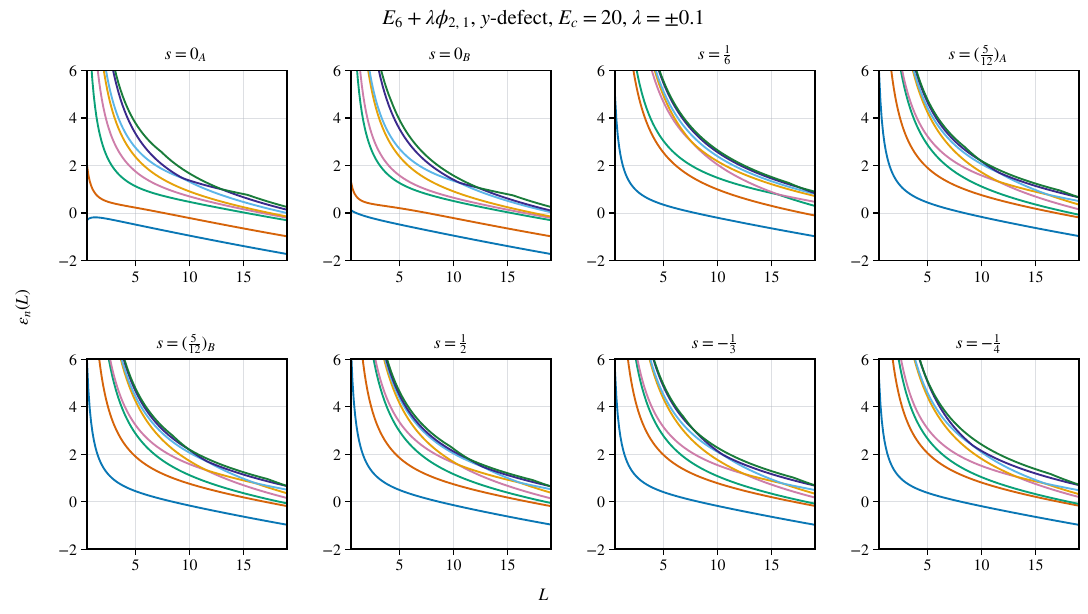}
	\caption{Defect spectrum of $E_6+\lambda\phi_{2,1}$ on the $y$ circle at $\lambda=\pm0.1$ and $\Ec=20$. The two signs give the same spectrum. From left to right, the top row shows $0_A$, $0_B$, $\frac{1}{6}$, $(\frac{5}{12})_A$ and the bottom row shows $(\frac{5}{12})_B$, $\frac{1}{2}$, $-\frac{1}{3}$, $-\frac{1}{4}$. The lowest levels in the first two panels are the defect vacua. We consider the first excited level in each zero momentum panel and the lowest level in each of the six fractional momentum panels.}
	\label{fig:e6-defect}
\end{figure}

\paragraph{The spectra on the defect circle.}
On the $y$-defect circle the two lowest degenerate states have zero momentum. Above them, eight light excitations appear in Figure~\ref{fig:e6-defect}, one in each of the channels $P_\xi=\frac{2\pi }{L}s_\xi$,
\begin{align}
s_\xi:\qquad	0_A,\quad 0_B,\quad \frac{1}{6},\quad
	\left(\frac{5}{12}\right)_A,\quad
	\left(\frac{5}{12}\right)_B,\quad
	\frac{1}{2},\quad -\frac{1}{3},\quad-\frac{1}{4}\,,
	\label{eq:e6-channels}
\end{align}
where, for $s_\xi=0,\frac{5}{12}$, the eigenvalue of the spatial $z$ line resolves the two channels, as given in Eq.~\eqref{eq:app-e6-channel-charges} of Appendix~\ref{app:B}. The lowest states in $0_A$ and $0_B$ are the two defect vacua. We consider the first excited state in each of the two zero momentum channels and the lowest state in each of the six fractional momentum channels in Figure~\ref{fig:e6-defect}. We measure their gaps from the same defect vacuum and obtain their rest masses using Eq.~\eqref{eq:results-mass}. At $L=18$ and $\Ec=20$, this gives masses $0.7456$, $0.7447$, $0.7451$, $0.7427$, $0.7427$, $0.7427$, $0.7432$ and $0.7447$ in the channel order above. The eight states thus have a common rest mass within the numerical resolution, in agreement with the particle masses on the periodic circle.

\paragraph{From vacuum pairs to quivers.}
We now measure the vacuum pair probabilities $(E_{ab})_{\xi\xi}$ of these eight states using the projectors in Section~\ref{subsec:measurement-endpoints}. The probabilities for each channel and their sums $\widehat D_{ab}$ are listed in Tables~\ref{tab:e6-ordered-raw} and~\ref{tab:e6-disordered-raw}. Combining $\widehat D_{ab}$ with the junction multiplicities of Eq.~\eqref{eq:e6-closures} via Eq.~\eqref{eq:measurement-quiver-count}, we set $\widehat Q_{ab}= \tfrac{\widehat{D}_{ab}}{j_{ab}}$ for $C_{ab}\gtrsim 1$ and display $\widehat D_{ab}$ itself whenever $C_{ab}\simeq 0$:
\begin{align}
	\widehat Q_{\rm ord}&=\begin{pmatrix}
		0.0608 & 0.9641 & 0.0045 \\
		0.9641 & 2.0066 & 0.9641 \\
		0.0045 & 0.9641 & 0.0608
	\end{pmatrix}\,,&
	\widehat Q_{\rm dis}&=\begin{pmatrix}
		1.0147 & 0.9923 & 0.9470 \\
		0.9923 & 0.1075 & 0.9923 \\
		0.9470 & 0.9923 & 1.0147
	\end{pmatrix}\,.
	\label{eq:e6-measured-Q}
\end{align}
All off-diagonal entries with $j_{ab}>0$ are close to one. On the diagonal terms, the ordered phase has $\widehat D_{11}=4.0131$, corresponding to four circle states with vacuum pair $(1,1)$. Divided by $j_{11}=2$ in Eq.~\eqref{eq:e6-closures}, we have $\widehat Q_{11}=2.0066$, implying two neutral particles above vacuum $1$. In the disordered phase, $\widehat Q_{00}$ and $\widehat Q_{22}$ are each close to one with a single self-junction, giving one particle above vacuum $0$ and one above vacuum $2$. Overall, the measured quiver matrices $\widehat Q$ in Eq.~\eqref{eq:e6-measured-Q} reproduce our orbifolding results \eqref{eq:orbifold-e6-quivers}, and thus reconstruct the following quiver diagrams,
\begin{align}
	Q_{\rm ord}=\begin{pmatrix}0&1&0\\1&2&1\\0&1&0\end{pmatrix}
	\quad\Longrightarrow\quad
	\vcenter{\hbox{%
			\begin{tikzpicture}[exceptional reference body,baseline={(current bounding box.center)}]
				\draw[kink arrow] (2.07400,-0.05500) .. controls (1.73808,-0.30000) and (0.67592,-0.30000) .. (0.34000,-0.05500);
				\draw[kink arrow] (0.34000,0.05500) .. controls (0.67592,0.30000) and (1.73808,0.30000) .. (2.07400,0.05500);
				\draw[kink arrow] (-0.34000,-0.05500) .. controls (-0.67592,-0.30000) and (-1.73808,-0.30000) .. (-2.07400,-0.05500);
				\draw[kink arrow] (-2.07400,0.05500) .. controls (-1.73808,0.30000) and (-0.67592,0.30000) .. (-0.34000,0.05500);
				\coordinate (v0) at (2.41400,0.00000);
				\coordinate (v1) at (0.00000,0.00000);
				\exceptionalreferencecircleloop[particle loop]{v1}{0.3}{90}{0.36}
				\exceptionalreferencecircleloop[particle loop]{v1}{0.3}{-90}{0.36}
				\coordinate (v2) at (-2.41400,0.00000);
				\node[vacuum] at (2.41400,0.00000) {$0$};
				\node[vacuum] at (0.00000,0.00000) {$1$};
				\node[vacuum] at (-2.41400,0.00000) {$2$};
			\end{tikzpicture}
	}}\,.
	\label{eq:e6-ordered-Q}
\end{align}
\begin{align}
	Q_{\rm dis}=\begin{pmatrix}1&1&1\\1&0&1\\1&1&1\end{pmatrix}
	\quad\Longrightarrow\quad
	\vcenter{\hbox{%
			\begin{tikzpicture}[exceptional reference body,baseline={(current bounding box.center)}]
				\draw[kink arrow] (-0.98937,0.77834) .. controls (-0.60923,0.60993) and (-0.07815,-0.30993) .. (-0.12237,-0.72334);
				\draw[kink arrow] (-0.21763,-0.77834) .. controls (-0.59777,-0.60993) and (-1.12885,0.30993) .. (-1.08463,0.72334);
				\draw[kink arrow] (-0.86700,1.10029) .. controls (-0.53108,1.34529) and (0.53108,1.34529) .. (0.86700,1.10029);
				\draw[kink arrow] (0.86700,0.99029) .. controls (0.53108,0.74529) and (-0.53108,0.74529) .. (-0.86700,0.99029);
				\draw[kink arrow] (0.12237,-0.72334) .. controls (0.07815,-0.30993) and (0.60923,0.60993) .. (0.98937,0.77834);
				\draw[kink arrow] (1.08463,0.72334) .. controls (1.12885,0.30993) and (0.59777,-0.60993) .. (0.21763,-0.77834);
				\coordinate (v0) at (-1.20700,1.04529);
				\exceptionalreferencecircleloop[particle loop]{v0}{0.3}{180}{0.36}
				\coordinate (v1) at (0.00000,-1.04529);
				\coordinate (v2) at (1.20700,1.04529);
				\exceptionalreferencecircleloop[particle loop]{v2}{0.3}{0}{0.36}
				\node[vacuum] at (-1.20700,1.04529) {$0$};
				\node[vacuum] at (0.00000,-1.04529) {$1$};
				\node[vacuum] at (1.20700,1.04529) {$2$};
			\end{tikzpicture}
	}}\,.
	\label{eq:e6-disordered-Q}
\end{align}

\FloatBarrier
\vspace{\baselineskip}

\subsection{The two phases of \texorpdfstring{$E_7$}{E7}}
\label{subsec:e7-results}

\begin{figure}[!htbp]
	\centering
	\includegraphics[width=0.90\linewidth]{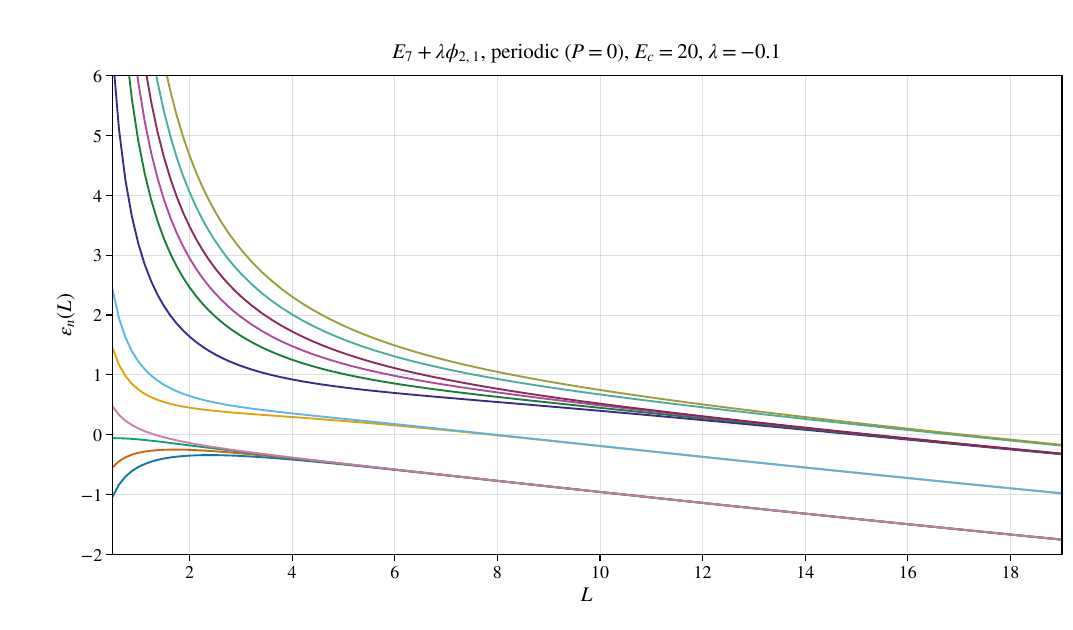}
	\par\medskip
	{\small
		\begin{tabular}{@{}crrrrr@{}}
			\toprule
			$L$ & $\varepsilon_1-\varepsilon_0$ & $\varepsilon_2-\varepsilon_0$ & $\varepsilon_3-\varepsilon_0$ & $\varepsilon_4-\varepsilon_0$ & $\varepsilon_5-\varepsilon_0$ \\
			\midrule
			10 & 0.000158 & 0.000172 & 0.000176 & 0.767312 & 0.769453 \\
			14 & 0.000017 & 0.000073 & 0.000083 & 0.770652 & 0.770778 \\
			18 & 0.000004 & 0.000024 & 0.000156 & 0.772011 & 0.772921 \\
			\bottomrule
		\end{tabular}
	}
	\caption{The zero momentum periodic energies $\varepsilon_n(L)$ of $E_7+\lambda\phi_{2,1}$ at $\Ec=20$ and $\lambda=-0.1$ in the ordered phase. We plot the twelve lowest levels.}
	\label{fig:e7-periodic}
\end{figure}

\begin{figure}[!htbp]
	\centering
	\includegraphics[width=0.90\linewidth]{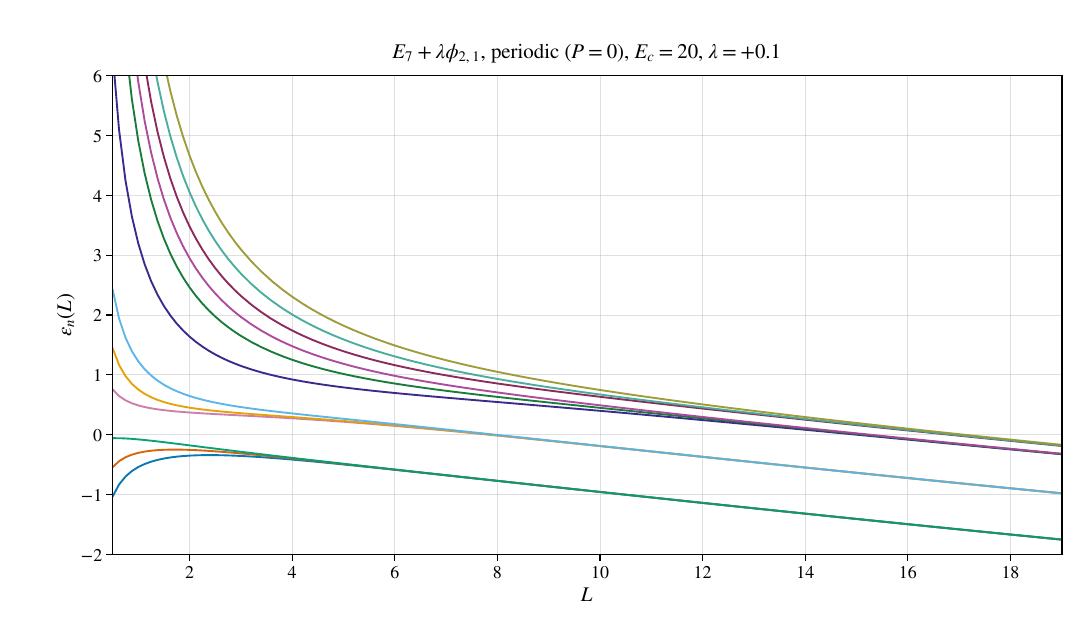}
	\par\medskip
	{\small
		\begin{tabular}{@{}crrrrr@{}}
			\toprule
			$L$ & $\varepsilon_1-\varepsilon_0$ & $\varepsilon_2-\varepsilon_0$ & $\varepsilon_3-\varepsilon_0$ & $\varepsilon_4-\varepsilon_0$ & $\varepsilon_5-\varepsilon_0$ \\
			\midrule
			10 & 0.000158 & 0.000176 & 0.766437 & 0.767312 & 0.769453 \\
			14 & 0.000056 & 0.000067 & 0.770604 & 0.770636 & 0.770762 \\
			18 & 0.000024 & 0.000156 & 0.772011 & 0.772755 & 0.772921 \\
			\bottomrule
		\end{tabular}
	}
	\caption{The zero momentum periodic energies $\varepsilon_n(L)$ of $E_7+\lambda\phi_{2,1}$ at $\Ec=20$ and $\lambda=+0.1$ in the disordered phase. We plot the twelve lowest levels.}
	\label{fig:e7-periodic-disordered}
\end{figure}

\paragraph{Periodic spectrum and vacuum labels.}
Consider next $E_7$. Figures~\ref{fig:e7-periodic} and~\ref{fig:e7-periodic-disordered} show the periodic spectra at momentum $P_\xi=0$ of the ordered and disordered phases and tabulate the lowest gaps $\varepsilon_n(L)-\varepsilon_0(L)$. At the two largest circumferences in the tables, the splittings among the lowest four ordered states and the lowest three disordered states are less than one percent of the first particle gap, so we identify them as the massive vacua. The next two gaps in the ordered phase and the next three gaps in the disordered phase stay close and indicate two and three neutral particles at this mass, respectively.

To distinguish the vacua, we follow the prescription in Section~\ref{subsec:calibration} to diagonalize the interval averaged form factor in Eq.~\eqref{eq:measurement-probe} for operator $\mathcal O=\phi_{5,5}$, with $h_{5,5}=\tfrac{1}{51}$, in the basis of the nearly degenerate ground states in the ordered and disordered phase.
At $L=18$, we order the basis of the periodic reference matrices by its $y$ charges,
\begin{align}
 d_1&=1+2\cos\left(\frac{\pi}{9}\right)\,,&
 d_2&=1+2\cos\left(\frac{5\pi}{9}\right)\,,\notag\\
 d_3&=1+2\cos\left(\frac{7\pi}{9}\right)\,,&
 d_4&=-1\,.
 \label{eq:e7-vacuum-charges}
\end{align}
The ordered basis contains all four states, while the disordered basis
contains the three states with charges $d_1,d_2,d_3$.
Choosing their phases so that the off diagonal probe matrix elements
are real and positive, the vacuum matrices in the cylinder normalization
of Eq.~\eqref{eq:measurement-probe} are
\begin{align}
 \mathcal O_{\rm vac}^{\rm ord}&\simeq \begin{pmatrix}
 0 & 0.972124 & 0 & 0\\
 0.972124 & 0.972322 & 0.972161 & 0.973199\\
 0 & 0.972161 & 0.000705 & 0.972981\\
 0 & 0.973199 & 0.972981 & 0.972842
 \end{pmatrix}\,,\notag\\[6pt]
 \mathcal O_{\rm vac}^{\rm dis}&\simeq \begin{pmatrix}
 0 & 0.972124 & 0\\
 0.972124 & 0.972322 & 0.972161\\
 0 & 0.972161 & -0.973172
 \end{pmatrix}\,.
 \label{eq:e7-vacuum-probe}
\end{align}
Diagonalizing these matrices gives
\begin{align}
 (\mu_0,\mu_1,\mu_2,\mu_3)_{\rm ord}
 &\simeq (-0.971848,-0.517624,0.634507,2.800834)\,,\notag\\
 (\mu_0,\mu_1,\mu_2)_{\rm dis}
 &\simeq (-1.490115,-0.337799,1.827065)\,.
 \label{eq:e7-vacuum-readings}
\end{align}
The corresponding normalized eigenvectors, written as columns in the
same bases, are
\begin{align}
 \bigl(u^{(0)},u^{(1)},u^{(2)},u^{(3)}\bigr)_{\rm ord}
 &\simeq \begin{pmatrix}
0.577658 & 0.427867 & 0.656765 & 0.227819\\
-0.577494 & -0.227825 & 0.428672 & 0.656382\\
0.576898 & -0.656854 & -0.228152 & 0.428577\\
0.000363 & 0.577554 & -0.576929 & 0.577567
\end{pmatrix}\,,\notag\displaybreak[1]\\[6pt]
 \bigl(u^{(0)},u^{(1)},u^{(2)}\bigr)_{\rm dis}
 &\simeq \begin{pmatrix}
0.292862 & 0.844122 & 0.449099\\
-0.448912 & -0.293321 & 0.844062\\
0.844221 & -0.448800 & 0.293034
\end{pmatrix}\,.
 \label{eq:e7-reference-vectors}
\end{align}
Each column defines a reference vacuum
$\ket{\Omega_a}=\sum_i u_i^{(a)}\ket{e_i}$, with the labels $a$
ordered by increasing $\mu_a$.
We now express the spatial line $y$ in these reference vacuum bases.
Its matrix is $y_{\rm vac}=C$, whose entries follow from
Eq.~\eqref{eq:measurement-closure-matrix} as
\begin{align}
 C_{ab}=\bra{\Omega_a}y\ket{\Omega_b}
 =\sum_{i=1}^{n_{\rm vac}}\bigl(u_i^{(a)}\bigr)^*d_i u_i^{(b)}\,,
 \label{eq:e7-vacuum-line}
\end{align}
where $n_{\rm vac}=4$ and $3$ in the ordered and disordered phases,
respectively. Using the eigenvectors before rounding gives
\begin{align}
 C_{\rm ord}&\simeq\begin{pmatrix}
1.001409 & 0.998964 & 1.001059 & -0.000246\\
0.998964 & -0.002136 & 0.998853 & -0.000720\\
1.001059 & 0.998853 & 1.001389 & 0.999720\\
-0.000246 & -0.000720 & 0.999720 & -0.000663
\end{pmatrix}\,,\notag\\[6pt]
 C_{\rm dis}&\simeq\begin{pmatrix}
-0.000732 & 0.999362 & -0.000239\\
0.999362 & 2.000666 & 0.999939\\
-0.000239 & 0.999939 & 1.000066
\end{pmatrix}\,.
 \label{eq:E7_C}
\end{align}
Rounding these entries to the nearest integers gives the junction matrices
\begin{align}
	j_{\rm ord}&=\begin{pmatrix}1&1&1&0\\1&0&1&0\\1&1&1&1\\0&0&1&0\end{pmatrix}\,,&
	j_{\rm dis}&=\begin{pmatrix}0&1&0\\1&2&1\\0&1&1\end{pmatrix}\,.
	\label{eq:e7-closures}
\end{align}
In the ordered phase vacua $0$ and $2$ each admit a single junction closing onto themselves through $y$. In the disordered phase vacuum $1$ has two independent self junctions and vacuum $2$ has one. The two phases therefore have two and three defect vacua, respectively.
Using these reference vacua, we calibrate the probe thresholds by maximizing the mean identification accuracy in Eq.~\eqref{eq:measurement-accuracy}. This gives $97.50\%$ in the ordered phase and $99.10\%$ in the disordered phase.

\begin{figure}[!htbp]
	\centering
	\includegraphics[width=0.98\linewidth]{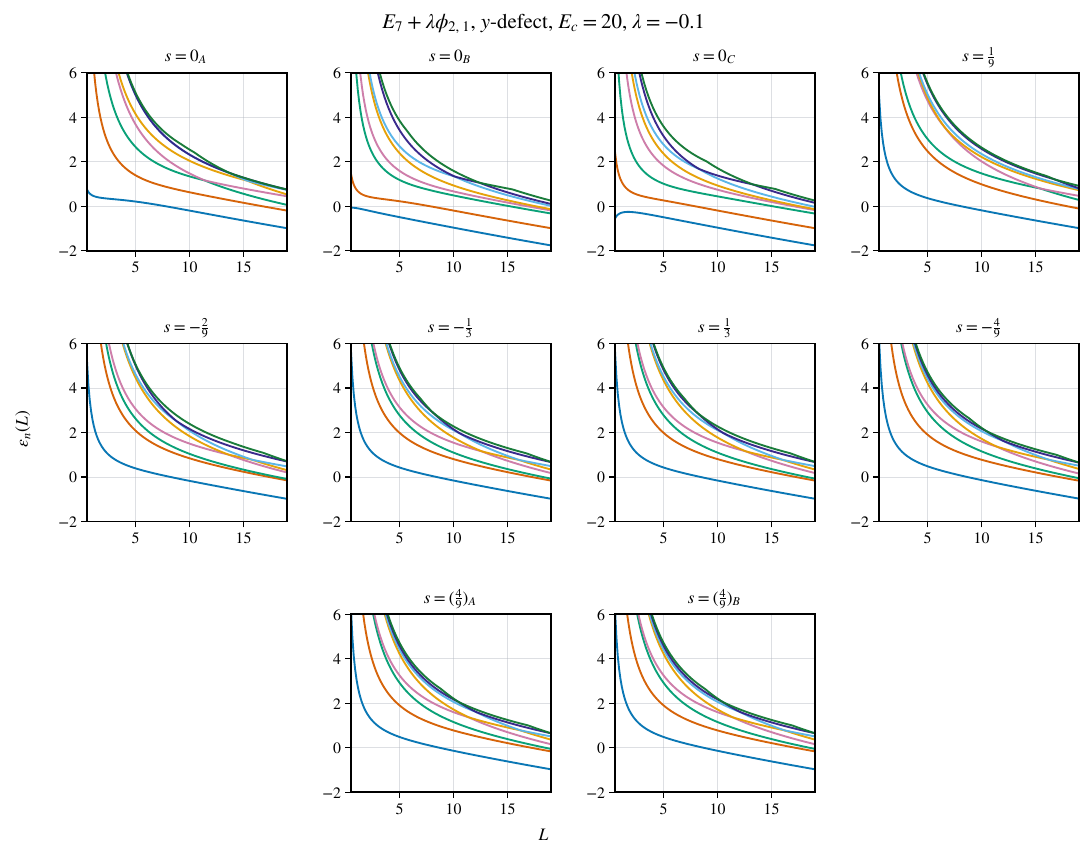}
	\caption{Defect spectrum of $E_7+\lambda\phi_{2,1}$ on the $y$ circle at $\lambda=-0.1$ and $\Ec=20$ in the ordered phase. From left to right, the top row shows $0_A$, $0_B$, $0_C$, $\frac{1}{9}$, the middle row shows $-\frac{2}{9}$, $-\frac{1}{3}$, $\frac{1}{3}$, $-\frac{4}{9}$, and the bottom row shows $(\frac{4}{9})_A$, $(\frac{4}{9})_B$. The lowest levels in $0_B$ and $0_C$ are the two defect vacua. We consider the first excited level in $0_B$ and $0_C$ and the lowest level in $0_A$ and each of the seven fractional momentum panels.}
	\label{fig:e7-defect}
\end{figure}

\begin{figure}[!htbp]
	\centering
	\includegraphics[width=0.98\linewidth]{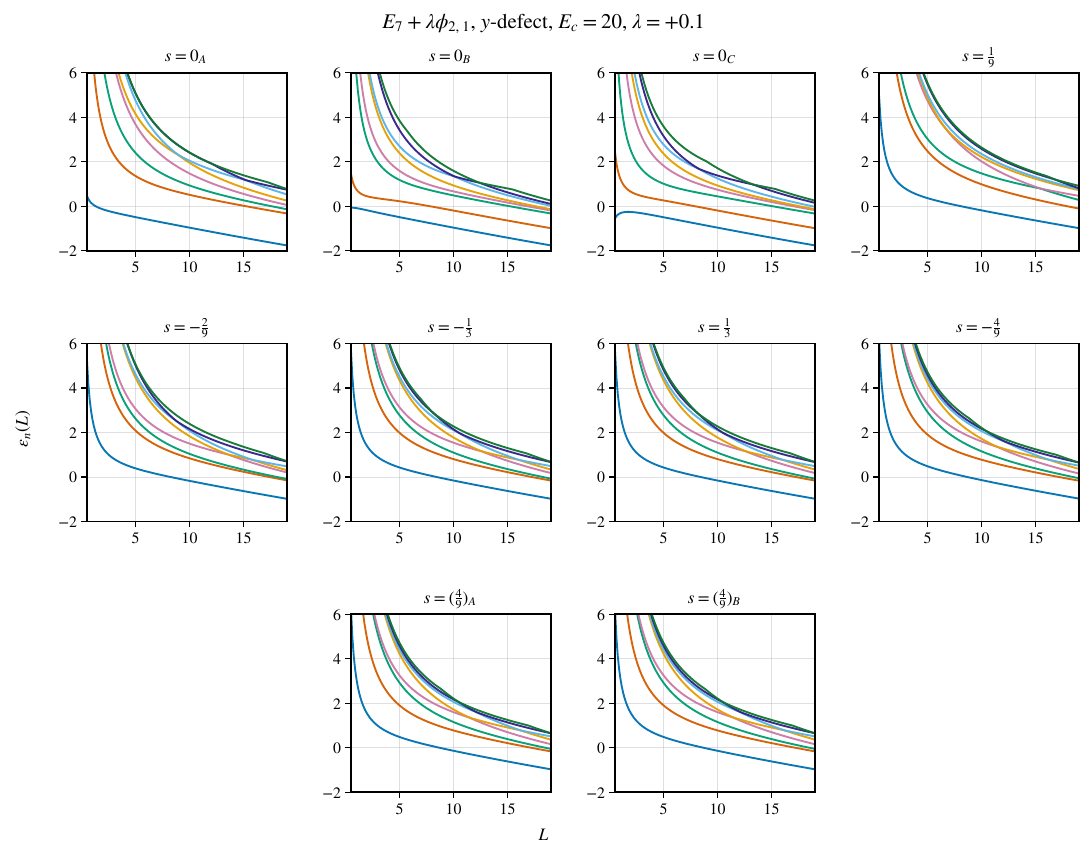}
	\caption{Defect spectrum of $E_7+\lambda\phi_{2,1}$ on the $y$ circle at $\lambda=+0.1$ and $\Ec=20$ in the disordered phase, in the same panel order as Figure~\ref{fig:e7-defect}. The lowest levels in $0_A$, $0_B$, and $0_C$ are the three defect vacua, with no light excitation in $0_A$. We consider the first excited level in $0_B$ and $0_C$ and the lowest level in each of the seven fractional momentum panels.}
	\label{fig:e7-defect-disordered}
\end{figure}

\paragraph{The spectra on the defect circle.}
On the $y$ defect circle the two lowest states in the ordered phase and the three lowest states in the disordered phase have zero momentum. Above them, ten and nine light excitations appear in Figures~\ref{fig:e7-defect} and~\ref{fig:e7-defect-disordered}, respectively, in the channels $P_\xi=\frac{2\pi}{L}s_\xi$,
\begin{align}
s_\xi:\qquad	0_A,\quad 0_B,\quad 0_C,\quad \frac{1}{9},\quad -\frac{2}{9},\quad
	-\frac{1}{3},\quad \frac{1}{3},\quad -\frac{4}{9},\quad
	\left(\frac{4}{9}\right)_A,\quad \left(\frac{4}{9}\right)_B\,,
	\label{eq:e7-channels}
\end{align}
where, for $s_\xi=0,\frac{4}{9}$, the eigenvalue of the spatial $u$ line resolves the repeated momentum channels, as given in Eq.~\eqref{eq:app-e7-channel-charges} of Appendix~\ref{app:B}. The lowest states in $0_B$ and $0_C$ are defect vacua in both phases. The lowest state in $0_A$ is a light excitation in the ordered phase and a third defect vacuum in the disordered phase. In both phases, we consider the first excited state in each of $0_B$ and $0_C$ and the lowest state in each of the seven fractional momentum channels. We also include the lowest state in $0_A$ in the ordered phase.

We measure their gaps from the lowest defect vacuum in each phase and obtain their rest masses using Eq.~\eqref{eq:results-mass}. At $L=18$ and $\Ec=20$, this gives masses $0.7727$, $0.7729$, $0.7720$, $0.7737$, $0.7728$, $0.7714$, $0.7714$, $0.7710$, $0.7710$ and $0.7710$ in the ordered phase, and $0.7729$, $0.7720$, $0.7737$, $0.7728$, $0.7714$, $0.7714$, $0.7710$, $0.7710$ and $0.7710$ in the disordered phase, in the channel order above with $0_A$ omitted in the latter. The light states in both phases thus have a common rest mass within the numerical resolution, in agreement with the particle masses on the periodic circle.

\paragraph{From vacuum pairs to quivers.}
We now measure the vacuum pair probabilities $(E_{ab})_{\xi\xi}$ of these ten ordered states and nine disordered states using the projectors in Section~\ref{subsec:measurement-endpoints}. The probabilities for each channel and their sums $\widehat D_{ab}$ are listed in Tables~\ref{tab:e7-ordered-raw} and~\ref{tab:e7-disordered-raw}. Combining $\widehat D_{ab}$ with the junction multiplicities of Eq.~\eqref{eq:e7-closures} via Eq.~\eqref{eq:measurement-quiver-count}, we set $\widehat Q_{ab}=\tfrac{\widehat D_{ab}}{j_{ab}}$ for $C_{ab}\gtrsim 1$ and display $\widehat D_{ab}$ itself whenever $C_{ab}\simeq 0$:
\begin{align}
	\widehat Q_{\rm ord}&=\begin{pmatrix}
		0.9032 & 0.9411 & 0.9247 & 0.0010 \\
		0.9411 & 0.1730 & 1.0196 & 0.0035 \\
		0.9247 & 1.0196 & 1.2073 & 0.9364 \\
		0.0010 & 0.0035 & 0.9364 & 0.0636
	\end{pmatrix}\,,&
	\widehat Q_{\rm dis}&=\begin{pmatrix}
		0.0639 & 0.9723 & 0.0064 \\
		0.9723 & 1.9687 & 0.9984 \\
		0.0064 & 0.9984 & 1.0445
	\end{pmatrix}\,.
	\label{eq:e7-measured-Q}
\end{align}

All off diagonal entries with $j_{ab}>0$ are close to one. On the diagonal terms, the ordered phase has $\widehat Q_{00}$ and $\widehat Q_{22}$ close to one with a single self junction, giving one particle above vacuum $0$ and one above vacuum $2$. In the disordered phase, $\widehat D_{11}=3.9374$ corresponds to four circle states with vacuum pair $(1,1)$. Divided by $j_{11}=2$ in Eq.~\eqref{eq:e7-closures}, we have $\widehat Q_{11}=1.9687$, implying two neutral particles above vacuum $1$. The entry $\widehat Q_{22}$ is close to one with a single self junction, giving one particle above vacuum $2$. Overall, the measured quiver matrices $\widehat Q$ in Eq.~\eqref{eq:e7-measured-Q} reproduce our orbifolding results in Eq.~\eqref{eq:theory-e7-quivers}, and thus reconstruct the following quiver diagrams,
\begin{align}
	Q_{\rm ord}&=%
	\begin{pmatrix}
		1 & 1 & 1 & 0 \\
		1 & 0 & 1 & 0 \\
		1 & 1 & 1 & 1 \\
		0 & 0 & 1 & 0
	\end{pmatrix}
	\quad\Longrightarrow\quad
	\vcenter{\hbox{%
			\begin{tikzpicture}[exceptional reference body,baseline={(current bounding box.center)}]
				\draw[kink arrow] (1.80600,0.78872) .. controls (2.05100,0.49664) and (2.05100,-0.49664) .. (1.80600,-0.78872);
				\draw[kink arrow] (1.69600,-0.78872) .. controls (1.45100,-0.49664) and (1.45100,0.49664) .. (1.69600,0.78872);
				\draw[kink arrow] (1.48405,0.91109) .. controls (1.35360,0.55287) and (0.49340,0.05623) .. (0.11795,0.12237);
				\draw[kink arrow] (0.06295,0.21763) .. controls (0.19340,0.57585) and (1.05360,1.07249) .. (1.42905,1.00635);
				\draw[kink arrow] (1.42905,-1.00635) .. controls (1.05360,-1.07249) and (0.19340,-0.57585) .. (0.06295,-0.21763);
				\draw[kink arrow] (0.11795,-0.12237) .. controls (0.49340,-0.05623) and (1.35360,-0.55287) .. (1.48405,-0.91109);
				\draw[kink arrow] (-0.54400,-0.05500) .. controls (-0.85136,-0.30000) and (-1.86864,-0.30000) .. (-2.17600,-0.05500);
				\draw[kink arrow] (-2.17600,0.05500) .. controls (-1.86864,0.30000) and (-0.85136,0.30000) .. (-0.54400,0.05500);
				\coordinate (v0) at (1.75100,1.12872);
				\exceptionalreferencecircleloop[particle loop]{v0}{0.3}{0}{0.36}
				\coordinate (v1) at (1.75100,-1.12872);
				\coordinate (v2) at (-0.20400,0.00000);
				\exceptionalreferencecircleloop[particle loop]{v2}{0.3}{94}{0.36}
				\coordinate (v3) at (-2.51600,0.00000);
				\node[vacuum] at (1.75100,1.12872) {$0$};
				\node[vacuum] at (1.75100,-1.12872) {$1$};
				\node[vacuum] at (-0.20400,0.00000) {$2$};
				\node[vacuum] at (-2.51600,0.00000) {$3$};
			\end{tikzpicture}
	}}\,.
	\label{eq:e7-ordered-quiver}
\end{align}
\begin{align}
	Q_{\rm dis}&=%
	\begin{pmatrix}
		0 & 1 & 0 \\
		1 & 2 & 1 \\
		0 & 1 & 1
	\end{pmatrix}
	\quad\Longrightarrow\quad
	\vcenter{\hbox{%
			\begin{tikzpicture}[exceptional reference body,baseline={(current bounding box.center)}]
				\draw[kink arrow] (-2.07400,0.05500) .. controls (-1.73808,0.30000) and (-0.67592,0.30000) .. (-0.34000,0.05500);
				\draw[kink arrow] (-0.34000,-0.05500) .. controls (-0.67592,-0.30000) and (-1.73808,-0.30000) .. (-2.07400,-0.05500);
				\draw[kink arrow] (0.34000,0.05500) .. controls (0.67592,0.30000) and (1.73808,0.30000) .. (2.07400,0.05500);
				\draw[kink arrow] (2.07400,-0.05500) .. controls (1.73808,-0.30000) and (0.67592,-0.30000) .. (0.34000,-0.05500);
				\coordinate (v0) at (-2.41400,0.00000);
				\coordinate (v1) at (0.00000,0.00000);
				\exceptionalreferencecircleloop[particle loop]{v1}{0.3}{90}{0.36}
				\exceptionalreferencecircleloop[particle loop]{v1}{0.3}{-90}{0.36}
				\coordinate (v2) at (2.41400,0.00000);
				\exceptionalreferencecircleloop[particle loop]{v2}{0.3}{0}{0.36}
				\node[vacuum] at (-2.41400,0.00000) {$0$};
				\node[vacuum] at (0.00000,0.00000) {$1$};
				\node[vacuum] at (2.41400,0.00000) {$2$};
			\end{tikzpicture}
	}}\,.
	\label{eq:e7-disordered-quiver}
\end{align}

\vspace{\baselineskip}

\subsection{The two phases of \texorpdfstring{$E_8$}{E8}}
\label{subsec:e8-results}

\begin{figure}[!htbp]
	\centering
	\includegraphics[width=0.90\linewidth]{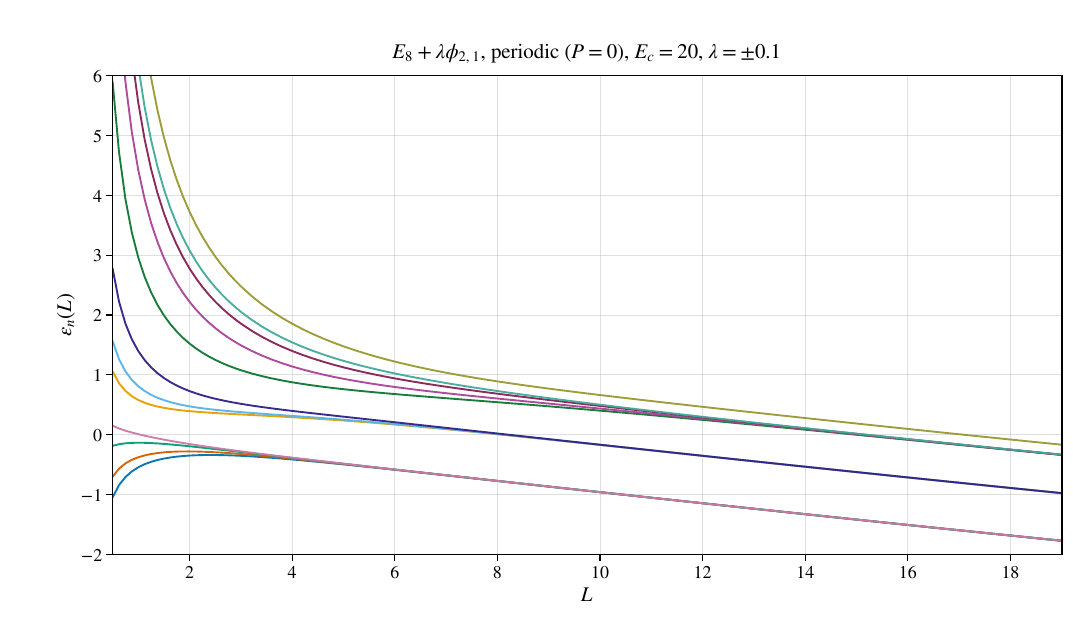}
	\par\medskip
	{\small
		\begin{tabular}{@{}crrrrrr@{}}
			\toprule
			$L$ & $\varepsilon_1-\varepsilon_0$ & $\varepsilon_2-\varepsilon_0$ & $\varepsilon_3-\varepsilon_0$ & $\varepsilon_4-\varepsilon_0$ & $\varepsilon_5-\varepsilon_0$ & $\varepsilon_6-\varepsilon_0$ \\
			\midrule
			10 & 0.000027 & 0.000028 & 0.000056 & 0.789156 & 0.789909 & 0.792633 \\
			14 & 0.000084 & 0.000142 & 0.000287 & 0.792850 & 0.793151 & 0.793208 \\
			18 & 0.000269 & 0.000286 & 0.000660 & 0.794266 & 0.794505 & 0.795450 \\
			\bottomrule
		\end{tabular}
	}
	\caption{The zero momentum periodic energies $\varepsilon_n(L)$ of $E_8+\lambda\phi_{2,1}$ at $\Ec=20$ and $\lambda=\pm0.1$. The two signs give the same spectrum. We plot the twelve lowest levels.}
	\label{fig:e8numerical-periodic}
\end{figure}

\paragraph{Periodic spectrum and vacuum labels.}
Consider finally $E_8$. Figure~\ref{fig:e8numerical-periodic} shows the periodic spectrum at momentum $P_\xi=0$ of the two phases and tabulates the lowest gaps $\varepsilon_n(L)-\varepsilon_0(L)$. At the two largest circumferences in the table, the first three gaps are less than one percent of $\varepsilon_4-\varepsilon_0$, so we identify the lowest four states as the massive vacua. The gaps $\varepsilon_4-\varepsilon_0$, $\varepsilon_5-\varepsilon_0$ and $\varepsilon_6-\varepsilon_0$ stay close and indicate three neutral particles at this mass in either phase.

To distinguish the four vacua, we follow the prescription in Section~\ref{subsec:calibration} to diagonalize the interval averaged form factor in Eq.~\eqref{eq:measurement-probe} for operator $\mathcal O=\phi_{7,7}$, with $h_{7,7}=\tfrac{2}{145}$, in the basis of the nearly degenerate ground states in the ordered and disordered phases.
At $L=18$, we order the basis of the periodic reference matrices by its $y$ charges,
\begin{align}
 d_1&=1+2\cos\left(\frac{\pi}{15}\right)\,,&
 d_2&=1+2\cos\left(\frac{7\pi}{15}\right)\,,\notag\\
 d_3&=1+2\cos\left(\frac{11\pi}{15}\right)\,,&
 d_4&=1+2\cos\left(\frac{13\pi}{15}\right)\,.
 \label{eq:e8-vacuum-charges}
\end{align}
Both phases contain all four states.
Choosing their phases so that the probe matrices are real, the vacuum
matrices in the cylinder normalization of Eq.~\eqref{eq:measurement-probe}
are
\begin{align}
 \mathcal O_{\rm vac}^{\rm ord}&\simeq \begin{pmatrix}
 0 & 0.981151 & 0 & 0\\
 0.981151 & -1.267368 & 0.982647 & -0.567456\\
 0 & 0.982647 & 0 & 1.590052\\
 0 & -0.567456 & 1.590052 & -1.485682
 \end{pmatrix}\,,\notag\\[6pt]
 \mathcal O_{\rm vac}^{\rm dis}&\simeq \begin{pmatrix}
 0 & 0.981151 & 0 & 0\\
 0.981151 & -1.267368 & 0.982647 & -0.567456\\
 0 & 0.982647 & 0 & 0.606754\\
 0 & -0.567456 & 0.606754 & 0.217121
 \end{pmatrix}\,.
 \label{eq:e8-vacuum-probe}
\end{align}
Diagonalizing these matrices gives
\begin{align}
 (\mu_0,\mu_1,\mu_2,\mu_3)_{\rm ord}
 &\simeq (-3.162179,-1.291927,0.493262,1.207794)\,,\notag\\
 (\mu_0,\mu_1,\mu_2,\mu_3)_{\rm dis}
 &\simeq (-2.361435,-0.251744,0.660929,0.902002)\,.
 \label{eq:e8-vacuum-readings}
\end{align}
The corresponding normalized eigenvectors, written as columns in the
same bases, are
\begin{align}
 \bigl(u^{(0)},u^{(1)},u^{(2)},u^{(3)}\bigr)_{\rm ord}
 &\simeq \begin{pmatrix}
0.168100 & 0.498070 & 0.806006 & 0.272071\\
-0.541774 & -0.655832 & 0.405210 & 0.334918\\
0.498124 & -0.168029 & -0.272115 & 0.805973\\
-0.655817 & 0.541827 & -0.334833 & 0.405234
\end{pmatrix}\,,\notag\\[6pt]
 \bigl(u^{(0)},u^{(1)},u^{(2)},u^{(3)}\bigr)_{\rm dis}
 &\simeq \begin{pmatrix}
0.334781 & 0.656154 & 0.404832 & 0.541751\\
-0.805752 & -0.168356 & 0.272706 & 0.498048\\
0.405361 & -0.541877 & -0.334578 & 0.655827\\
-0.272704 & 0.497481 & -0.806102 & 0.168358
\end{pmatrix}\,.
 \label{eq:e8-reference-vectors}
\end{align}
Each column defines a reference vacuum
$\ket{\Omega_a}=\sum_{i=1}^{4}u_i^{(a)}\ket{e_i}$, with
$a=0,1,2,3$ ordered by increasing $\mu_a$.
We now express the spatial line $y$ in these reference vacuum bases.
Its matrix is $y_{\rm vac}=C$, whose entries follow from
Eq.~\eqref{eq:measurement-closure-matrix} as
\begin{align}
 C_{ab}=\bra{\Omega_a}y\ket{\Omega_b}
 =\sum_{i=1}^{4}\bigl(u_i^{(a)}\bigr)^*d_i u_i^{(b)}\,.
 \label{eq:e8-vacuum-line}
\end{align}
Using the eigenvectors before rounding gives
\begin{align}
 C_{\rm ord}&\simeq\begin{pmatrix}
-0.001242 & 0.999321 & -0.000649 & -0.000173\\
0.999321 & 1.001051 & 1.000077 & -0.000753\\
-0.000649 & 1.000077 & 2.001290 & 0.998783\\
-0.000173 & -0.000753 & 0.998783 & -0.001100
\end{pmatrix}\,,\notag\\[6pt]
 C_{\rm dis}&\simeq\begin{pmatrix}
0.999210 & 0.999925 & -0.000943 & -0.000973\\
0.999925 & 1.003047 & 1.000132 & 1.000439\\
-0.000943 & 1.000132 & -0.000889 & 0.999055\\
-0.000973 & 1.000439 & 0.999055 & 0.998632
\end{pmatrix}\,.
 \label{eq:E8_C}
\end{align}
Rounding these entries to the nearest integers gives the junction matrices
\begin{align}
	j_{\rm ord}&=\begin{pmatrix}0&1&0&0\\1&1&1&0\\0&1&2&1\\0&0&1&0\end{pmatrix}\,,&
	j_{\rm dis}&=\begin{pmatrix}1&1&0&0\\1&1&1&1\\0&1&0&1\\0&1&1&1\end{pmatrix}\,.
	\label{eq:e8-closures}
\end{align}
In the ordered phase vacuum $2$ can close onto itself through $y$ with two independent junctions, while vacuum $1$ admits a single self junction. In the disordered phase vacua $0$, $1$ and $3$ each admit a single self junction. Both phases therefore have three defect vacua, but with different local vacuum content.
Using these reference vacua, we calibrate the probe thresholds by maximizing the mean identification accuracy in Eq.~\eqref{eq:measurement-accuracy}. This gives $99.23\%$ in the ordered phase and $97.36\%$ in the disordered phase.

\begin{figure}[!htbp]
	\centering
	\includegraphics[width=0.98\linewidth]{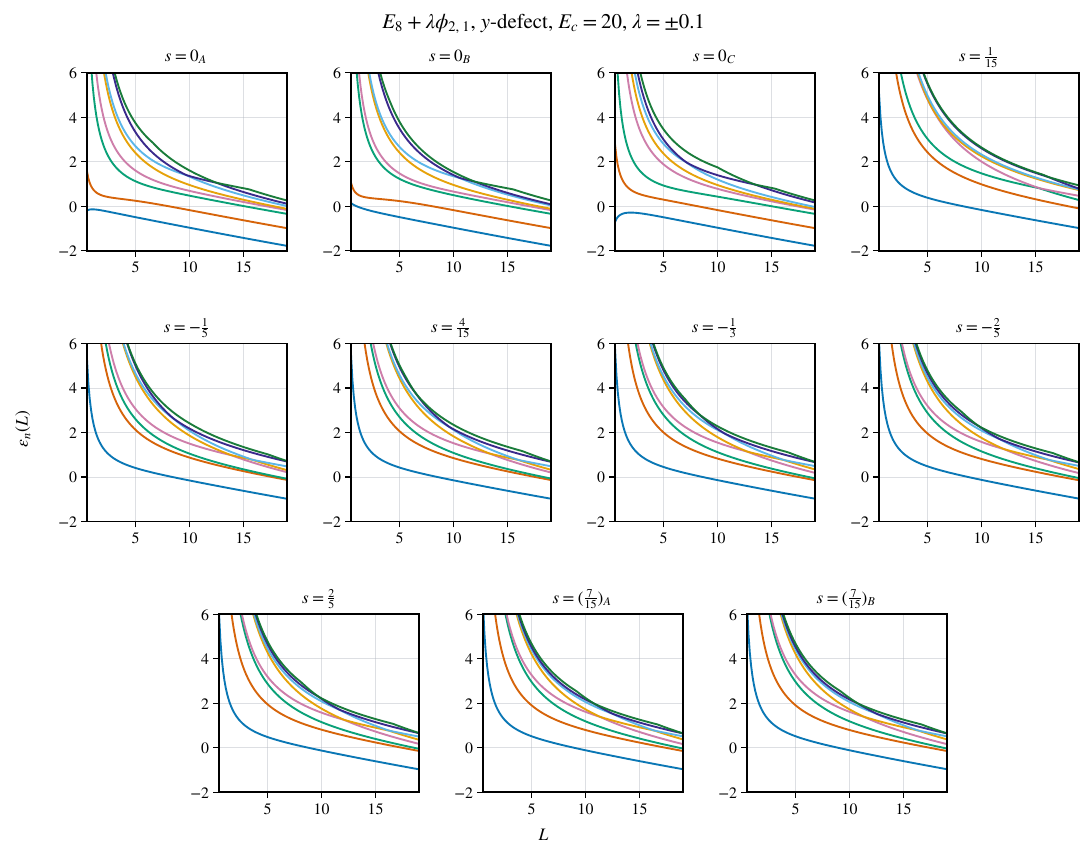}
	\caption{Defect spectrum of $E_8+\lambda\phi_{2,1}$ on the $y$ circle at $\lambda=\pm0.1$ and $\Ec=20$. The two signs give the same spectrum. From left to right, the top row shows $0_A$, $0_B$, $0_C$, $\frac{1}{15}$, the middle row shows $-\frac{1}{5}$, $\frac{4}{15}$, $-\frac{1}{3}$, $-\frac{2}{5}$, and the bottom row shows $\frac{2}{5}$, $(\frac{7}{15})_A$, $(\frac{7}{15})_B$. The lowest levels in the first three panels are the defect vacua. We consider the first excited level in each zero momentum panel and the lowest level in each of the eight fractional momentum panels.}
	\label{fig:e8numerical-defect}
\end{figure}

\paragraph{The spectra on the defect circle.}
On the $y$ defect circle the three lowest degenerate states have zero momentum. Above them, eleven light excitations appear in Figure~\ref{fig:e8numerical-defect}, one in each of the channels $P_\xi=\frac{2\pi}{L}s_\xi$,
\begin{align}
	s_\xi:\qquad 0_A,\quad 0_B,\quad 0_C,\quad \frac{1}{15},\quad -\frac{1}{5},\quad
	\frac{4}{15},\quad -\frac{1}{3},\quad -\frac{2}{5},\quad \frac{2}{5},\quad
	\left(\frac{7}{15}\right)_A,\quad \left(\frac{7}{15}\right)_B\,,
	\label{eq:e8-channels}
\end{align}
where, for $s_\xi=0,\frac{7}{15}$, the eigenvalue of the spatial line $(p,p)$ resolves the repeated momentum channels, as given in Eq.~\eqref{eq:app-e8-channel-charges} of Appendix~\ref{app:B}. The lowest states in $0_A$, $0_B$ and $0_C$ are the three defect vacua. We consider the first excited state in each of the three zero momentum channels and the lowest state in each of the eight fractional momentum channels in Figure~\ref{fig:e8numerical-defect}. We measure their gaps from the same defect vacuum and obtain their rest masses using Eq.~\eqref{eq:results-mass}. At $L=18$ and $\Ec=20$, this gives masses $0.7943$, $0.7954$, $0.7945$, $0.7965$, $0.7953$, $0.7952$, $0.7940$, $0.7938$, $0.7938$, $0.7937$ and $0.7937$ in the channel order above. The eleven states thus have a common rest mass within the numerical resolution, in agreement with the particle masses on the periodic circle.

\paragraph{From vacuum pairs to quivers.}
We now measure the vacuum pair probabilities $(E_{ab})_{\xi\xi}$ of these eleven states using the projectors in Section~\ref{subsec:measurement-endpoints}. The probabilities for each channel and their sums $\widehat D_{ab}$ are listed in Tables~\ref{tab:e8numerical-raw-ordered} and~\ref{tab:e8numerical-raw-disordered}. Combining $\widehat D_{ab}$ with the junction multiplicities of Eq.~\eqref{eq:e8-closures} via Eq.~\eqref{eq:measurement-quiver-count}, we set $\widehat Q_{ab}=\tfrac{\widehat D_{ab}}{j_{ab}}$ for $C_{ab}\gtrsim1$ and display $\widehat D_{ab}$ itself whenever $C_{ab}\simeq0$:
\begin{align}
	\widehat Q_{\rm ord} &= \begin{pmatrix}
		0.0688 & 0.9314 & 0.0069 & 0.0000 \\
		0.9314 & 1.0788 & 0.9902 & 0.0074 \\
		0.0069 & 0.9902 & 1.9707 & 0.9873 \\
		0.0000 & 0.0074 & 0.9873 & 0.0649
	\end{pmatrix}\,,
	\label{eq:e8numerical-Q-ordered}\\[6pt]
	\widehat Q_{\rm dis} &= \begin{pmatrix}
		1.0388 & 0.9481 & 0.0196 & 0.0011 \\
		0.9481 & 1.0294 & 1.0475 & 0.8755 \\
		0.0196 & 1.0475 & 0.3649 & 0.9498 \\
		0.0011 & 0.8755 & 0.9498 & 0.8837
	\end{pmatrix}\,.
	\label{eq:e8numerical-Q-disordered}
\end{align}

The off diagonal entries with $j_{ab}>0$ are close to one. On the diagonal, the ordered phase has $\widehat D_{22}=3.9413$, corresponding to four circle states with vacuum pair $(2,2)$. Dividing by $j_{22}=2$ in Eq.~\eqref{eq:e8-closures} gives $\widehat Q_{22}=1.9707$, indicating two neutral particles above vacuum $2$. Together with $\widehat Q_{11}=1.0788$ and $j_{11}=1$, we find a third particle above vacuum $1$. In the disordered phase, $\widehat Q_{00}$, $\widehat Q_{11}$ and $\widehat Q_{33}$ are each close to one with a single self junction, giving one particle above each of vacua $0$, $1$ and $3$. However, $\widehat Q_{22}=0.3649$ in the disordered phase displays the raw
sum $\widehat D_{22}$ at $j_{22}=0$, where $y$ admits no self junction.
This unwanted readout is much larger than its change of $0.0364$ when
the probe cutoff increases from $7$ to $8$.
Moreover, $\widehat Q_{13}=\widehat Q_{31}=0.8755$ falls below one by
$12.45\%$. To improve the measurement, we compute the form factors of the linear combination
\begin{align}
  \mathcal O_{\rm mix}
  &=\phi_{7,7}^{}
  -\frac{1}{2}
  \phi_{11,11}^{}\,,
  \label{eq:e8numerical-mixed}
\end{align}
Keeping the reference vacua fixed and
recalibrating the thresholds for this probe at the same parameters gives,
in the same vacuum order $(0,1,2,3)$,
\begin{align}
  \widehat Q_{\rm dis}^{\rm mix}&\simeq
  \begin{pmatrix}
    1.0389 & 0.9462 & 0.0113 & 0.0036 \\
    0.9462 & 1.0257 & 0.9735 & 0.9343 \\
    0.0113 & 0.9735 & 0.1329 & 1.0088 \\
    0.0036 & 0.9343 & 1.0088 & 1.0473
  \end{pmatrix}\,.
  \label{eq:e8numerical-Q-disordered-mixed}
\end{align}
The mixed probe reduces the unwanted $(2,2)$ sum to $0.1329$ and brings
the $(1,3)$ and $(3,1)$ entries to $0.9343$, closer to one. Combining the ordered result in Eq.~\eqref{eq:e8numerical-Q-ordered}
with the improved disordered result in
Eq.~\eqref{eq:e8numerical-Q-disordered-mixed}, we reconstruct the
following quivers, in agreement with Eq.~\eqref{eq:theory-e8-quivers},
\begin{align}
	Q_{\rm ord} = \begin{pmatrix}
		0 & 1 & 0 & 0\\
		1 & 1 & 1 & 0\\
		0 & 1 & 2 & 1\\
		0 & 0 & 1 & 0
	\end{pmatrix}
	\quad\Longrightarrow\quad
	\vcenter{\hbox{%
			\begin{tikzpicture}[exceptional reference body,baseline={(current bounding box.center)}]
				\draw[kink arrow] (2.72000,-0.05500) .. controls (2.48880,-0.30000) and (1.59120,-0.30000) .. (1.36000,-0.05500);
				\draw[kink arrow] (1.36000,0.05500) .. controls (1.59120,0.30000) and (2.48880,0.30000) .. (2.72000,0.05500);
				\draw[kink arrow] (0.68000,-0.05500) .. controls (0.44880,-0.30000) and (-0.44880,-0.30000) .. (-0.68000,-0.05500);
				\draw[kink arrow] (-0.68000,0.05500) .. controls (-0.44880,0.30000) and (0.44880,0.30000) .. (0.68000,0.05500);
				\draw[kink arrow] (-1.36000,-0.05500) .. controls (-1.59120,-0.30000) and (-2.48880,-0.30000) .. (-2.72000,-0.05500);
				\draw[kink arrow] (-2.72000,0.05500) .. controls (-2.48880,0.30000) and (-1.59120,0.30000) .. (-1.36000,0.05500);
				\coordinate (v0) at (3.06000,0.00000);
				\coordinate (v1) at (1.02000,0.00000);
				\exceptionalreferencecircleloop[particle loop]{v1}{0.3}{85}{0.36}
				\coordinate (v2) at (-1.02000,0.00000);
				\exceptionalreferencecircleloop[particle loop]{v2}{0.3}{90}{0.36}
				\exceptionalreferencecircleloop[particle loop]{v2}{0.3}{-90}{0.36}
				\coordinate (v3) at (-3.06000,0.00000);
				\node[vacuum] at (3.06000,0.00000) {$0$};
				\node[vacuum] at (1.02000,0.00000) {$1$};
				\node[vacuum] at (-1.02000,0.00000) {$2$};
				\node[vacuum] at (-3.06000,0.00000) {$3$};
			\end{tikzpicture}
	}}\,.
	\label{eq:e8numerical-integerQ-ordered}
\end{align}
\begin{align}
	Q_{\rm dis} = \begin{pmatrix}
		1 & 1 & 0 & 0\\
		1 & 1 & 1 & 1\\
		0 & 1 & 0 & 1\\
		0 & 1 & 1 & 1
	\end{pmatrix}
	\quad\Longrightarrow\quad
	\vcenter{\hbox{%
			\begin{tikzpicture}[exceptional reference body,baseline={(current bounding box.center)}]
				\draw[kink arrow] (-2.17600,0.05500) .. controls (-1.86864,0.30000) and (-0.85136,0.30000) .. (-0.54400,0.05500);
				\draw[kink arrow] (-0.54400,-0.05500) .. controls (-0.85136,-0.30000) and (-1.86864,-0.30000) .. (-2.17600,-0.05500);
				\draw[kink arrow] (0.11795,-0.12237) .. controls (0.49340,-0.05623) and (1.35360,-0.55287) .. (1.48405,-0.91109);
				\draw[kink arrow] (1.42905,-1.00635) .. controls (1.05360,-1.07249) and (0.19340,-0.57585) .. (0.06295,-0.21763);
				\draw[kink arrow] (0.06295,0.21763) .. controls (0.19340,0.57585) and (1.05360,1.07249) .. (1.42905,1.00635);
				\draw[kink arrow] (1.48405,0.91109) .. controls (1.35360,0.55287) and (0.49340,0.05623) .. (0.11795,0.12237);
				\draw[kink arrow] (1.69600,-0.78872) .. controls (1.45100,-0.49664) and (1.45100,0.49664) .. (1.69600,0.78872);
				\draw[kink arrow] (1.80600,0.78872) .. controls (2.05100,0.49664) and (2.05100,-0.49664) .. (1.80600,-0.78872);
				\coordinate (v0) at (-2.51600,0.00000);
				\exceptionalreferencecircleloop[particle loop]{v0}{0.3}{180}{0.36}
				\coordinate (v1) at (-0.20400,0.00000);
				\exceptionalreferencecircleloop[particle loop]{v1}{0.3}{94}{0.36}
				\coordinate (v2) at (1.75100,-1.12872);
				\coordinate (v3) at (1.75100,1.12872);
				\exceptionalreferencecircleloop[particle loop]{v3}{0.3}{0}{0.36}
				\node[vacuum] at (-2.51600,0.00000) {$0$};
				\node[vacuum] at (-0.20400,0.00000) {$1$};
				\node[vacuum] at (1.75100,-1.12872) {$2$};
				\node[vacuum] at (1.75100,1.12872) {$3$};
			\end{tikzpicture}
	}}\,.
	\label{eq:e8numerical-integerQ-disordered}
\end{align}

\FloatBarrier

\subsection{The heavy multiplets}
\label{subsec:heavy-results}

In addition to the light multiplet respect to $y$-defect, the diagonal parent theories also contain neutral bound states of two light excitations, with exact mass ratio \cite{Dorey:2002gj}
\begin{align}
	\frac{m_H}{m_L}=2\cos\left(\frac{\pi(k-4)}{6k}\right),\qquad k=10,16,28\,,
	\label{eq:results-heavy-ratio}
\end{align}
giving $2\cos(\tfrac{\pi}{10})\simeq 1.90211$, $2\cos(\tfrac{\pi}{8})\simeq 1.84776$, and $2\cos(\tfrac{\pi}{7})\simeq 1.80194$ for $E_6$, $E_7$, and $E_8$, respectively. The generalized orbifolding construction of Section~\ref{sec:orb} gives the exceptional heavy quiver $Q_0^E=\mathbf 1$, with one neutral particle above every vacuum. We therefore expect a group of nearly degenerate periodic levels between the light particle gap and the two-particle threshold. Subtracting the lowest ground-state energy, $\Delta \varepsilon_n=\varepsilon_n-\varepsilon_0$, we define
\begin{align}
	B_i&=2\overline m_L-\Delta \varepsilon_{H_i}\,, &
	\overline m_H&=\frac{1}{N_H}\sum_{i=1}^{N_H}\Delta \varepsilon_{H_i}\,,
	\label{eq:results-heavy-binding}
\end{align}
where $\overline m_L$ is the mean light periodic gap, $N_H$ is the number of heavy levels below $2\overline m_L$, and $i=1,\ldots,N_H$ orders the heavy branches by increasing energy.

\begin{figure}[!htbp]
	\centering
	\includegraphics[width=0.86\linewidth]{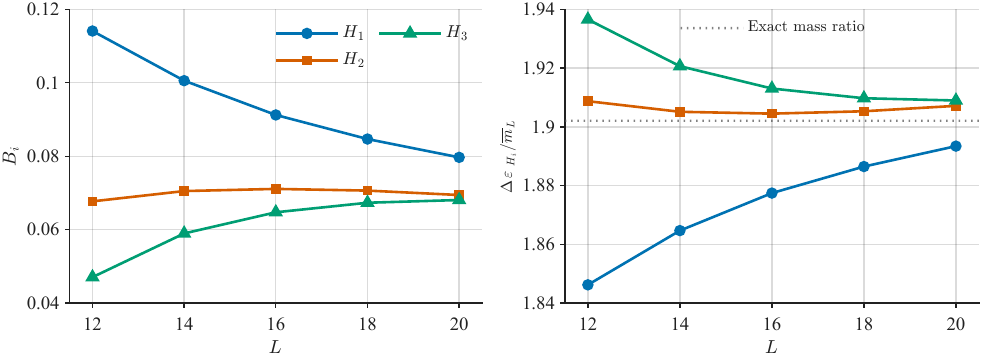}
	\caption{The heavy periodic levels of $E_6$ at $\Ec=20$ and $\lambda=\pm0.1$. The left panel shows $B_i=2\overline m_L-\Delta \varepsilon_{H_i}$ and the right panel shows the heavy gaps in units of the mean light gap. The two phases have the same gaps. The dotted line is the bound-state mass ratio in Eq.~\eqref{eq:results-heavy-ratio}.}
	\label{fig:e6numerical-heavy}
\end{figure}

\begin{figure}[!htbp]
	\centering
	\includegraphics[width=0.86\linewidth]{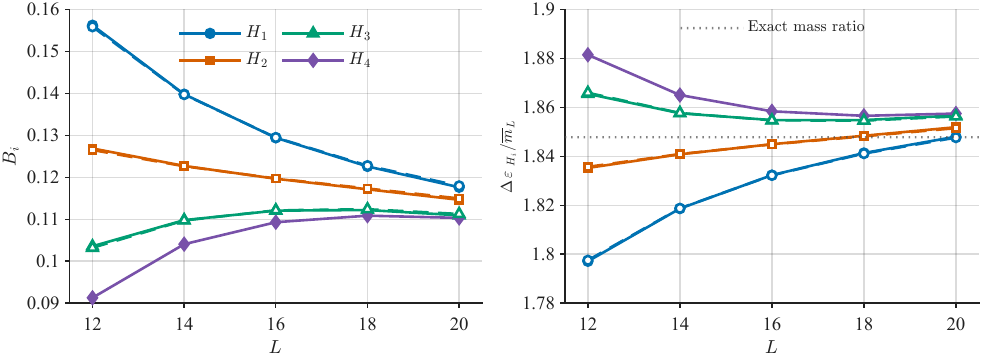}
	\caption{The heavy periodic levels of $E_7$ at $\Ec=20$ and $|\lambda|=0.1$. Solid lines show the ordered phase and dashed lines the disordered phase. Panels and dotted line are as in Figure~\ref{fig:e6numerical-heavy}.}
	\label{fig:e7numerical-heavy}
\end{figure}

\begin{figure}[!htbp]
	\centering
	\includegraphics[width=0.86\linewidth]{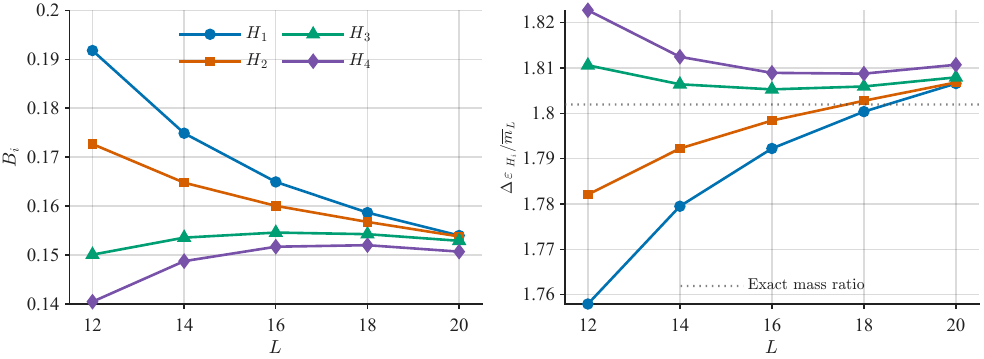}
	\caption{The heavy periodic levels of $E_8$ at $\Ec=20$ and $\lambda=\pm0.1$. The two phases have the same gaps. Panels and dotted line are as in Figure~\ref{fig:e6numerical-heavy}.}
	\label{fig:e8numerical-heavy}
\end{figure}

At fixed cutoff $\Ec=20$ and $|\lambda|=0.1$, Figures~\ref{fig:e6numerical-heavy}--\ref{fig:e8numerical-heavy} show the binding differences $B_i$ and individual mass ratios $\Delta \varepsilon_{H_i}/\overline m_L$ over $L\in[12,20]$, while Table~\ref{tab:app-heavy-gaps} lists the heavy periodic energy gaps at $L=18$ together with $\overline m_L$. In every phase, $N_H$ equals the number of massive vacua, the splittings among the heavy levels decrease steadily as $L$ increases, and the ratios $\Delta \varepsilon_{H_i}/\overline m_L$ approach the exact bound-state value in Eq.~\eqref{eq:results-heavy-ratio}. In $E_7$, the additional heavy state in the ordered phase belongs to the $(9,9)$ conformal sector, while the corresponding excitation in the disordered phase lies above the two-particle threshold.

\begin{table}[!htbp]
	\centering\small
	\begin{tabular}{llrrrrr}
		\toprule
		Model & Phase & $\overline m_L$ & $\Delta \varepsilon_{H_1}$ & $\Delta \varepsilon_{H_2}$ & $\Delta \varepsilon_{H_3}$ & $\Delta \varepsilon_{H_4}$ \\
		\midrule
		$E_6$ & Both & $0.7461$ & $1.4075$ & $1.4216$ & $1.4249$ & $\text{--}$ \\
		$E_7$ & Ordered & $0.7725$ & $1.4224$ & $1.4278$ & $1.4327$ & $1.4341$ \\
		$E_7$ & Disordered & $0.7726$ & $1.4224$ & $1.4278$ & $1.4327$ & $\text{--}$ \\
		$E_8$ & Both & $0.7947$ & $1.4308$ & $1.4327$ & $1.4352$ & $1.4375$ \\
		\bottomrule
	\end{tabular}
	\caption{Heavy periodic energy gaps at $\Ec=20$, $L=18$, and $|\lambda|=0.1$, together with the mean light gap $\overline m_L$.}
	\label{tab:app-heavy-gaps}
\end{table}

\FloatBarrier

\begin{figure}[!htbp]
	\centering
	\begingroup\footnotesize
	\setlength{\tabcolsep}{6.5pt}
	\renewcommand{\arraystretch}{1.06}
	\begin{tabular}{@{}llrrrrrrr@{}}
		\toprule
		Model & Phase & $\Ec$ & $B_1$ & $B_2$ & $B_3$ & $B_4$ & $\frac{\overline m_H}{\overline m_L}$ & Exact \\
		\midrule
		$E_6$ & Both & 16 & 0.08478 & 0.06869 & 0.06577 & \text{--} & 1.90240 & 1.90211 \\
		      &      & 18 & 0.08458 & 0.06977 & 0.06666 & \text{--} & 1.90140 & \\
		      &      & 20 & 0.08469 & 0.07065 & 0.06734 & \text{--} & 1.90052 & \\
		      &      & 22 & 0.08465 & 0.07112 & 0.06768 & \text{--} & 1.90006 & \\
		\midrule
		$E_7$ & Ordered & 16 & 0.12028 & 0.11497 & 0.10873 & 0.10738 & 1.85423 & 1.84776 \\
		      &         & 18 & 0.12163 & 0.11626 & 0.11080 & 0.10951 & 1.85184 & \\
		      &         & 20 & 0.12257 & 0.11712 & 0.11218 & 0.11084 & 1.85025 & \\
		      &         & 22 & 0.12313 & 0.11768 & 0.11305 & 0.11168 & 1.84923 & \\
		\midrule
		$E_7$ & Disordered & 16 & 0.12065 & 0.11534 & 0.10910 & \text{--} & 1.85144 & 1.84776 \\
		      &            & 18 & 0.12190 & 0.11653 & 0.11107 & \text{--} & 1.84934 & \\
		      &            & 20 & 0.12276 & 0.11732 & 0.11238 & \text{--} & 1.84793 & \\
		      &            & 22 & 0.12328 & 0.11784 & 0.11320 & \text{--} & 1.84701 & \\
		\midrule
		$E_8$ & Both & 16 & 0.15523 & 0.15408 & 0.15169 & 0.14824 & 1.80874 & 1.80194 \\
		      &      & 18 & 0.15728 & 0.15572 & 0.15329 & 0.15056 & 1.80612 & \\
		      &      & 20 & 0.15867 & 0.15676 & 0.15426 & 0.15201 & 1.80443 & \\
		      &      & 22 & 0.15953 & 0.15745 & 0.15489 & 0.15294 & 1.80334 & \\
		\bottomrule
	\end{tabular}
	\endgroup
	\caption{Heavy periodic branches of $E_6$, $E_7$, and $E_8$ at zero momentum, $|\lambda|=0.1$, and $L=18$, ordered by increasing energy ($H_1,\ldots,H_{N_H}$). The table lists the binding differences $B_i$ defined in Eq.~\eqref{eq:results-heavy-binding}, the mean ratio $\overline m_H/\overline m_L$, and the exact integrable mass ratio for each phase.}
	\label{fig:heavy-cutoff-L18}
\end{figure}

\begin{figure}[!htbp]
	\centering
	\includegraphics[width=\linewidth]{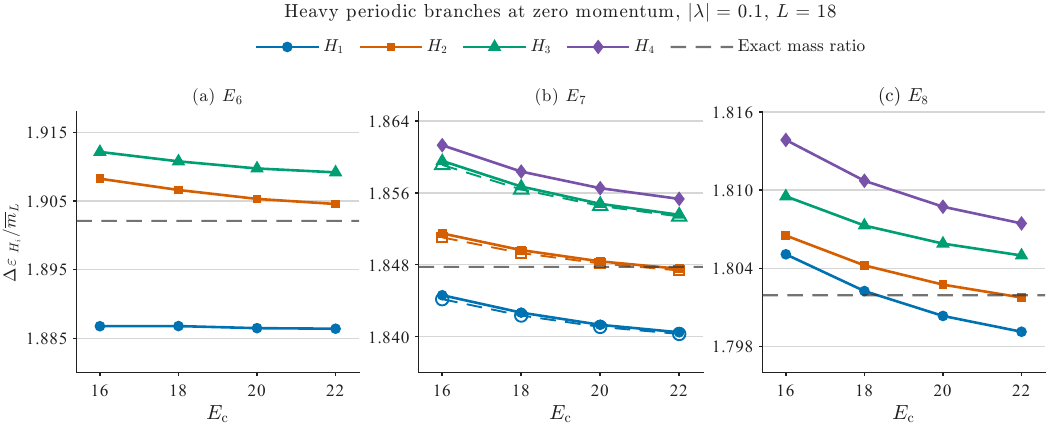}
	\makeatletter
	\par\begingroup\@parboxrestore\normalsize
	\@makecaption{\fnum@figure\space(continued)}{The three panels show the individual mass ratios $\Delta \varepsilon_{H_i}/\overline m_L$ as functions of $\Ec$. In panel~(b) for $E_7$, solid lines with filled markers denote the ordered phase ($H_1$--$H_4$) and dashed lines with open markers denote the disordered phase ($H_1$--$H_3$); in $E_6$ and $E_8$, the two phases have identical periodic spectra. The dashed horizontal lines mark the exact bound-state mass ratios $2\cos(\frac{\pi}{10})$, $2\cos(\frac{\pi}{8})$, and $2\cos(\frac{\pi}{7})$ from Eq.~\eqref{eq:results-heavy-ratio}.}
	\par\endgroup
	\makeatother
\end{figure}

\FloatBarrier

To verify that these bound states remain stable against truncation, we fix $L=18$ at $|\lambda|=0.1$ and increase the conformal energy cutoff through $\Ec=16,18,20,22$. Figure~\ref{fig:heavy-cutoff-L18} shows that all differences $B_i$ remain clearly positive as $\Ec$ increases, changing by less than $0.8\%$ between $\Ec=20$ and $\Ec=22$, which supports the bound-state interpretation of the heavy multiplets. At $L=18$ and $\Ec=22$, the mean gaps $(\overline m_L,\overline m_H)$ are $(0.74526,1.41603)$ in $E_6$; $(0.77194,1.42749)$ in ordered $E_7$; $(0.77201,1.42592)$ in disordered $E_7$; and $(0.79425,1.43230)$ in $E_8$. Across all six phases, $\overline m_H/\overline m_L$ shifts by at most $0.0011$ between $\Ec=20$ and $\Ec=22$, and agrees with Eq.~\eqref{eq:results-heavy-ratio} to within $0.11\%$ at $\Ec=22$. Together, the periodic multiplicities, volume convergence, and mass ratios are consistent with the heavy quiver $Q_0^E=\mathbf 1$, with one neutral particle above each vacuum.

\FloatBarrier
\section*{Acknowledgments}
\begingroup\emergencystretch=2em
J.~Chen is supported by
the Fujian Provincial Natural Science Foundation of China (No.~2025J01004)
and the National Natural Science Foundation of China (Grant No.~12247103).
T.C. is supported by the Xiamen University ``Undergraduate Innovation
Training Program'' -- Innovation Training (No.~2025X1372). We acknowledge assistance from generative AI with drafting, numerical coding and checking the manuscript. The authors take full responsibility for the contents of this paper.
\par\endgroup

\section*{Data availability}
The ancillary package provides the complex measurement matrices,
vacuum calibration data and energy spectra used in Section~\ref{sec:results}.
The accompanying code reproduces the vacuum labels, junction matrices,
vacuum pair probabilities and quiver matrices for all six phases,
including the mixed probe comparison in $E_8$.
It also reconstructs the $E_6$ ordered measurement matrices from the
saved actions of the two probes and reproduces the spectral plots and
heavy mass comparisons. The README describes the inputs, software
requirements and commands for reproducing these results and checking
them against the reported values.

\FloatBarrier
\appendix
\section{Numerical probability}
\label{app:numerics}
\begingroup
\setlength{\intextsep}{6pt plus 2pt minus 2pt}
\setlength{\floatsep}{10pt plus 2pt minus 2pt}
\setlength{\abovecaptionskip}{6pt}

Tables~\ref{tab:e6-ordered-raw}--\ref{tab:e8numerical-raw-disordered} list the diagonal vacuum pair probabilities $(E_{ab})_{\xi\xi}$ and their sums $\widehat D_{ab}=\sum_\xi(E_{ab})_{\xi\xi}$ for all six phases, including pairs with $j_{ab}=0$. For each momentum state $\ket{\xi}$, $(E_{ab})_{\xi\xi}$ estimates $p_{ab,\xi}=\sum_\mu |U_{ab;\mu,\xi}|^2$ in Eq.~\eqref{eq:measurement-internal}. Numerically, we find $(E_{ab})_{\xi\xi}=(E_{ba})_{\xi\xi}$ for all six phases, so only pairs $a\leq b$ are shown. Each momentum column sums to one before rounding when off-diagonal pairs are counted twice.

\begin{table}[!htbp]\centering\small
	\setlength{\tabcolsep}{4.0pt}
	\renewcommand{\arraystretch}{1.12}
	\begin{tabular}{crrrrrrrrr}\toprule
		$(a,b)$ & $0_A$ & $0_B$ & $\frac{1}{6}$ & $(\frac{5}{12})_A$ & $(\frac{5}{12})_B$ & $\frac{1}{2}$ & $-\frac{1}{3}$ & $-\frac{1}{4}$ & $\widehat D_{ab}$ \\ \midrule
		$(0,0)$ & 0.0073 & 0.0058 & 0.0060 & 0.0162 & 0.0002 & 0.0108 & 0.0077 & 0.0068 & 0.0608 \\
		$(0,1)$ & 0.1736 & 0.1392 & 0.1035 & 0.1958 & 0.0097 & 0.1359 & 0.1025 & 0.1040 & 0.9641 \\
		$(0,2)$ & 0.0005 & 0.0004 & 0.0005 & 0.0010 & 0.0003 & 0.0007 & 0.0005 & 0.0005 & 0.0045 \\
		$(1,1)$ & 0.2900 & 0.4310 & 0.5731 & 0.1822 & 0.9603 & 0.4333 & 0.5735 & 0.5697 & 4.0131 \\
		$(1,2)$ & 0.1736 & 0.1392 & 0.1035 & 0.1958 & 0.0097 & 0.1359 & 0.1025 & 0.1040 & 0.9641 \\
		$(2,2)$ & 0.0073 & 0.0058 & 0.0060 & 0.0162 & 0.0002 & 0.0108 & 0.0077 & 0.0068 & 0.0608 \\
		\bottomrule\end{tabular}
	\caption{Measured vacuum-pair probabilities $(E_{ab})_{\xi\xi}$ in the ordered phase of $E_6$ at $\Ec=20$, $\Eop=8$, $L=18$, and $\lambda=-0.1$. Column labels match Eq.~\eqref{eq:e6-channels}.}
	\label{tab:e6-ordered-raw}\end{table}

\begin{table}[!htbp]\centering\small
	\setlength{\tabcolsep}{4.0pt}
	\renewcommand{\arraystretch}{1.12}
	\begin{tabular}{crrrrrrrrr}\toprule
		$(a,b)$ & $0_A$ & $0_B$ & $\frac{1}{6}$ & $(\frac{5}{12})_A$ & $(\frac{5}{12})_B$ & $\frac{1}{2}$ & $-\frac{1}{3}$ & $-\frac{1}{4}$ & $\widehat D_{ab}$ \\ \midrule
		$(0,0)$ & 0.0716 & 0.0246 & 0.1460 & 0.2736 & 0.0164 & 0.1913 & 0.1470 & 0.1441 & 1.0147 \\
		$(0,1)$ & 0.1732 & 0.1431 & 0.1065 & 0.0398 & 0.1747 & 0.1409 & 0.1068 & 0.1074 & 0.9923 \\
		$(0,2)$ & 0.0646 & 0.1821 & 0.1368 & 0.1442 & 0.1249 & 0.0223 & 0.1361 & 0.1360 & 0.9470 \\
		$(1,1)$ & 0.0350 & 0.0140 & 0.0083 & 0.0053 & 0.0188 & 0.0091 & 0.0068 & 0.0102 & 0.1075 \\
		$(1,2)$ & 0.1732 & 0.1431 & 0.1065 & 0.0398 & 0.1747 & 0.1409 & 0.1068 & 0.1074 & 0.9923 \\
		$(2,2)$ & 0.0716 & 0.0246 & 0.1460 & 0.2736 & 0.0164 & 0.1913 & 0.1470 & 0.1441 & 1.0147 \\
		\bottomrule\end{tabular}
	\caption{Measured vacuum-pair probabilities $(E_{ab})_{\xi\xi}$ in the disordered phase of $E_6$ at $\Ec=20$, $\Eop=8$, $L=18$, and $\lambda=+0.1$. Column labels match Eq.~\eqref{eq:e6-channels}.}
	\label{tab:e6-disordered-raw}\end{table}

\begin{table}[!htbp]
	\centering\small
	\setlength{\tabcolsep}{3.0pt}
	\renewcommand{\arraystretch}{1.08}
	\resizebox{\linewidth}{!}{%
	\begin{tabular}{c*{11}{r}}
		\toprule
		$(a,b)$ & $0_A$ & $0_B$ & $0_C$ & $\frac{1}{9}$ & $-\frac{2}{9}$ & $-\frac{1}{3}$ & $\frac{1}{3}$ & $-\frac{4}{9}$ & $(\frac{4}{9})_A$ & $(\frac{4}{9})_B$ & $\widehat D_{ab}$ \\
		\midrule
		$(0,0)$ & 0.0023 & 0.0385 & 0.0383 & 0.1044 & 0.0562 & 0.2392 & 0.1029 & 0.0016 & 0.3146 & 0.0053 & 0.9032 \\
		$(0,1)$ & 0.1076 & 0.0217 & 0.1978 & 0.0837 & 0.0460 & 0.1204 & 0.0120 & 0.0552 & 0.1232 & 0.1735 & 0.9411 \\
		$(0,2)$ & 0.0687 & 0.1217 & 0.0296 & 0.1209 & 0.1131 & 0.1301 & 0.1082 & 0.0381 & 0.0770 & 0.1174 & 0.9247 \\
		$(0,3)$ & 0.0001 & 0.0001 & 0.0001 & 0.0001 & 0.0001 & 0.0001 & 0.0002 & 0.0001 & 0.0001 & 0.0001 & 0.0010 \\
		$(1,1)$ & 0.0168 & 0.0126 & 0.0245 & 0.0131 & 0.0109 & 0.0204 & 0.0048 & 0.0142 & 0.0185 & 0.0373 & 0.1730 \\
		$(1,2)$ & 0.0313 & 0.2069 & 0.0802 & 0.1016 & 0.1121 & 0.0381 & 0.0631 & 0.1609 & 0.0536 & 0.1719 & 1.0196 \\
		$(1,3)$ & 0.0001 & 0.0002 & 0.0002 & 0.0004 & 0.0005 & 0.0002 & 0.0007 & 0.0007 & 0.0003 & 0.0002 & 0.0035 \\
		$(2,2)$ & 0.3373 & 0.1227 & 0.0359 & 0.1673 & 0.1983 & 0.0615 & 0.1677 & 0.0572 & 0.0284 & 0.0310 & 1.2073 \\
		$(2,3)$ & 0.1116 & 0.0611 & 0.1398 & 0.0497 & 0.0924 & 0.0486 & 0.1710 & 0.1995 & 0.0624 & 0.0003 & 0.9364 \\
		$(3,3)$ & 0.0049 & 0.0027 & 0.0060 & 0.0025 & 0.0061 & 0.0040 & 0.0141 & 0.0179 & 0.0055 & 0.0000 & 0.0636 \\
		\bottomrule
	\end{tabular}}
	\caption{Measured vacuum-pair probabilities $(E_{ab})_{\xi\xi}$ in the ordered phase of $E_7$ at $\Ec=20$, $\Eop=8$, $L=18$, and $\lambda=-0.1$. Column labels match Eq.~\eqref{eq:e7-channels}.}
	\label{tab:e7-ordered-raw}
\end{table}

\begin{table}[!htbp]
	\centering\small
	\setlength{\tabcolsep}{3.5pt}
	\renewcommand{\arraystretch}{1.08}
	\resizebox{\linewidth}{!}{%
	\begin{tabular}{c*{10}{r}}
		\toprule
		$(a,b)$ & $0_B$ & $0_C$ & $\frac{1}{9}$ & $-\frac{2}{9}$ & $-\frac{1}{3}$ & $\frac{1}{3}$ & $-\frac{4}{9}$ & $(\frac{4}{9})_A$ & $(\frac{4}{9})_B$ & $\widehat D_{ab}$ \\
		\midrule
		$(0,0)$ & 0.0104 & 0.0029 & 0.0041 & 0.0029 & 0.0154 & 0.0066 & 0.0001 & 0.0212 & 0.0003 & 0.0639 \\
		$(0,1)$ & 0.2366 & 0.0712 & 0.0831 & 0.0492 & 0.1896 & 0.0814 & 0.0051 & 0.2431 & 0.0131 & 0.9723 \\
		$(0,2)$ & 0.0004 & 0.0007 & 0.0007 & 0.0007 & 0.0010 & 0.0008 & 0.0004 & 0.0008 & 0.0009 & 0.0064 \\
		$(1,1)$ & 0.4513 & 0.4043 & 0.4900 & 0.3995 & 0.4027 & 0.1605 & 0.3846 & 0.2973 & 0.9473 & 3.9374 \\
		$(1,2)$ & 0.0248 & 0.1731 & 0.1300 & 0.1799 & 0.0644 & 0.2083 & 0.1553 & 0.0509 & 0.0117 & 0.9984 \\
		$(2,2)$ & 0.0147 & 0.1030 & 0.0784 & 0.1380 & 0.0719 & 0.2518 & 0.2938 & 0.0920 & 0.0010 & 1.0445 \\
		\bottomrule
	\end{tabular}}
	\caption{Measured vacuum-pair probabilities $(E_{ab})_{\xi\xi}$ in the disordered phase of $E_7$ at $\Ec=20$, $\Eop=8$, $L=18$, and $\lambda=+0.1$. Column labels match Eq.~\eqref{eq:e7-channels} (omitting $0_A$).}
	\label{tab:e7-disordered-raw}
\end{table}

\begin{table}[!htbp]
	\centering\small
	\setlength{\tabcolsep}{2.7pt}
	\renewcommand{\arraystretch}{1.06}
	\resizebox{\linewidth}{!}{%
		\begin{tabular}{c*{12}{r}}
			\toprule
			$(a,b)$ & $0_A$ & $0_B$ & $0_C$ & $\frac{1}{15}$ & $-\frac{1}{5}$ & $\frac{4}{15}$ & $-\frac{1}{3}$ & $-\frac{2}{5}$ & $\frac{2}{5}$ & $(\frac{7}{15})_A$ & $(\frac{7}{15})_B$ & $\widehat D_{ab}$ \\
			\midrule
			$(0,0)$ & 0.0037 & 0.0065 & 0.0043 & 0.0013 & 0.0047 & 0.0093 & 0.0061 & 0.0116 & 0.0081 & 0.0133 & 0.0000 & 0.0688 \\
			$(0,1)$ & 0.0792 & 0.1394 & 0.0919 & 0.0270 & 0.0703 & 0.1127 & 0.0688 & 0.1244 & 0.0823 & 0.1355 & 0.0000 & 0.9314 \\
			$(0,2)$ & 0.0001 & 0.0002 & 0.0005 & 0.0005 & 0.0009 & 0.0013 & 0.0007 & 0.0011 & 0.0007 & 0.0010 & 0.0000 & 0.0069 \\
			$(0,3)$ & 0.0000 & 0.0000 & 0.0000 & 0.0000 & 0.0000 & 0.0000 & 0.0000 & 0.0000 & 0.0000 & 0.0000 & 0.0000 & 0.0000 \\
			$(1,1)$ & 0.1113 & 0.3565 & 0.0246 & 0.0878 & 0.1562 & 0.1704 & 0.0626 & 0.0514 & 0.0349 & 0.0220 & 0.0013 & 1.0788 \\
			$(1,2)$ & 0.0042 & 0.0185 & 0.1112 & 0.1317 & 0.1795 & 0.1741 & 0.0712 & 0.1045 & 0.0691 & 0.1143 & 0.0118 & 0.9902 \\
			$(1,3)$ & 0.0003 & 0.0002 & 0.0006 & 0.0005 & 0.0008 & 0.0007 & 0.0009 & 0.0005 & 0.0009 & 0.0004 & 0.0017 & 0.0074 \\
			$(2,2)$ & 0.2950 & 0.2101 & 0.4701 & 0.4449 & 0.2688 & 0.1484 & 0.2697 & 0.3656 & 0.1970 & 0.3432 & 0.9284 & 3.9413 \\
			$(2,3)$ & 0.2068 & 0.0541 & 0.0455 & 0.0718 & 0.0328 & 0.0455 & 0.1819 & 0.0531 & 0.2173 & 0.0572 & 0.0214 & 0.9873 \\
			$(3,3)$ & 0.0088 & 0.0023 & 0.0016 & 0.0032 & 0.0018 & 0.0034 & 0.0148 & 0.0042 & 0.0197 & 0.0048 & 0.0003 & 0.0649 \\
			\bottomrule
	\end{tabular}}
	\caption{Measured vacuum-pair probabilities $(E_{ab})_{\xi\xi}$ in the ordered phase of $E_8$ at $\Ec=20$, $\Eop=8$, $L=18$, and $\lambda=-0.1$. Column labels match Eq.~\eqref{eq:e8-channels}.}
	\label{tab:e8numerical-raw-ordered}
\end{table}

\begin{table}[!htbp]
	\centering\small
	\setlength{\tabcolsep}{2.7pt}
	\renewcommand{\arraystretch}{1.06}
	\resizebox{\linewidth}{!}{%
		\begin{tabular}{c*{12}{r}}
			\toprule
			$(a,b)$ & $0_A$ & $0_B$ & $0_C$ & $\frac{1}{15}$ & $-\frac{1}{5}$ & $\frac{4}{15}$ & $-\frac{1}{3}$ & $-\frac{2}{5}$ & $\frac{2}{5}$ & $(\frac{7}{15})_A$ & $(\frac{7}{15})_B$ & $\widehat D_{ab}$ \\
			\midrule
			$(0,0)$ & 0.0240 & 0.0120 & 0.0827 & 0.0417 & 0.1060 & 0.1677 & 0.1015 & 0.1829 & 0.1210 & 0.1992 & 0.0001 & 1.0388 \\
			$(0,1)$ & 0.0383 & 0.0181 & 0.1404 & 0.0708 & 0.1397 & 0.1773 & 0.0821 & 0.1107 & 0.0734 & 0.0971 & 0.0003 & 0.9481 \\
			$(0,2)$ & 0.0009 & 0.0005 & 0.0023 & 0.0021 & 0.0033 & 0.0034 & 0.0013 & 0.0021 & 0.0011 & 0.0021 & 0.0004 & 0.0196 \\
			$(0,3)$ & 0.0001 & 0.0001 & 0.0001 & 0.0001 & 0.0002 & 0.0002 & 0.0001 & 0.0001 & 0.0001 & 0.0001 & 0.0001 & 0.0011 \\
			$(1,1)$ & 0.2746 & 0.1896 & 0.0975 & 0.1429 & 0.1433 & 0.0819 & 0.0150 & 0.0174 & 0.0128 & 0.0362 & 0.0181 & 1.0294 \\
			$(1,2)$ & 0.1747 & 0.0971 & 0.0454 & 0.0982 & 0.0975 & 0.0723 & 0.0169 & 0.1280 & 0.0142 & 0.1361 & 0.1669 & 1.0475 \\
			$(1,3)$ & 0.0550 & 0.1288 & 0.0111 & 0.1044 & 0.0870 & 0.0812 & 0.1231 & 0.0436 & 0.1102 & 0.0201 & 0.1110 & 0.8755 \\
			$(2,2)$ & 0.0426 & 0.0442 & 0.0441 & 0.0294 & 0.0181 & 0.0118 & 0.0247 & 0.0284 & 0.0239 & 0.0304 & 0.0675 & 0.3649 \\
			$(2,3)$ & 0.0384 & 0.1291 & 0.1750 & 0.0743 & 0.0217 & 0.0081 & 0.0936 & 0.0706 & 0.0862 & 0.0780 & 0.1750 & 0.9498 \\
			$(3,3)$ & 0.0441 & 0.0069 & 0.0272 & 0.0864 & 0.0338 & 0.0536 & 0.2247 & 0.0612 & 0.2720 & 0.0671 & 0.0069 & 0.8837 \\
			\bottomrule
	\end{tabular}}
	\caption{Measured vacuum-pair probabilities $(E_{ab})_{\xi\xi}$ in the disordered phase of $E_8$ at $\Ec=20$, $\Eop=8$, $L=18$, and $\lambda=+0.1$. Column labels match Eq.~\eqref{eq:e8-channels}.}
	\label{tab:e8numerical-raw-disordered}
\end{table}
\FloatBarrier
\endgroup

\section{Preserved symmetries and their action on the vacua}
\label{app:preserved-symmetry}
\label{app:B}

We collect the symmetry data needed to distinguish the defect labels
and their action on the exceptional vacua. Throughout this appendix,
$1$ denotes the identity defect, juxtaposition denotes fusion, and
addition in a fusion rule denotes a direct sum. The full preserved
category $\mathcal C^E$ is given by
Eq.~\eqref{eq:orbifold-preserved-category}, with the algebra in
Table~\ref{tab:theory-models}. Its product is fusion over $\mathcal A$,
and its identity object is the bimodule $\mathcal A$.
The algebra includes its multiplication junctions.
For $E_7$ it is a noncommutative Frobenius algebra, whose construction
is given in~\cite{Fuchs:2002cm}.

\subsection{Preserved defects and vacuum modules}

For a simple UV defect $L$, let $\widehat L$ be its spatial loop
operator and $d_L$ its quantum dimension. The transparency criterion
for the perturbation is~\cite{Chang:2018iay}
\begin{align}
 \widehat L\ket{\phi_{2,1}}=d_L\ket{\phi_{2,1}}\,.
 \label{eq:app-preserved-transparency}
\end{align}
Applying this criterion to the Virasoro preserving defects of the
exceptional CFT gives the same bimodule category as gauging the
preserved parent symmetry~\cite{Coquereaux:2001di,Fuchs:2002cm,Diatlyk:2023fwf,Tanaka:2025qou}.

The object count can also be read from the exceptional defect fusion
data~\cite{Coquereaux:2001di}. Before including the first Kac label, these data contain
$12$, $17$ and $32$ simple objects for $E_6$, $E_7$ and $E_8$.
Each object has a parity inherited from integer or half integer parent
spin. Write this parity as $p(\omega)\in\{0,1\}$ for a defect label
$\omega$, and choose the first Kac label
$r=1,3,\ldots,m-2$ to remove the Kac identification.
If $\lambda_L(\phi_{2,1})$ denotes the loop eigenvalue on the
perturbing field, the minimal model modular data give
\begin{align}
 \frac{\lambda_L(\phi_{2,1})}{d_L}
 =(-1)^{p(\omega)}
   \frac{\cos\left(\frac{\pi r}{m}\right)}
        {\cos\left(\frac{\pi}{m}\right)}\,.
 \label{eq:app-preserved-parity}
\end{align}
Thus Eq.~\eqref{eq:app-preserved-transparency} selects $r=1$ and
even parity, leaving $6$, $9$ and $16$ simple lines.
The automorphism used in the $E_7$ construction preserves this
parity~\cite{Coquereaux:2001di}. The selection applies to either sign of the
coupling.

To describe symmetry breaking, we use the vacuum modules
$\mathcal M^E_{\rm ord.}$ and $\mathcal M^E_{\rm dis.}$ of
Eq.~\eqref{eq:orbifold-vacuum-modules}. For a line $L$, define its
vacuum action matrix by
\begin{align}
 L\otimes_{\mathcal A}w_b
 &\simeq\bigoplus_a(N_L)_{ab}\,w_a\,,
 &
 (N_L)_{ab}
 &=\dim\operatorname{Hom}_{\mathcal A}
       (L\otimes_{\mathcal A}w_b,w_a)\,.
 \label{eq:app-vacuum-action}
\end{align}
Rows label the output vacuum and columns the input vacuum.
For the defect used in this paper, Eq.~\eqref{eq:def_y} gives
$N_y=(Q_1^E)^{\mathsf T}=Q_1^E$, since the displayed quivers are
symmetric. These matrices have nonnegative integer entries and obey
the defect fusion rules.

A regular vacuum module is the category acting on itself and
describes complete spontaneous symmetry breaking~\cite{Cordova:2024iti}.
The full exceptional categories act on fewer vacua than their
numbers of simple lines, so their vacuum modules are nonregular.
It is useful to restrict to the fusion subcategory generated by $y$,
which we denote by $\mathcal S_n\subset\mathcal C^{E_n}$.
Table~\ref{tab:app-symmetry-modules} summarizes this restriction.
In particular, a subcategory can act regularly even when the action
of the full preserved category is nonregular.

\begin{table}[tbp]
\centering\small
\setlength{\tabcolsep}{6pt}
\renewcommand{\arraystretch}{1.15}
\begin{tabular}{@{}lcccl@{}}
\toprule
Model & $\#\mathcal C^{E_n}$ & $\#\mathcal S_n$
& Ordered module & Disordered module\\
\midrule
$E_6$ & $6$ & $3$ & regular, $3$ vacua & nonregular, $3$ vacua\\
$E_7$ & $9$ & $6$ & nonregular, $4$ vacua & nonregular, $3$ vacua\\
$E_8$ & $16$ & $4$ & regular, $4$ vacua & nonregular, $4$ vacua\\
\bottomrule
\end{tabular}
\caption{Preserved symmetry categories and their restrictions to
$\mathcal S_n$, the fusion subcategory generated by $y$.
The symbol $\#$ counts simple objects. The last two columns describe
the vacuum modules of $\mathcal S_n$. All six vacuum modules are
nonregular for the full category $\mathcal C^{E_n}$.}
\label{tab:app-symmetry-modules}
\end{table}

\subsection{\texorpdfstring{$E_6$}{E6}}

The full preserved category has simple lines
\begin{align}
 \operatorname{Irr}(\mathcal C^{E_6})
 &=\{1,x,y,\widetilde y,z,xz\}\,,
 \notag\\
 (d_1,d_x,d_y,d_{\widetilde y},d_z,d_{xz})
 &=(1,1,1+\sqrt3,1+\sqrt3,2+\sqrt3,2+\sqrt3)\,.
 \label{eq:app-e6-lines}
\end{align}
Here $\operatorname{Irr}$ denotes the set of simple objects.
In the notation $L_{(r,a,b)}$ of~\cite{Tanaka:2025qou}, $r$ is the first
Kac label and $a=1,\ldots,6$, $b=1,2$ label the two exceptional
graph factors. The dictionary is
\begin{align}
 (1,x,y,\widetilde y,z,xz)
 =\bigl(L_{(1,1,1)},L_{(1,5,1)},L_{(1,6,2)},
        L_{(1,3,1)},L_{(1,2,2)},L_{(1,4,2)}\bigr)\,.
 \label{eq:app-e6-labels}
\end{align}
The two lines $y$ and $\widetilde y$ arise from the two choices of
braiding in the induction of $X_1$.
They generate two half $E_6$ subcategories sharing the line $x$.
With $y$ fixed as in Eq.~\eqref{eq:def_y}, the full fusion rules are
\begin{align}
 x^2&=1\,,&
 xy&=y\,,&
 x\widetilde y&=\widetilde y\,,\notag\\
 y^2&=1+x+2y\,,&
 \widetilde y^2&=1+x+2\widetilde y\,,&
 y\widetilde y&=z+xz\,,\notag\\
 yz=y(xz)&=\widetilde y+z+xz\,,&
 \widetilde yz=\widetilde y(xz)&=y+z+xz\,,\notag\\
 z^2=(xz)^2&=1+y+\widetilde y+z+xz\,,&
 z(xz)&=x+y+\widetilde y+z+xz\,.
 \label{eq:app-e6-fusion}
\end{align}
Products commute at the level of these fusion rules, and
$x(xz)=z$. In particular, $xz$ denotes the simple line $x\otimes z$,
whereas $xy=y$~\cite{Chang:2018iay,Tanaka:2025qou,Diatlyk:2023fwf}.
The subcategory generated by the chosen $y$ is
$\mathcal S_6=\{1,x,y\}$.

In the ordered vacuum order $(w_1,w_3,w_5)$,
the identification $(w_1,w_3,w_5)\leftrightarrow(1,y,x)$ makes
$\mathcal M^{E_6}_{\rm ord.}$ the regular $\mathcal S_6$ module.
The disordered module has vacuum order $(w_2,w_4,w_6)$ and is
the odd component of the $E_6$ graph module.
Its $N_y$ in Eq.~\eqref{eq:orbifold-e6-quivers} differs from the regular
fusion matrix, even though both phases have three vacua.

The full six line action can be expressed using
$Y=N_y=Q_1^{E_6}$ and $X=N_x$:
\begin{align}
 X&=Y^2-2Y-\mathbf1\,,&
 N_{\widetilde y}&=Y\,,&
 N_z&=\mathbf1+Y\,,&
 N_{xz}&=X+Y\,.
 \label{eq:app-e6-full-action}
\end{align}
These formulas apply in either phase, in its stated vacuum order.
They show explicitly that the lines outside $\mathcal S_6$ also
connect distinct vacua. Thus the three vacua realize the full
preserved category through a nonregular module, with the half $E_6$
restriction described above.

The spatial insertion of $z$ on the $y$ defect circle also distinguishes
the repeated momentum channels in Eq.~\eqref{eq:e6-channels}.
In the induced defect convention of Eq.~\eqref{eq:def_y},
the corresponding line eigenvalues are~\cite{Chui:2002bp,Pearce:2024udz}
\begin{align}
 \begin{array}{c|cccc}
 \text{state}&\ket{0_A}&\ket{0_B}&
       \ket{\left(\frac{5}{12}\right)_A}&
       \ket{\left(\frac{5}{12}\right)_B}\\[3pt]\hline
 \widehat z\text{ eigenvalue}&1&2-\sqrt3&1&-1
 \end{array}\,.
 \label{eq:app-e6-channel-charges}
\end{align}
The joint labels of momentum and the spatial $z$ eigenvalue therefore
resolve both pairs in either massive phase.

\subsection{\texorpdfstring{$E_7$}{E7}}

The preserved category has nine simple lines. We name them
\begin{align}
 \operatorname{Irr}(\mathcal C^{E_7})
 =\{1,y_1,y_2,y_3,b,c,u,v,w\}\,,\qquad y_1=y\,.
 \label{eq:app-e7-lines}
\end{align}
The six lines $\mathcal S_7=\{1,y_1,y_2,y_3,b,c\}$ form the even
$D_{10}$ subcategory in the exceptional defect construction~\cite{Coquereaux:2001di}.
The three $y_i$ are permuted by the triality of its fusion rules.
Write $S=y_1+y_2+y_3$. Besides multiplication by the identity,
their fusion table is specified by
\begin{align}
 y_i^2&=1+y_i+b\,,&
 y_i y_j&=c+y_k\quad(\{i,j,k\}=\{1,2,3\})\,,\notag\\
 y_i b&=y_i+b+c\,,&
 y_i c&=\sum_{j\ne i}y_j+b+c\,,\notag\\
 b^2&=1+S+b+c\,,&
 bc&=S+b+2c\,,\notag\\
 c^2&=1+S+2b+2c\,.
 \label{eq:app-e7-subfusion}
\end{align}
All products commute at the level of the fusion rules.
Each $y_i$ generates this subcategory.
The remaining three lines are also preserved by the perturbation.
Some products that connect them to $\mathcal S_7$ are
\begin{align}
 y_i u&=u+v\,,&
 y_i v&=u+2v+w\,,&
 y_i w&=v\,,\notag\\
 u^2&=1+S+c\,,&
 uw&=b+c\,,&
 w^2&=1+c\,.
 \label{eq:app-e7-mixed-fusion}
\end{align}
This notation specifies the nine objects through the fusion
presentation of~\cite{Coquereaux:2001di}.

For completeness, their quantum dimensions can be written using
$\theta=\frac{\pi}{18}$ and $\delta=2\cos\theta$:
\begin{align}
 d_{y_i}&=\frac{\sin(3\theta)}{\sin\theta}\,,&
 d_b&=\frac{\sin(5\theta)}{\sin\theta}\,,&
 d_c&=\frac{\sin(7\theta)}{\sin\theta}\,,\notag\\
 d_u&=\delta^2\,,&
 d_v&=\delta\,\frac{\sin(4\theta)}{\sin\theta}\,,&
 d_w&=\delta\,
 \frac{\sin(6\theta)-\sin(4\theta)}{\sin\theta}\,.
 \label{eq:app-e7-dimensions}
\end{align}
Only the identity has dimension one.

To make the vacuum action explicit, put $Y=Q_1^{E_7}$ in each phase,
using the vacuum orders of Eq.~\eqref{eq:theory-e7-quivers}.
We choose the names of $y_2,y_3$ so that
\begin{align}
 N_{y_1}=N_{y_2}&=Y\,,&
 N_b&=Y^2-Y-\mathbf1\,,\notag\\
 N_c&=Y^3-2Y^2-Y+\mathbf1\,,&
 N_{y_3}&=Y^2-N_c\,.
 \label{eq:app-e7-action}
\end{align}
These nonnegative integer matrices obey
Eq.~\eqref{eq:app-e7-subfusion}. In the disordered phase
$N_{y_3}=Y$ as well. The ordered and disordered modules have four
and three vacua, respectively, so both are nonregular modules
of the six object category $\mathcal S_7$.

The spatial insertion of $u$ on the $y$ defect circle distinguishes
the repeated momentum channels in Eq.~\eqref{eq:e7-channels}.
In the induced defect convention of Eq.~\eqref{eq:def_y},
the corresponding line eigenvalues are~\cite{Coquereaux:2001di,Chui:2002bp}
\begin{align}
 \begin{array}{c|ccccc}
 \text{state}&\ket{0_A}&\ket{0_B}&\ket{0_C}&
       \ket{\left(\frac{4}{9}\right)_A}&
       \ket{\left(\frac{4}{9}\right)_B}\\[3pt]\hline
 \widehat u\text{ eigenvalue}&0&
 4\cos^2\left(\frac{7\pi}{18}\right)&
 4\cos^2\left(\frac{5\pi}{18}\right)&
 4\cos^2\left(\frac{7\pi}{18}\right)&
 -4\cos^2\left(\frac{7\pi}{18}\right)
 \end{array}\,.
 \label{eq:app-e7-channel-charges}
\end{align}
The joint labels of momentum and the spatial $u$ eigenvalue therefore
resolve both groups in either massive phase.

\subsection{\texorpdfstring{$E_8$}{E8}}

The subcategory generated by $y$ has four simple lines,
\begin{align}
 \mathcal S_8=\{1,t,y,ty\}\,,\qquad
 d_t=\frac{1+\sqrt5}{2}\,,\qquad
 d_y=\frac{\sin\left(\frac{3\pi}{30}\right)}
           {\sin\left(\frac{\pi}{30}\right)}\,,\qquad
 d_{ty}=d_t d_y\,.
 \label{eq:app-e8-subcategory}
\end{align}
Here $ty$ is a simple object. Its fusion table is
\begin{align}
 t^2&=1+t\,,&
 t\,y&=ty\,,&
 t(ty)&=y+ty\,,\notag\\
 y^2&=1+y+ty\,,&
 y(ty)&=t+y+2ty\,,&
 (ty)^2&=1+t+2y+3ty\,.
 \label{eq:app-e8-subfusion}
\end{align}

The full sixteen object category has a convenient presentation in
terms of two copies of the $E_8$ graph fusion data~\cite{Coquereaux:2001di}.
Use auxiliary graph labels $\{1,t,a,ta,p,tp,q,tq\}$, whose even
part $\{1,t,a,ta\}$ obeys Eq.~\eqref{eq:app-e8-subfusion} with
$y$ replaced by $a$. The remaining products are fixed by
\begin{align}
 p^2&=1+a\,,&
 pq&=ta\,,&
 q^2&=1+ta\,,\notag\\
 ap&=p+tq\,,&
 aq&=tp+tq\,.
 \label{eq:app-e8-auxiliary}
\end{align}
Write $(r,s)$ for the product of the two graph copies, with
$(tr,s)=(r,ts)$ identifying their common Fibonacci line $t$.
The sixteen simple preserved lines are
\begin{align}
 \operatorname{Irr}(\mathcal C^{E_8})
 &=\mathcal Z\ \cup\ t\mathcal Z\,,\notag\\
 \mathcal Z
 &=\{(1,1),(a,1),(1,a),(a,a),
       (p,p),(p,q),(q,p),(q,q)\}\,,
 \label{eq:app-e8-lines}
\end{align}
where $t\mathcal Z$ denotes the eight further simple objects obtained
by multiplying the first entry by $t$.
Products are computed by
$(r,s)(r',s')=(rr',ss')$, expanding each graph product and using
$(tr,s)=(r,ts)$.
The first graph copy is chosen so that the defect used in this paper
is $y=(a,1)$. The auxiliary odd labels $p,q,tp,tq$ enter the preserved
category through the pairs in Eq.~\eqref{eq:app-e8-lines}.

For either phase, put $Y=Q_1^{E_8}$ with the orders of
Eq.~\eqref{eq:theory-e8-quivers}. The other two nontrivial actions in
$\mathcal S_8$ are
\begin{align}
 N_t&=Y^3-3Y^2+2\mathbf1\,,&
 N_{ty}&=Y^2-Y-\mathbf1\,.
 \label{eq:app-e8-action}
\end{align}
They obey Eq.~\eqref{eq:app-e8-subfusion}.
In the ordered phase,
$(w_1,w_3,w_5,w_7)\leftrightarrow(t,ty,y,1)$ identifies the
vacuum module with the regular module of $\mathcal S_8$.
The disordered module is the odd graph component, with
$(w_2,w_4,w_6,w_8)\leftrightarrow(tp,tq,p,q)$.
Its four vacua carry a different action: the diagonal entries
of $N_y$ are $(1,1,1,0)$, compared with $(0,2,1,0)$ in the ordered
phase. This gives two different symmetry breaking patterns with
the same number of vacua.

The spatial insertion of the simple line $(p,p)$ on the $y$ defect
circle distinguishes the repeated momentum channels in
Eq.~\eqref{eq:e8-channels}. In the induced defect convention of
Eq.~\eqref{eq:def_y}, the corresponding line eigenvalues are~\cite{Coquereaux:2001di,Chui:2002bp}
\begin{align}
 \begin{array}{c|ccccc}
 \text{state}&\ket{0_A}&\ket{0_B}&\ket{0_C}&
       \ket{\left(\frac{7}{15}\right)_A}&
       \ket{\left(\frac{7}{15}\right)_B}\\[3pt]\hline
 \widehat{(p,p)}\text{ eigenvalue}&
 4\cos^2\left(\frac{11\pi}{30}\right)&
 4\cos^2\left(\frac{13\pi}{30}\right)&
 4\cos^2\left(\frac{7\pi}{30}\right)&
 4\cos^2\left(\frac{13\pi}{30}\right)&
 -4\cos^2\left(\frac{13\pi}{30}\right)
 \end{array}\,.
 \label{eq:app-e8-channel-charges}
\end{align}
The joint labels of momentum and the spatial $(p,p)$ eigenvalue therefore
resolve both groups in either massive phase.

\FloatBarrier

\end{document}